%% file: main_revised.tex
\documentclass[aoas, preprint]{imsart}

\RequirePackage{amsthm,amsmath,amsfonts,amssymb}
\RequirePackage[authoryear]{natbib}
\RequirePackage[colorlinks,citecolor=blue,urlcolor=blue]{hyperref}%% uncomment this for coloring bibliography citations and linked URLs
\RequirePackage{graphicx}%% uncomment this for including figures
\usepackage[dvipsnames]{xcolor}
\usepackage{bm}
\usepackage{bbm}
\usepackage{float}
\usepackage{enumitem}
\definecolor{mdgreen}{HTML}{2cbe4e}

\startlocaldefs
\theoremstyle{remark}

\newtheorem{proposition}{proposition}

\newtheorem{remark}{Remark}
\endlocaldefs

\begin{document}

\begin{frontmatter}
%\title{Simulation-based Unfolding of Particle Distributions in the Presence of Nuisance Parameters}
%\title{Simultaneous Deconvolutuon and Parameter Estimation for Sets}
\title{Machine Learning-based Unfolding for Cross Section Measurements in the Presence of Nuisance Parameters}
\runtitle{Unfolding in the Presence of Nuisance Parameters}
%\thankstext{T1}{A sample of additional note to the title.}

\begin{aug}
%%%%%%%%%%%%%%%%%%%%%%%%%%%%%%%%%%%%%%%%%%%%%%%
%% Only one address is permitted per author. %%
%% Only division, organization and e-mail is %%
%% included in the address.                  %%
%% Additional information can be included in %%
\author[A]{\fnms{Huanbiao}~\snm{Zhu}\ead[label=e1]{huanbiaz@andrew.cmu.edu}},
\author[B]{\fnms{Krish}~\snm{Desai}\ead[label=e2]{krish.desai@berkeley.edu}},
\author[A]{\fnms{Mikael}~\snm{Kuusela}\ead[label=e3]{mkuusela@andrew.cmu.edu}},
\author[C]{\fnms{Vinicius}~\snm{Mikuni}\ead[label=e4]{vmikuni@hepl.phys.nagoya-u.ac.jp}},
\author[D]{\fnms{Benjamin}~\snm{Nachman}\ead[label=e5]{nachman@stanford.edu}}
\and
\author[A]{\fnms{Larry}~\snm{Wasserman}\ead[label=e6]{larry@stat.cmu.edu}}
%%%%%%%%%%%%%%%%%%%%%%%%%%%%%%%%%%%%%%%%%%%%%%
%% Addresses                                %%
%%%%%%%%%%%%%%%%%%%%%%%%%%%%%%%%%%%%%%%%%%%%%%
\address[A]{Department of Statistics and Data Science, Carnegie Mellon University\printead[presep={,\ }]{e1,e3,e6}}

\address[B]{Department of Physics, University of California, Berkeley\printead[presep={,\ }]{e2}}

\address[C]{Nagoya University, Kobayashi-Maskawa Institute, Japan\printead[presep={,\ }]{e4}}

\address[D]{Department of Particle Physics and Astrophysics, Stanford University; Fundamental Physics Directorate, SLAC National Laboratory\printead[presep={,\ }]{e5}}

\end{aug}
\begin{abstract}
%final version
Statistically correcting measured cross sections for detector effects is an important step across many applications. In particle physics, this inverse problem is known as \textit{unfolding}. In cases with complex instruments, the distortions they introduce are often known only implicitly through simulations of the detector. Modern machine learning has enabled efficient simulation-based approaches for unfolding high-dimensional data. Among these, one of the first methods successfully deployed on experimental data is the \textsc{OmniFold} algorithm, a classifier-based Expectation-Maximization procedure. In practice, however, the forward model is only approximately specified, and the corresponding uncertainty is encoded through nuisance parameters. Building on the well-studied \textsc{OmniFold} algorithm, we show how to extend machine learning-based unfolding to incorporate nuisance parameters. Our new algorithm, called Profile \textsc{OmniFold}, is demonstrated using a Gaussian example as well as a particle physics case study using simulated data from the CMS Experiment at the Large Hadron Collider.

\end{abstract}

\begin{keyword}
\kwd{Ill-posed inverse problems}
\kwd{Fredholm integral equation of the first kind}
\kwd{Simulation-based Inference}
\kwd{Expectation-Maximization algorithm}
\end{keyword}

\end{frontmatter}
%%%%%%%%%%%%%%%%%%%%%%%%%%%%%%%%%%%%%%%%%%%%%%
%% Please use \tableofcontents for articles %%
%% with 50 pages and more                   %%
%%%%%%%%%%%%%%%%%%%%%%%%%%%%%%%%%%%%%%%%%%%%%%
%\tableofcontents

%%%%%%%%%%%%%%%%%%%%%%%%%%%%%%%%%%%%%%%%%%%%%%
%%%% Main text entry area:

\section{Introduction}
Detector effects distort spectra from their true values. Statistically removing these distortions is essential for comparing results across experiments and for facilitating broad, detector-independent analysis of the data. In particle and nuclear physics, this problem is known as \textit{unfolding}. While the problem is general, we will focus on this application area because particle detector responses are highly complex and are typically characterized only implicitly through detailed simulations, making simulation-based approaches particularly relevant. The objective is to recover the underlying distribution (called \emph{differential cross section} in physics) of some physical quantity $x$, referred to as particle-level
(or pre-detector level) truth, from observations of a smeared version $y$, known as detector-level (or reconstructed) data. 

In practice, both $x$ and $y$ can be high-dimensional and their probability densities are related by a Fredholm integral equation of the first kind
\begin{equation}
\label{eq:fred_first_kind}
    p_Y(y)=\int_{\mathcal{X}} k(y,x)p_X(x)dx,
\end{equation}
where $k(y,x)$ is the \textit{response kernel} that models the detector response. Without considering efficiency effects, it can be interpreted as the conditional density of observing smeared $y$ given true $x$, i.e., $k(y,x)=p(y|x)$.
The goal of unfolding is to estimate the true density function $p_X$ given an i.i.d. sample of smeared observations $Y_1,...,Y_n\sim p_Y$. 

Recently, a number of machine learning-based approaches have been proposed to address this problem \citep{Arratia:2021otl,Huetsch:2024quz}, and among those, one of the earliest methods proposed is \textsc{OmniFold}~\citep{Andreassen2020, Andreassen:2021zzk}. \textsc{OmniFold} (OF) is a classifier-based algorithm that iteratively reweights simulated data to match the experimental data. At the population level, \textsc{OmniFold} is an Expectation-Maximization (EM) algorithm \citep{10.2307/2984875}, provided an infinite sample size and optimal classifier. \textsc{OmniFold} has been shown to be effective in high-dimensional settings and successfully applied to experimental data from the Large Hadron Collider (LHC) at CERN and other particle and nuclear physics experiments~\citep{H1:2022LeptonJetCorrelation,H1prelim-22-031,Komiske:2022vxg,LHCb:2023ZTaggedJets,H1:2023fzk,H1prelim-21-031,Song:2023sxb,Pani:2024mgy,CMS-PAS-SMP-23-008,ATLAS:2024xxl,ATLAS:2024jrp, H1:2024leptonjet, ATLAS:2025JetTrackFunctions,  H1:2025unfolding, Badea2025UnbinnedThrust, Huang:2025T2KNeutrino, Canelli2026UnbinnedUnfolding}.

However, one limitation in \textsc{OmniFold}, and all other current machine learning-based methods, is the assumption that the detector response is accurately modeled in the simulation. In practice, this is only approximately true, with the simulation potentially depending on a number of nuisance parameters that can be constrained by the observed data. Specifically, this means we have the following forward model
\begin{equation}
    \label{eq:forward_model_nuisance}
    p_{Y,\theta}(y)=\int_{\mathcal{X}} k_{\theta}(y,x)p_X(x)dx,
\end{equation}
where the response kernel depends on some nuisance parameters $\theta$. In our current experiments, $\theta$ is one- or two-dimensional, but it can be extended to higher dimensions as well. Without a correctly specified response kernel, the solution by \textsc{OmniFold} and other machine learning methods will be biased. Such misspecification can arise from several sources, including incorrect modeling of the detector resolution and biases in the detector response. Traditionally, the impact of these systematic uncertainties has been assessed by repeating the measurement under a set of alternative detector-response models. These alternatives are typically generated by varying nuisance parameters that encode plausible deviations from the nominal response, which is computationally expensive and often conservative. In this paper, we approach this problem by proposing a new algorithm, called Profile \textsc{OmniFold} (POF), for unfolding in the presence of nuisance parameters. POF can be seen as an extension to the original OF algorithm, which iteratively reweights the simulated data, but at the same time simultaneously updates the nuisance parameters. This paper builds upon and expands the preliminary results presented in the NeurIPS Machine Learning and the Physical Sciences (ML4PS) workshop paper \cite{zhu2024multidimensional}.

The rest of this paper is organized as follows. Section~\ref{sec:data} introduces the motivating application based on CMS Open Data and describes the limitations of the existing methods in this setting. In Section \ref{sec:EM_unfolding}, we provide an overview of EM algorithms applied to the unfolding problem, along with recent developments in simulation-based machine learning approaches for this class of problems. Building on this foundation, Section \ref{sec:pof} introduces our new methodology, which addresses the challenge posed by nuisance parameters in the response kernel—a scenario where existing machine learning methods do not apply. In Section \ref{sec:gaussian}, we demonstrate the proposed method using a simulated Gaussian example. Section \ref{sec:open_data} presents an application to publicly available simulated data from the CMS Experiment at the LHC. Finally, Section \ref{sec:discussion} discusses the limitations of our approach and outlines directions for future research. The proofs of the main propositions as well as additional experimental results are provided in the supplementary material.

\section{Data and Motivation}
\label{sec:data}
We use publicly available simulated data from the CMS Experiment at the Large Hadron Collider (LHC)~\citep{CMS:QCDsim1000-1400,CMS:QCDsim1400-1800,CMS:QCDsim1800} to motivate and benchmark the proposed methodology. In high-energy proton-proton collisions, quarks and gluons scatter and produce collimated sprays of particles called jets. The quantity of interest in our setting is the sum of the transverse momenta of the two leading jets per event, where transverse momentum is momentum perpendicular to the beam direction and the leading jets are those with the largest transverse momenta. This quantity is central to many downstream analyses, including searches for new particles.

The particle detector records smeared versions of the jet momenta rather than the true particle-level quantities. The response kernel $k(y,x) = p(y|x)$ encoding this smearing is not available in closed form; it is instead characterized through Monte Carlo simulations, which provide paired samples $\{(X_i', Y_i')\}$ from an approximate forward model $q(y|x)$. The amount and type of smearing is approximately known from these simulations, but the simulations and their data-based calibrations contain a number of nuisance parameters. In our motivating setting, we are primarily sensitive to the effective jet energy resolution, which governs the overall amount of momentum smearing. We introduce a scalar nuisance parameter $\theta$ that multiplicatively scales the smearing in the data relative to the simulation, so that the true response is $p(y|x,\theta)$ while the nominal simulation corresponds to $\theta=1$. A full description of the dataset and nuisance parameter construction is given in Section~\ref{sec:open_data}.

Applying \textsc{OmniFold} with the nominal simulation $(\theta=1)$ when the true $\theta \neq 1$ leads to a biased estimate of $p_X$. In this case, the algorithm recovers the particle-level distribution that best matches the observed data under the misspecified kernel $q(y\mid x)$. This limitation is not unique to \textsc{OmniFold} but applies to all existing machine learning-based unfolding methods~\citep{Datta2019, Bellagente2020, Shmakov2023, Backes2024, Diefenbacher2024, Butter2025_DistributionMappingUnfolding, Butter2025_SPINUP, Butter2025_AnalysisReadyGenUnfold, Barman2025_SparticleML, Petitjean2025_GenerativeJets}, which assume a correctly specified response kernel. In the broader statistical literature, this problem can be viewed as an ill-posed inverse problem with a misspecified forward operator. The presence of nuisance parameters in the forward operator is a common challenge in many scientific applications. Profile \textsc{OmniFold} addresses this issue by treating $\theta$ as an additional unknown in the EM objective and updating it jointly with the particle-level reweighting at each iteration, as described in Section~\ref{sec:pof}.

\section{EM Algorithm for Unfolding}
\label{sec:EM_unfolding}
There is a rich body of literature on solving the Fredholm integral equation of the first kind in Equation \eqref{eq:fred_first_kind}. In particular, an EM algorithm has been widely used to solve this problem. Starting from an initial density estimate $f^{(0)}$, the algorithm constructs a sequence of intermediate density estimates $\{f^{(k)}\}_{k\geq 1}$, where $f^{(k)}$ denotes the current density estimate at iteration $k$. Under suitable conditions, this sequence is expected to converge to a solution of Eq.~\eqref{eq:fred_first_kind}. The update takes the form
\begin{equation}
\label{eq:continuous_EM}
    f^{(k+1)}(x) = f^{(k)}(x)\int\frac{p_Y(y)}{p_Y^{(k)}(y)}p(y|x)dy,
\end{equation}
where
\begin{equation}
    p^{(k)}_Y(y)=\int p(y|x')f^{(k)}(x')dx',
\end{equation}
is the reconstructed density in the observation space induced by the current estimate $f^{(k)}$. The initial estimate is chosen to satisfy $f^{(0)}(x)\geq 0$ for any $x\in\mathcal{X}$ and $\int_{\mathcal{X}} f^{(0)}(x)dx=1$, so that each $f^{(k)}$ remains a valid density function. After $k$ iterations, the unfolded solution is given by $f^{(k)}$. To the best of our knowledge, the earliest description of this algorithm appears in the work of \citet{kondor1983}, which referred to it as the method of convergent weights. They did not derive the algorithm from the EM perspective, but instead based on the intuition that
\begin{equation}
    \left[1-\frac{1}{r^{(k)}(y)}\right]p_Y(y) = \int p(y|x)(p_X(x)-f^{(k)}(x))dx, \;\;\; y\in\mathcal{Y},
\end{equation}
where $r^{(k)}(y):=\frac{p_Y(y)}{p_Y^{(k)}(y)}$ is the ratio of the smearing density and the updated density from the algorithm after $k$ iteration. Therefore, 
by constructing a sequence $r^{(k)}(y)$ that converges to one, the hope is that the corresponding $f^{(k)}$ will converge to $p_X$. Kondor presented the description of the algorithm along with a few examples, but did not establish the convergence property of the algorithm. Subsequently, \citet{multhei1987, multhei1987b, multhei1989, multhei1992} connected the algorithm to the maximization of a concave functional, namely the population-level log-likelihood for a density function $f$ on $\mathcal{X}$
\begin{equation}
    \label{eq:population_log_likelihood}
        \ell(f)=\int p_Y(y)\log\left(\int p(y|x)f(x)dx\right)dy,
\end{equation}
or equivalently, minimization of the Kullback-Leibler divergence \citep{multhei1989}
\begin{equation}
    \label{eq:KL_divergence}
    \text{KL}\left(p_Y,\int p(\cdot|x)f(x)dx\right)=\int p_Y(y)\log\frac{p_Y(y)}{\int p(y|x)f(x)dx}dy,
\end{equation}
with respect to $f$.
In particular, under the assumption that the kernel $p(y|x)$ is strictly positive on the compact support $[0,1]^2$, \citet[Theorem 8]{multhei1987} showed that if $f^{(k)}$ converges to some $\tilde{f}$ with respect to the $L^1$ norm, then $\tilde{f}$ is a maximizer of \eqref{eq:population_log_likelihood}. Subsequently, \citet[Theorem 5]{multhei1992} showed that $p_Y^{(k)}$ converges uniformly to $p_Y$ if $p_X$ is strictly positive. Without assuming compact support, \citet[Theorem 1]{Chae2019} proved that $\text{KL}(p_Y,p_Y^{(k)})\rightarrow \inf_f \text{KL}(p_Y,\int p(y|x)f(x)dx)$, provided there exists a convergent sequence $(f_*^{(k)})_{k\geq 1}$ such that $\text{KL}(p_Y,p_Y^{(k)})\rightarrow \text{KL}(p_Y,\int p(y|x)f_*^{(k)}(x)dx)$. Beyond these results, related work has studied smoothed or regularized variants of the EM algorithm. Specifically, \citet{silverman1990} proposed the expectation-maximization smoothing (EMS) scheme as a smoothed version of the EM algorithm, where a smoothing step is added after each iteration. \citet{Crucinio2023} analyzed the theoretical properties of the EMS scheme and proposed a particle algorithm as a sequential Monte Carlo method to approximate the EMS iteration. In a related line of work, \citet{eggemontLariccia1995, eggemontLariccia1997, eggemont1999} proposed the nonlinearly smoothed EM (NEMS) algorithm, for which the resulting estimator can be characterized as a maximizer of a smoothed likelihood function, and established similar convergence results. \citet{liu2009} developed the functional EM (FEM) algorithm for maximizing a penalized likelihood function and showed that its M-steps can be obtained by solving a set of nonlinear ordinary differential equations.

Before these developments, part of the motivation for understanding the algorithm \eqref{eq:continuous_EM} comes from the well-known results by \citet{shepp1982maximum, doi:10.1080/01621459.1985.10477119}, where they studied the discretized version of \eqref{eq:fred_first_kind} as a model for image reconstruction in Positron Emission Tomography (PET). In particle physics, similar models have also been applied in binned unfolding, where measurements are binned into a histogram (or are naturally represented as discrete, e.g., in images) and the particle-level spectrum is likewise represented as a histogram. 

\subsection{Binned unfolding}
In the binned setting, the particle-level and detector-level spaces are each partitioned into bins, and these two binning schemes need not be the same.
The particle-level bins $\mathcal{X}_1,...,\mathcal{X}_B$ partition the particle-level space in which the target distribution $p_X$ is defined. The detector-level bins, by contrast, partition the observed space after detector effects, including smearing inefficiency and acceptance, have acted on the events. Thus, an event generated in a given particle-level bin may be observed in a different detector-level bin, or may fail to be observed at all. The goal is to estimate the unknown particle-level histogram mean $\bm{\lambda}=[\lambda_1,...,\lambda_B]$, where $\lambda_j=\int_{\mathcal{X}_j}p_X(x)dx$ denotes the expected event count in particle-level bin $\mathcal{X}_j$. The observed data are the detector-level histogram $\mathbf{n}^*=[n_1^*,...,n_D^*]^T$, where $D$ is the number of bins at the detector level. Since events can be modeled as Poisson point processes, each bin count independently follows a Poisson distribution \citep{Kuusela2012StatisticalII, Blobel:2203257}.

Therefore, the likelihood function for $\bm{\lambda}$ given the the observed data $\mathbf{n}^*$ is
\begin{equation}
L(\bm{\lambda}|\mathbf{n}^*)=\prod_{i=1}^D\frac{\left(\sum_{j=1}^BK_{ij}\lambda_j\right)^{n^*_i}}{n_i^*!}e^{-\sum_{j=1}^BK_{ij}\lambda_j},
\end{equation}
where $K$ is a $D\times B$ \textit{response matrix} with entries representing the bin-to-bin smearing probabilities, i.e., $K_{ij}=P(\text{observation in bin } i \;\vert\; \text{true value in bin }j)$.

To obtain the maximum likelihood estimate, a classical approach has been the D'Agostini iteration \citep{DAgostini:1994fjx}. D'Agostini iteration can be viewed as an EM algorithm with early stopping \citep{Kuusela2012StatisticalII}, which is equivalent to the procedure originally proposed in \cite{shepp1982maximum}. The same algorithm has also been known as the Richardson-Lucy algorithm \citep{Richardson:72, 1974AJ.....79..745L}. Specifically, starting from an initial guess $\bm{\lambda}^{(0)}>\bm{0}$, each component of $\hat{\bm{\lambda}}^{(k+1)}$ is updated iteratively by
\begin{equation}
    \label{eq:discrete_EM}
    \hat{\lambda}_j^{(k+1)} = \frac{\hat{\lambda}_j^{(k)}}{\sum_i{K}_{ij}}\sum_{i}\frac{{K}_{ij}{n}^*_i}{\sum_{l}{K}_{il}\hat{\lambda}_l^{(k)}}, \;\;\; j=1,...,B.
\end{equation}
After $k$ iterations, the solution is given by $\hat{\bm{\lambda}}^{(k)}=(\hat{\lambda}_1^{(k)},...,\hat{\lambda}_B^{(k)})$. As $k\rightarrow\infty$, it can be shown that $\hat{\bm{\lambda}}^{(k)}$ converges to the maximum likelihood estimate of $\bm{\lambda}$~\citep{doi:10.1080/01621459.1985.10477119}. Also, since each step of the iteration increases the likelihood monotonically, stopping early in the iterations regularizes the solution. Moreover, comparing with the update rule in \eqref{eq:discrete_EM}, the algorithm \eqref{eq:continuous_EM} can be viewed as a continuous analog of the discrete update in \eqref{eq:discrete_EM}, assuming the full efficiency (i.e. $\sum_i K_{ij}=1$ for all $j$).

\subsection{Unbinned unfolding}
Binned unfolding has been the classical approach in particle and nuclear physics for decades. However, discretization requires pre-specifying the number of bins, which itself is a tuning parameter and can vary between different experiments. Additionally, binning limits the number of observables that can be simultaneously unfolded. This motivates the development of unbinned unfolding, which turns out to be closely related to algorithm~\eqref{eq:continuous_EM}.

As mentioned above, the iterative algorithm~\eqref{eq:continuous_EM} can be derived as an EM algorithm that aims to maximize the population-level log-likelihood.
The idea is that we treat the set of smeared observations $Y$ as the observed variables and the target truth quantities $X$ as the unobserved latent variables. The parameter (function) of interest $f$ is the density function of $X$. The corresponding population-level Q-function is the expected complete-data log-likelihood conditioning on the observed variables $Y$ and the current estimate $f^{(k)}$ integrating with respect to $p_Y$, i.e.,
\begin{equation}
\begin{split}
    Q(f,f^{(k)}) &= \int p_Y(y) \int p(x|y,f^{(k)})\log [p(y|x)f(x)]dxdy \\
    &= \int p_Y(y) \int \frac{p(y|x)f^{(k)}(x)}{\int p(y|x')f^{(k)}(x')dx'}\log [p(y|x)f(x)]dxdy.
\end{split}
\end{equation}
In the EM algorithm, the expectation (E) step computes the function $Q(f,f^{(k)})$ and the maximization (M) step updates the estimate by solving $f^{(k+1)}=\arg\max_{f}Q(f,f^{(k)})$ subject to the constraint $\int f(x)dx=1$.

Proposition \ref{thm:omnifold_update} establishes that solving the optimization problem above yields the algorithm~\eqref{eq:continuous_EM}. This result (and similar variants) has been presented in \cite{multhei1987,Andreassen2020, Falcao2025HighDimensionalUnfolding}. We provide the proof in supplementary material for completeness.
\begin{proposition}
\label{thm:omnifold_update}
    Let $f^{(k+1)}=\arg\max_{f}Q(f,f^{(k)})$ subject to the constraint that $\int f(x)dx=1$. Then
    \begin{equation}
        f^{(k+1)}(x) = f^{(k)}(x)\int\frac{p_Y(y)}{\int p(y|x')f^{(k)}(x')dx'}p(y|x)dy.
    \end{equation}
\end{proposition}

\subsection{Machine-learning based method for unfolding: \textsc{OmniFold}}
Although the EM algorithm \eqref{eq:continuous_EM} provides a principled approach to solving the inverse problem in \eqref{eq:fred_first_kind}, this is challenging to implement in practice for two key reasons: (i) the analytic forms of $p_Y$ and $p(y|x)$ are typically unknown in particle and nuclear physics experiments, and (ii) both distributions may be high-dimensional, making them difficult to estimate. Recently, however, a line of machine learning-based unfolding methods has independently developed variants of algorithm~\eqref{eq:continuous_EM} that circumvent these challenges and enable obtaining solutions to~\eqref{eq:fred_first_kind} even in high-dimensional settings. The first (and so far, only) one to be deployed to experimental data is \textsc{OmniFold}~\citep{Andreassen2020,Andreassen:2021zzk}. The core idea of \textsc{OmniFold} is to use neural network classifiers to estimate the density ratios involved in the EM algorithm~\eqref{eq:continuous_EM}, thereby avoiding explicit estimation of $p_Y$ or $p(y|x)$. This is made possible by access to a set of Monte Carlo (MC) simulated data $\{X_i',Y_i'\}_{i=1}^n\sim q_{X,Y}$, which in particle and nuclear physics is routinely available from high-fidelity detector simulations that mimic the data-generating process. The key assumption here is that $q(y|x)=p(y|x)$, meaning the response kernel (or the forward operator) stays the same between the MC and experimental data. However, it should be noted that the marginal distributions $q_X$ and $p_X$ (and hence $q_Y$ and $p_Y$) are not assumed to be the same. For simplicity, we will omit the subscript $X$ and $Y$ in what follows if there is no confusion. Under this setting, \textsc{OmniFold} approaches the unfolding task as follows: 

Provided pairs of MC simulations $\{X_i',Y_i'\}_{i=1}^n\sim q_{X,Y}$ and a set of observed detector-level data $\{Y_i\}_{i=1}^m \sim p_Y$, let $\nu(x)$ be a reweighting function on the MC particle-level density $q(x)$. The goal is to estimate the true reweighting function $\nu^*(x)=\frac{p(x)}{q(x)}$. By this reparameterization, the population log-likelihood for a reweighting function $\nu$ is
\begin{equation}
    \label{eq:nu_log_likelihood}
    \ell(\nu)=\int p(y)\log\left(\int p(y|x)\nu(x)q(x)dx\right)dy,
\end{equation}
and the corresponding Q-function is
\begin{equation}
    \label{eq:nu_Q_function}
    Q(\nu,\nu^{(k)}) = \int p(y) \int \frac{p(y|x)\nu^{(k)}(x)q(x)}{\int p(y|x')\nu^{(k)}(x')q(x')dx'}\log [p(y|x)\nu(x)q(x)]dxdy.
\end{equation}
Subject to the normalization constraint $\int\nu(x)q(x)dx=1$, the EM update takes the form
\begin{equation}
    \label{eq:nu_update}
    \nu^{(k+1)}(x) = \nu^{(k)}(x)\int\frac{p(y)}{\int\nu^{(k)}(x')q(x',y)dx'}p(y|x)dy.
\end{equation}
\textsc{OmniFold} implements this update via a two-step procedure:
\begin{enumerate}
    \item Detector-level reweighting: \\
    $r^{(k)}(y) = \frac{p(y)}{\tilde{q}^{(k)}(y)}$,
    \text{where }$\tilde{q}^{(k)}(y)=\int \nu^{(k)}(x')q(x',y)dx'$.
    \item Particle-level reweighting: \\
    $\nu^{(k+1)}(x) = \nu^{(k)}(x)\frac{\tilde{q}^{(k)}(x)}{q(x)}$,
    \text{where } $\tilde{q}^{(k)}(x)=\int r^{(k)}(y')q(x,y')dy'$.
\end{enumerate}
Notice that combining these two steps yields the update \eqref{eq:nu_update}. Moreover, since each step involves a density ratio, it can be estimated using a binary classifier without estimation of the marginal densities separately; see Section \ref{sec:density_ratio_classifier} for details. These two steps also have intuitive interpretations: the first step learns to reweight the detector-level MC density $\tilde{q}^{(k)}(y)$ in the current iteration to match the detector-level experimental density $p(y)$. The second step pulls back this density ratio to the particle level and updates the corresponding particle-level weights. In the next iteration, the updated weights are pushed forward to the detector level, and the process repeats. Here, pushing forward (or pulling back) refers to transferring the corresponding weights between paired particle-level and detector-level events. Further details are available in~\cite{Andreassen2020}.

\subsection{Estimating density ratio through binary classification}
\label{sec:density_ratio_classifier}
As a key ingredient
in the two steps of \textsc{OmniFold}, we briefly describe how to estimate a density ratio using a classifier. Given i.i.d. samples $x_1,...,x_n\sim p_1$ and $x_1',...,x_m'\sim p_0$, assign class labels $c=1$ to $\{x_i\}_{i=1}^n$ and $c=0$ to $\{x_i'\}_{i=1}^m$. Additionally, let $w_i$ denote the weight associated with $x_i$. In the context of the algorithm, the weights $w_i$ correspond to the updated weights attached to the $i$-th MC data point during the EM iterations. The weighted density ratio we aim to estimate is
\begin{equation}
    r(x)=\frac{w(x)p_1(x)}{p_0(x)}.
\end{equation}
Using Bayes' rule, we can express
\begin{align*}
    \frac{p(c=1|x)}{p(c=0|x)} &= \frac{w(x)p_1(x)}{p_0(x)}\cdot \frac{p(c=1)}{p(c=0)}
\end{align*}
and hence
\begin{equation}
    r(x)=\frac{p(c=1|x)}{p(c=0|x)}\cdot \frac{p(c=0)}{p(c=1)}.
\end{equation}
A probabilistic classifier $\hat{f}:\mathcal{X}\rightarrow[0,1]$ is trained on the weighted dataset~$\{w_i,x_i, c=1\}_{i=1}^n$ and $\{x_i',c=0\}_{i=1}^m$. The output of the classifier approximates the conditional probability of class $c=1$, i.e., $\hat{f}(x)=\hat{p}(c=1|x)$. On the other hand, the prior odds can be estimated from the weighted sample sizes as 
\begin{equation}
    \frac{p(c=1)}{p(c=0)} \approx \frac{\sum_{i=1}^nw_i}{m}.
\end{equation}
Thus, the density ratio can be estimated as
\begin{equation}
    \hat{r}(x)=\frac{\hat{f}(x)}{1-\hat{f}(x)}\cdot\frac{\sum_{i=1}^nw_i}{m}.
\end{equation}
Under a Bayes optimal classifier and as the sample size tends to infinity, $\hat{r}(x) \to r(x)$. Similar results can also be shown if the weights are associated with $x_i'$ instead of $x_i$. For more details about density ratio estimation using classifiers, see, for example, \cite{Andreassen:2020nnm,Cranmer:2015bka}; Chapter 4 of \cite{sugiyama2012density}.

\subsection{Other machine learning approaches}
Although this work is primarily motivated by \textsc{OmniFold}, which is a classifier-based EM algorithm, other machine learning approaches have also been proposed for unbinned unfolding. Notably, another line of machine learning methods for unfolding uses generative models to learn the conditional distribution of the unfolded events given the observed data~\citep{Datta2019, Bellagente2020, Shmakov2023, Backes2024, Diefenbacher2024, Butter2025_DistributionMappingUnfolding, Butter2025_SPINUP, Butter2025_AnalysisReadyGenUnfold, Barman2025_SparticleML, Petitjean2025_GenerativeJets}. In particular, it proceeds as follows:
Initialize $p^{(0)}(x)=q(x),p^{(0)}(y)=q(y)$, then for iteration $k$,
\begin{enumerate}
    \item Train a generative model for $p^{(k)}(x|y)$ using the generated data at $(k-1)^{th}$ iteration. Conditioning on experimental data $Y_i\sim p_Y$, generate $\tilde{X}^{(k)}_i\sim p^{(k)}(\cdot|Y_i)$. Denote the distribution of $\tilde{X}^{(k)}_1,...,\tilde{X}^{(k)}_n$ as ${p}^{(k)}(x)$.
    \item Estimate $r^{(k)}(x)=\frac{{p}^{(k)}(x)}{q(x)}$. Then reweight the detector-level MC density by $p^{(k)}(y)=r^{(k)}_{push}(y)q(y)$, where $r^{(k)}_{push}(y)=\int r^{(k)}(x)q(x|y)dx$.
\end{enumerate}
At the population level, this approach is equivalent to \textsc{OmniFold}. To see this, notice that
\begin{equation}
\begin{split}
    p^{(k+1)}(x) &= \int p^{(k)}(x|y)p(y)dy \\
    &= \int \frac{p(y|x)p^{(k)}(x)}{p^{(k)}(y)}p(y)dy \\
    &= p^{(k)}(x)\int \frac{p(y)}{p^{(k)}(y)}p(y|x)dy,
\end{split}
\end{equation}
where $p^{(k)}(y)=\int r^{(k)}(x)q(x|y)q(y)dx$. Since $p^{(0)}(x)=q(x)$ and $p^{(0)}(y)=q(y)$, by induction we have $p^{(k)}(y)=\tilde{q}^{(k)}(y)$ for all $k\geq 0$.
This shows that the update for $p^{(k)}(x)$ matches that of $\nu^{(k)}(x)$ in \eqref{eq:nu_update} with $p^{(k)}(x)=\nu^{(k)}(x)q(x)$. While the generative unfolding is still an EM algorithm in population level, most paper have been focusing on one step of training a generative model to learn $p(x|y)$ without iterating the procedure necessarily. Implementation-wise, various generative models have been explored in this context, including generative adversarial networks~\citep{Datta2019,Bellagente2020}, diffusion models \citep{Shmakov2023}, normalizing flows~\citep{Backes2024}, Schrödinger Bridges~\citep{Diefenbacher2024}, and conditional flow matching~\citep{Petitjean2025_GenerativeJets}.

\section{Unfolding in the Presence of Nuisance Parameters in the Forward Operator}
\label{sec:pof}
Although the response kernel is approximately the same between the simulated and experimental data, in practice it may still depend on one or more nuisance parameters. Consider the forward model \eqref{eq:forward_model_nuisance}, where $\theta\in\mathbb{R}^p$ denotes the nuisance parameter. For simplicity, we focus on the case $p=1$ or $p=2$. If the nuisance parameter is misspecified (i.e. $p(y|x)\neq q(y|x)$), the results obtained from \textsc{OmniFold} and other machine learning–based unfolding methods will be biased. To address this, we introduce the Profile \textsc{OmniFold} algorithm, which extends the original \textsc{OmniFold} algorithm to simultaneously update the nuisance parameter $\theta$ while iteratively reweighting the simulation data.

Related work includes the approach of \cite{Chan2023}, which also performs unbinned unfolding with nuisance parameter profiling. Their method uses neural networks to directly maximize the Poisson log-likelihood function.  While a significant step forward, this approach is limited to the case where the particle-level data are unbinned but the detector-level data are binned so that one can write down the explicit Poisson likelihood for bin counts. On the other hand, POF is completely unbinned at both the detector and particle levels, and does not assume any parametric model.

\subsection{Algorithm} 
\label{sec:algorithm}
As in \textsc{OmniFold}, let $\nu(x)$ be a reweighting function on the MC particle-level density $q(x)$. Also, assume $q(y|x)$ is specified by the nuisance parameter $\bar{\theta}$, i.e., $q(y|x)=p(y|x,\bar{\theta})$. Moreover, let $\mathcal{W}(y,x,\theta)$ be a reweighting function on the MC response kernel $q(y|x)$, i.e., $\mathcal{W}(y,x,\theta)=p(y|x,\theta)/q(y|x)$. Then the goal is to maximize the population log-likelihood
%$q(y|x)\hookrightarrow q(y|x)\times \mathcal{W}(y,x,\theta)$.
\begin{equation}
\label{eq:log_likelihood}
\begin{split}
    \ell(\nu,\theta) &= \int p(y)\log p(y|\nu,\theta)dy + \log p_0(\theta) \\
    &\text{subject to }\int \nu(x)q(x)dx=1,
\end{split}
\end{equation}
where $p_0(\theta)$ is a prior on $\theta$ to constrain the nuisance parameter, usually determined from auxiliary measurements. This is not strictly a Bayesian prior, but rather can be viewed as an optional likelihood term for the auxiliary measurements \citep{cranmer2015practical}. One example is the Gaussian likelihood, $\log p_0(\theta)=-\frac{(\theta-\bar{\theta})^2}{2\sigma_0^2}$.

The POF algorithm, like the original OF algorithm, is an EM algorithm. It iteratively updates the reweighting function $\nu(x)$ and nuisance parameter $\theta$ toward the maximum likelihood estimate.  For the log-likelihood specified in \eqref{eq:log_likelihood}, the $Q$ function is given by
\begin{equation}
\begin{split}
    Q(\nu,\theta|\nu^{(k)},\theta^{(k)}) &= \int p(y)\int p(x|y,\nu^{(k)},\theta^{(k)})\log p(x,y|\nu,\theta)dxdy + \log p_0(\theta) \\
    &= \int p(y)\int \frac{\mathcal{W}(y,x,\theta^{(k)})q(y|x)\nu^{(k)}(x)q(x)}{\int \mathcal{W}(y,x',\theta^{(k)})q(y|x')\nu^{(k)}(x')q(x')dx'}\log p(x,y|\nu,\theta)dxdy + \log p_0(\theta) \\
    &\text{subject to }\int \nu(x)q(x)dx=1.
\end{split}
\end{equation}

The E-step in the EM algorithm is to estimate the $Q$ function and M-step is to maximize over $\nu$ and $\theta$. The maximizer will then be used as the updated parameter values in the next iteration. Specifically, in the $k^{\text{th}}$ iteration, we obtain the update $(\nu^{(k+1)},\theta^{(k+1)})$ by solving $(\nu^{(k+1)},\theta^{(k+1)})={\arg\max}_{\nu,\theta}Q(\nu,\theta|\nu^{(k)},\theta^{(k)})$. This optimization problem can be solved separately for $\nu$ and $\theta$, which is described by Proposition \ref{thm:pof_update} below.

\begin{proposition}
\label{thm:pof_update}
    Let 
    \begin{equation}
    \label{eq:nu_update_nuisance}
        \nu^{(k+1)}(x) = \nu^{(k)}(x)\int\frac{p(y)}{\tilde{q}^{(k)}(y)}\mathcal{W}(y,x,\theta^{(k)})q(y|x)dy,
    \end{equation}
    \begin{equation}
    \label{eq:theta_update_nuisance}
        \theta^{(k+1)}=\arg\max_\theta\left[\int\int q(x,y) \nu^{(k)}(x)\mathcal{W}(y,x,\theta^{(k)})\frac{p(y)}{\tilde{q}^{(k)}(y)}\log[\mathcal{W}(y,x,\theta)]dxdy + \log p_0(\theta)\right],
    \end{equation}
    where $\tilde{q}^{(k)}(y)=\int \mathcal{W}(y,x',\theta^{(k)})\nu^{(k)}(x')q(x',y)dx'$. Then $(\nu^{(k+1)},\theta^{(k+1)})=\arg\max_{\nu,\theta}Q(\nu,\theta|\nu^{(k)},\theta^{(k)})$ subject to the constraint that $\int \nu(x)q(x)dx=1$.
\end{proposition}
The proof of the proposition is provided in the supplementary material.

\begin{remark}
    The reason that the optimization problem can be solved separately for $\nu$ and $\theta$ is that the $Q$ function is separable in terms of $\nu$ and $\theta$. Specifically, we can write
    \begin{equation}
    \begin{split}
        Q(\nu,\theta|\nu^{(k)},\theta^{(k)})
        &= \int p(y)\int p(x|y,\nu^{(k)},\theta^{(k)})\log [\nu(x)q(x)q(y|x)]dxdy \\
        &\quad + \int p(y)\int p(x|y,\nu^{(k)},\theta^{(k)})\log [\mathcal{W}(y,x,\theta)]dxdy + \log p_0(\theta) \\
        &= Q_1(\nu|\nu^{(k)},\theta^{(k)}) + Q_2(\theta|\nu^{(k)},\theta^{(k)}).
    \end{split}
    \end{equation}
    Therefore, Equations \eqref{eq:nu_update_nuisance} and \eqref{eq:theta_update_nuisance} correspond to maximizing $Q_1$ and $Q_2$ separately.
\end{remark}
More concretely, the algorithm is initialized with $\nu^{(0)}(x)=1$ for all $x$ (i.e., uniform weights setting the initial solution equal to the MC distribution) and a chosen starting value $\theta^{(0)}$. Each subsequent iteration then proceeds via the following three steps:
\begin{enumerate}
    \item Detector-level reweighting: \\
    $r^{(k)}(y) = \frac{p(y)}{\tilde{q}^{(k)}(y)}$,
    \text{where }$\tilde{q}^{(k)}(y)=\int \mathcal{W}(y,x',\theta^{(k)})\nu^{(k)}(x')q(x',y)dx'$.
    \item Particle-level reweighting: \\
    $\nu^{(k+1)}(x) = \nu^{(k)}(x)\frac{\tilde{q}^{(k)}(x)}{q(x)}$,
    \text{where } $\tilde{q}^{(k)}(x)=\int \mathcal{W}(y',x,\theta^{(k)})r^{(k)}(y')q(x,y')dy'$.
    \item Nuisance parameter update: \\
    $\theta^{(k+1)}=\arg\max_\theta Q_2(\theta|\nu^{(k)},\theta^{(k)})$.
\end{enumerate}

The first step is almost identical to the first step in the original OF algorithm, which involves computing the ratio of the detector-level experimental density and reweighted detector-level MC density using the push-forward weights of $\mathcal{W}(y,x,\theta^{(k)})\nu^{(k)}(x)$. The difference from OF is the presence of the additional term $\mathcal{W}(y,x,\theta^{(k)})$, which is the ratio of the response kernels parametrized by $\theta$. As in the original \textsc{OmniFold}, the density ratio $r^{(k)}(y)$ can still be estimated by training a classifier to distinguish between the experimental data distribution $p(y)$ and the reweighted MC distribution $\tilde{q}^{(k)}(y)$. 

The second step also closely mirrors the second step of the original OF algorithm, which involves computing the ratio of the reweighted particle-level MC density using the pull-back weights $\mathcal{W}(y,x,\theta^{(k)})r^{(k)}(y)$ and the particle-level MC density.

The third step updates the nuisance parameter $\theta$ by numerically optimizing the $Q_2$ function. Several strategies are possible: one can directly optimize the $Q_2$ function or alternatively solve for its stationary condition. The key aspect is that the $Q_2$ function as well as its gradient with respect to $\theta$ are computable up to a constant for different values of $\theta$, which makes the optimization feasible. More details are given in Section \ref{sec:nuisance_update}.

\begin{figure}[h]
  \centering
  \includegraphics[width=0.8\linewidth]{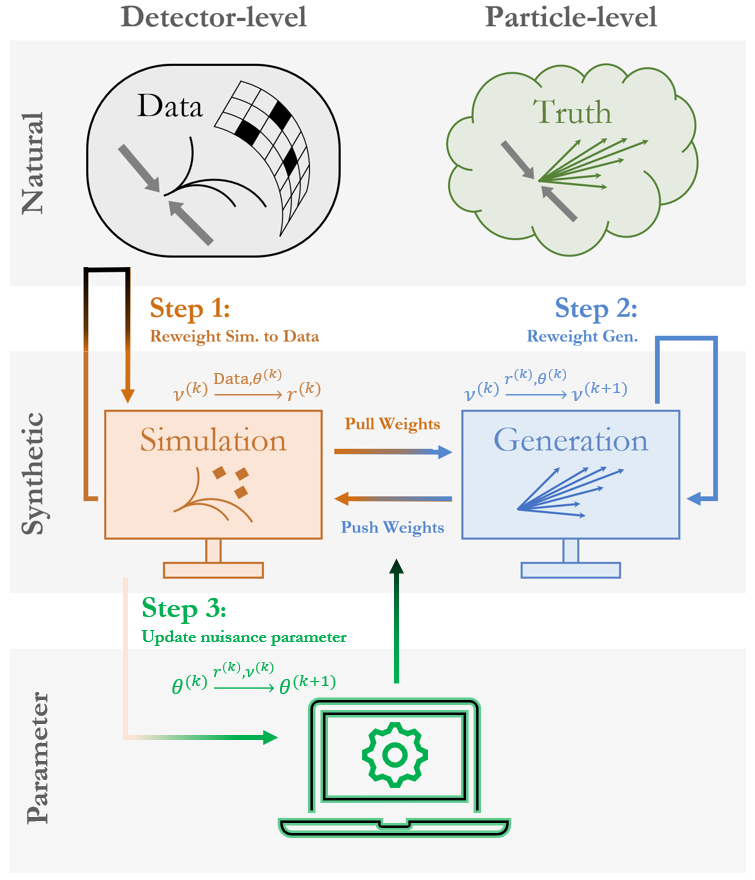}
  \caption{An overview of the POF algorithm. Portions of the image have been adapted from \cite{Andreassen2020} for the original \textsc{OmniFold} algorithm. In step 1, the current particle-level weights $\nu^{(k)}$ are pushed to the detector level with the current nuisance parameters $\theta^{(k)}$, which are used to compute the density ratio $r^{(k)}$. In step 2, the ratio $r^{(k)}$ is pulled pack to the particle level using the same nuisance parameters. In step 3, the nuisance parameters are updated based on the current weights $\nu^{(k)}$ and the density ratio $r^{(k)}$. The procedure is iterated for a fixed number of times.}
  \label{fig:POFoverview}
\end{figure}

In summary, the POF algorithm extends the original OF iteration by introducing an additional step for updating the nuisance parameter. This extension offers several advantages. First, as in OF, the key quantities estimated throughout the procedure are density ratios, which can be efficiently learned via classifiers without estimating each density separately. Second, POF preserves the EM structure, which guarantees that the likelihood is non-decreasing at each iteration under infinite sample size and optimal classifier. Finally, the first two steps closely resemble the OF update, making the algorithm easy to implement as an extension of existing software. An overview of the POF algorithm is illustrated in Figure \ref{fig:POFoverview}.

However, unlike OF, POF does not have a convergence guarantee since the likelihood function is generally not concave when there are nuisance parameters. Empirically, we observe that the algorithm can converge to different solutions depending on the initialization of the nuisance parameter $\theta^{(0)}$, raising the question of which solution should be selected.

To address this issue, we propose to use a goodness-of-fit statistic based on the weighted accuracy of the step-1 classifier. Recall that in Step 1, a classifier is trained to estimate the density ratio $r^{(k)}(y)=\frac{p(y)}{\tilde{q}^{(k)}(y)}$. If the iterative procedure converges to the correct solution, the reweighted distribution $\tilde{q}^{(k)}(y)$ should match the observed distribution $p(y)$. Consequently, the classifier will be trying to distinguish two distributions that are nearly identical. In that case, the classifier's accuracy should approach 0.5. Based on this observation, we propose the following goodness-of-fit statistic:
\begin{equation}
\label{eq:gof}
    V = 1 - 2\cdot \left\vert\frac{\sum_{i=1}^nw_i\mathbbm{1}\{\hat{c}_i=c_i\}}{\sum_{i=1}^nw_i}-0.5\right\vert
\end{equation}
where $w_i$ is the weight assigned to $i^{th}$ observation from the algorithm, $c_i$ is the true class label, and $\hat{c}_i$ is the predicted label. Given $b$ candidate solutions $(\hat{\nu}_1,\hat{\theta}_1),...,(\hat{\nu}_b,\hat{\theta}_b)$ initialized from different starting points, we select the solution $(\hat{\nu}_{i^*},\hat{\theta}_{i^*})$ with the highest statistic, i.e., $i^*=\arg\max_iV_i$. This selection rule should be understood only as a practical diagnostic for comparing repeated runs under the same classifier architecture and training procedure. It is not intended as a general model-selection criterion across different classifiers, architectures, or hyperparameter choices. We also emphasize that the statistic in Eq.~\eqref{eq:gof} evaluates agreement only at the detector level. That is, it tests whether the reweighted detector-level distribution $\tilde{q}^{(k)}(y)$ is consistent with the observed distribution $p(y)$, but it does not directly assess whether the corresponding particle-level distribution recovers the true $p(x)$. In the actual unfolding problem, $p(x)$ is unavailable, so such a direct validation is not possible in real applications. Therefore, we do not regard $V$ as a definitive validation of the unfolded particle-level result. Rather, $V$ is used as a heuristic diagnostic to identify runs that fail to reproduce the observed detector-level data. In our numerical studies, poor initializations often lead to visibly worse detector-level agreement, and the statistic provides a quantitative way to flag such failures. Since $V$ depends on the classifier predictions $\hat{c}_i$, it is also affected by classifier training uncertainties. A principled study of these uncertainties and of the statistical properties of $V$ is beyond the scope of the present work and is left for future investigation.

\subsection{More details on nuisance parameters update}
\label{sec:nuisance_update}

During each iteration, the nuisance parameters are updated according to Eq.~\eqref{eq:theta_update_nuisance}. Note that in the equation, $\mathcal{W}(y,x,\theta^{(k)}),\nu^{(k)}(x)$ were already computed in the previous iteration, $\frac{p(y)}{\tilde{q}^{(k)}(y)}$ is the density ratio being estimated by the step-1 classifier, and $\log p_0(\theta)$ is known. The only unknown term is $\log[\mathcal{W}(y,x,\theta)]$. Combined with the other quantities, this term is integrated with respect to the joint density $q(x,y)$, which can be approximated by a sample average over the MC data. Therefore, the only challenge lies in the estimation of the function   
 $\mathcal{W}(y,x,\theta)$. Recall that $\mathcal{W}(y,x,\theta)$ is defined as the ratio of two conditional densities $\frac{p(y|x,\theta)}{q(y|x)}$. \cite{Chan2023} proposed to learn $\mathcal{W}(y,x,\theta)$ by factorizing it into the product of two density ratios
 \begin{equation}
     \mathcal{W}(y,x,\theta)=\frac{p(y|x,\theta)}{q(y|x)}=\frac{p(x,y|\theta)}{q(x,y)}\cdot\frac{q(x)}{p(x|\theta)},
 \end{equation}
where $\frac{p(x,y|\theta)}{q(x,y)}$ and $\frac{q(x)}{p(x|\theta)}$ can be estimated separately using the classifiers. To learn this function, an additional set of synthetic data $\{X_i,\theta_i,Y_i\}$\footnote{We slightly abuse the notation here and use $X_i,Y_i$ to denote the synthetic data, but this should not be confused with the experimental data.} is required, where the density of $Y_i$ is given by $p_{Y_i,\theta_i}(y)=\int p(y|x,\theta_i)p_{X_i,\theta_i}(x)dx$. In this setup, the choice of the distribution for $X_i$ is flexible, as long as its support covers the data domain. One practical option is to use particle-level MC samples for $X_i$ and generate the corresponding $Y_i$ by applying forward models parametrized by different $\theta_i$. Proposition~\ref{thm:w_training} formalizes this procedure.

\begin{proposition}
\label{thm:w_training}
    Let $X_i\sim p_X, \theta_i\sim p_\theta, Y_i\sim p(\cdot|X_i,\theta_i)$, and $X_i'\sim q_X, \theta_i'\sim q_\theta, Y_i'\sim q(\cdot|X_i)$. Let $f_1:\mathcal{X}\times\mathcal{Y}\times\Theta\rightarrow[0,1]$ be the Bayes optimal classifier distinguishing dataset $\mathcal{D}_1=\{X_i,Y_i,\theta_i\}$ from $\mathcal{D}_2=\{X'_i,Y'_i,\theta_i'\}$. Let $f_2:\mathcal{X}\times\Theta\rightarrow[0,1]$ be the Bayes optimal classifier distinguishing dataset $\tilde{\mathcal{D}}_2=\{X'_i,\theta'_i\}$ from $\tilde{\mathcal{D}}_1=\{X_i,\theta_i\}$. Then the outputs of the classifiers satisfy: \begin{equation}
        \frac{f_1(x,y,\theta)f_2(x,\theta)}{(1-f_1(x,y,\theta))(1-f_2(x,\theta))} = \frac{p(y|x,\theta)}{q(y|x)}.
    \end{equation}
\end{proposition}
The proof of the proposition is provided in the supplementary material.
\begin{remark}
    The procedure outlined in Proposition~\ref{thm:w_training} was also used in \cite{Chan2023}, although a formal proof was not provided. In this setting, the auxiliary variable $\theta_i'$ does not influence the synthetic data $X_i',Y_i'$; rather, it functions solely as a supporting variable for classifier training. While the distributions of $\theta_i$ and $\theta_i'$ may differ, as allowed by the proposition, in practice, it might be preferable to set $p_\theta=q_\theta$ to avoid potential numerical instability. A convenient choice is to use a uniform distribution over the parameter space. Similarly, although the proposition specifies $X_i\sim p_X, X_i'\sim q_X$, there is no restriction against using the same distribution for both. 
    Therefore, a practical procedure can be summarized as follows:
    \begin{enumerate}
        \item Sample $X_i',X_i$ from the particle-level MC distribution $q_X$.
        \item Sample $\theta_i'$ from a chosen distribution $p_\theta$, e.g., uniform distribution over the parameter space. Sample $\theta_i$ from the same distribution $p_\theta$.
        \item Generate $Y_i'$ from the forward model $q(y|X_i')$. Generate $Y_i$ from the forward model $p(y|X_i,\theta_i)$.
        \item Train classifier $f_1$ to distinguish $\mathcal{D}_1=\{X_i,Y_i,\theta_i\}$ from $\mathcal{D}_2=\{X'_i,Y'_i,\theta_i'\}$, and train classifier $f_2$ to distinguish $\tilde{\mathcal{D}}_2=\{X'_i,\theta'_i\}$ from $\tilde{\mathcal{D}}_1=\{X_i,\theta_i\}$.
    \end{enumerate}
\end{remark}

\section{Simulation Study: Gaussian Example}
\label{sec:gaussian}
In this section, we illustrate the POF algorithm with a simple Gaussian example.  Consider a one-dimensional Gaussian distribution at the particle level and two Gaussian distributions at the detector level. The data are generated as follows:
\begin{equation}
\begin{split}
    Y_{i1} &= X_{i} + Z_{i1}, \\
    Y_{i2} &= X_{i} + Z_{i2},
\end{split}
\end{equation}
where $X_i\sim\mathcal{N}(\mu,\sigma^2), Z_{i1}\sim\mathcal{N}(0,1), Z_{i2}\sim\mathcal{N}(0,\theta^2)$. Here, $\theta$ is the nuisance parameter, which only affects the second coordinate of the detector-level data.  This is qualitatively similar to the physical case of being able to measure the same quantity twice. This is also an identifiable model since the characteristic function of $(Y_1,Y_2)$ satisfies
\begin{equation}
    \varphi_{Y_1,Y_2}(t_1,t_2) = \varphi_{X}(t_1+t_2)e^{-\frac{1}{2}(t_1^2+t_2^2\theta^2)},
\end{equation}
where $\varphi_X$ is the characteristic function of $X$. Since $e^{-\frac{1}{2}(t_1^2+t_2^2\theta^2)} > 0$ for all $t_1,t_2$, this uniquely determines $\varphi_X$, and hence also the distribution of $X$. Because the response kernel in this case is a Gaussian density, the analytic form of $p(y|x,\theta)$ is known and, consequently, $\mathcal{W}(y,x,\theta)$ as well. Thus, we can directly plug in the analytic form of $\mathcal{W}(y,x,\theta)$ into the POF algorithm without the need to estimate it using classifiers. As a comparison, we present results both using the analytic form and the estimated $\mathcal{W}$ function as described in Section~\ref{sec:nuisance_update}.

\subsection{Dataset} Based on the above data-generating process, Monte Carlo data are generated with $\mu=0,\sigma=1,\theta=1$ and experimental data are generated with $\mu=0.8,\sigma=1,\theta=1.5$. We simulate $10^5$ events each for the MC data and experimental data.

\begin{figure}[h]
  \centering
  \includegraphics[width=0.55\linewidth]{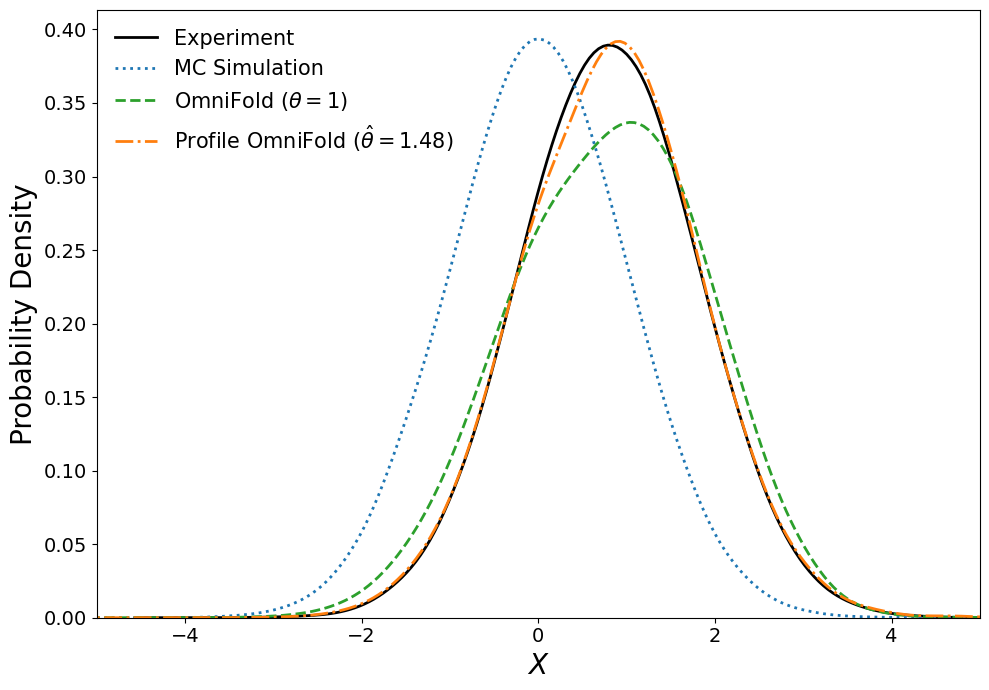}
  \includegraphics[width=0.4\linewidth]{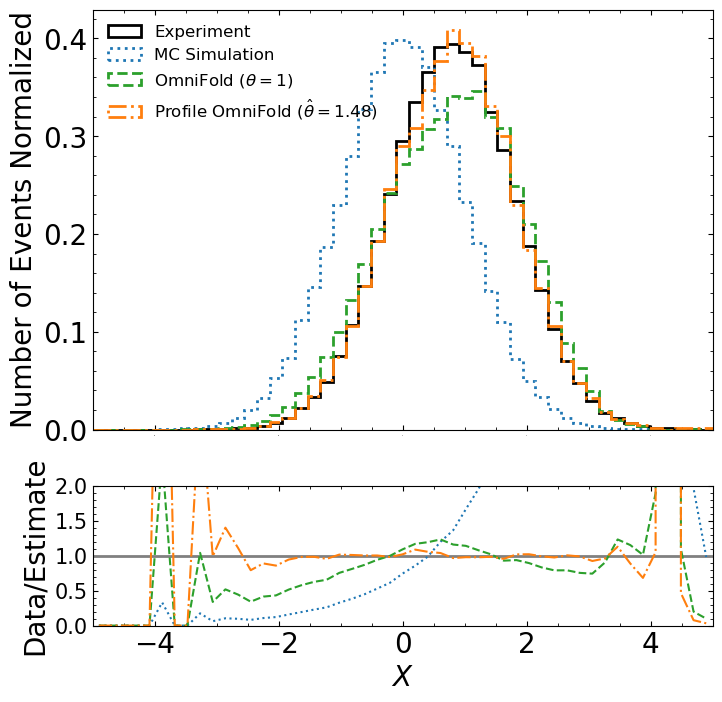}
  \caption{Results of unfolding the Gaussian example. Analytic $\mathcal{W}$ function is being used in the algorithm. \textbf{Left}: Particle-level kernel density estimates of the truth distribution (black), the MC distribution ({\color{blue}blue}), and the reweighted MC distributions obtained using the POF ({\color{orange}orange}) and OF ({\color{ForestGreen}green}) algorithms, each run for 10 iterations. \textbf{Top-right}: Histograms of the four corresponding spectra, aggregated into 50 bins. \textbf{Bottom-right}: The ratio of the truth spectrum to the unfolded spectra.} 
  \label{fig:Gaussian2DExample_X1}
\end{figure}

\begin{figure}[htbp]
    \centering
    \includegraphics[width=0.45\linewidth]{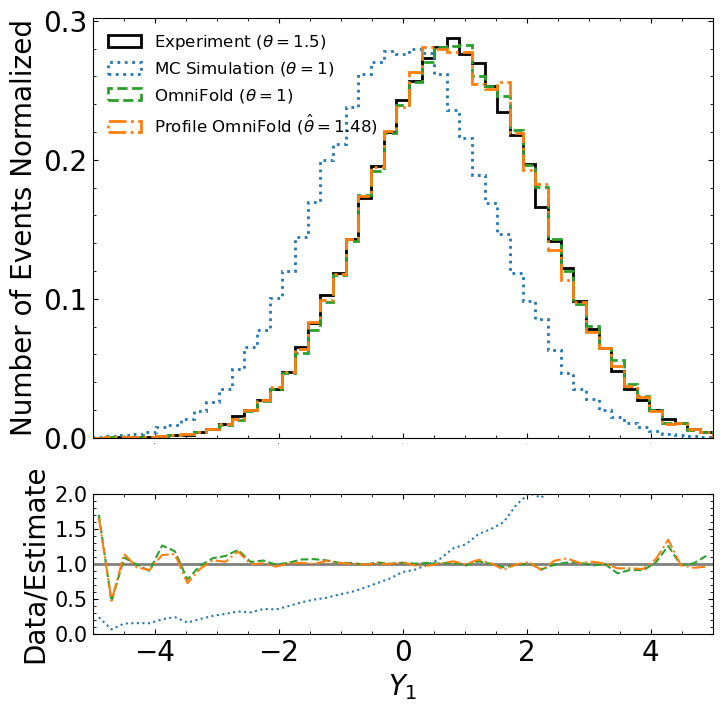}
    \includegraphics[width=0.45\linewidth]{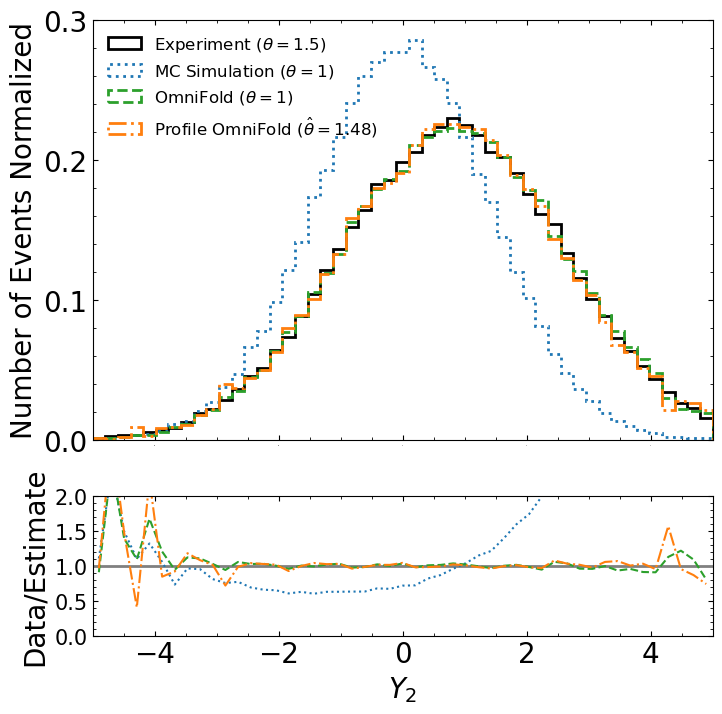}
    \caption{Results corresponding to Figure~\ref{fig:Gaussian2DExample_X1} in detector-level space. \textbf{Left}: Histograms of the corresponding spectra of $Y_1$. \textbf{Right}: Histograms of the corresponding spectra of $Y_2$.}
    \label{fig:Gaussian2DExample_Y1_Y2}
\end{figure}

\subsection{Neural network architecture and training}
\label{sec:gaussian_neural_network}
The neural network classifier for estimating density ratios during POF iteration is implemented in TensorFlow and Keras~\citep{tensorflow, keras}. The network contains three hidden layers with 50 nodes per layer and employs the ReLU activation function. The output layer consists of a single node with a sigmoid activation function. Training is performed with the Adam optimizer~\citep{kingma2017adammethodstochasticoptimization} with learning rate $\eta=0.001$ using a weighted binary cross-entropy loss. The model is trained for up to 20 epochs with a batch size of 10{,}000, and early stopping with a patience of 3 epochs is applied, that is, training stops if the validation loss does not improve for 3 consecutive epochs. None of the hyperparameters were optimized.

\begin{figure}[htbp]
    \centering
    \includegraphics[width=\linewidth]{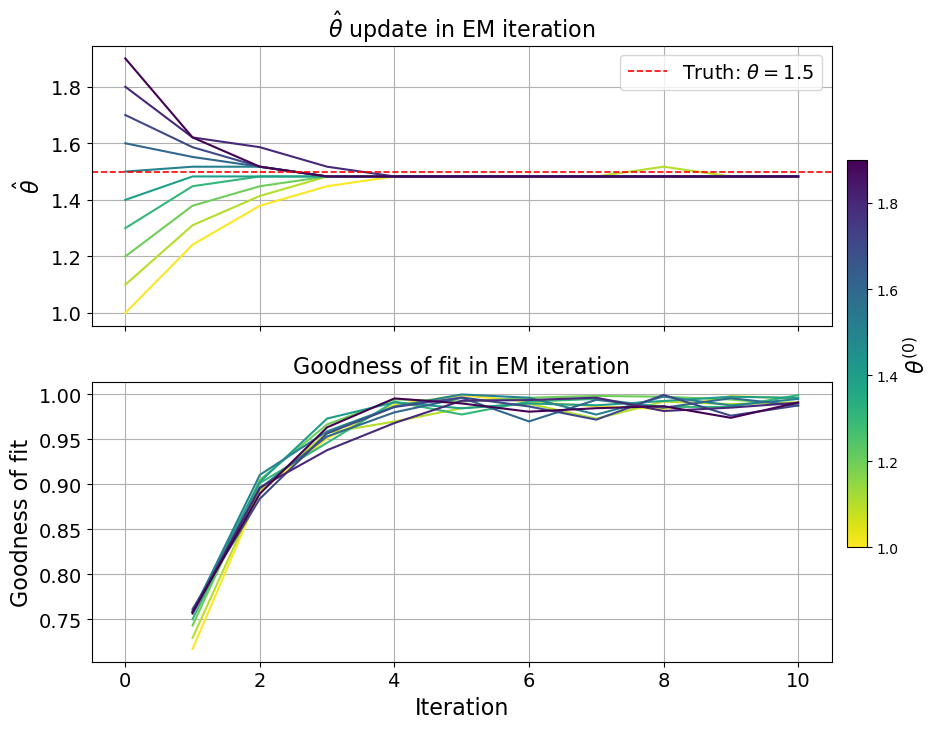}
    \caption{Evolution of the nuisance parameter and the step-1 classifier’s goodness-of-fit statistic for the POF algorithm under different initializations $\theta^{(0)}$. The results shown in Figures~\ref{fig:Gaussian2DExample_X1}--\ref{fig:Gaussian2DExample_Y1_Y2} correspond to the solution yielding the highest goodness-of-fit statistic.  \textbf{Top}: Updated estimates $\hat{\theta}$ across iterations for different initializations $\theta^{(0)}$. \textbf{Bottom}: Goodness-of-fit statistic of the step-1 classifier at each iteration.}
    \label{fig:Gaussian2DExample_theta_acc}
\end{figure}

The neural network classifiers for estimating the $\mathcal{W}$ function ($\mathcal{W}(y,x,\theta)=p(y|x,\theta)/q(y|x)$) are implemented in PyTorch~\citep{paszke2019pytorch}.\footnote{The reason we used two different frameworks is that the original implementation of \textsc{OmniFold} was in TensorFlow/Keras, while the codebase for learning $\mathcal{W}$ function from \cite{Chan2023} was in PyTorch. There is no technical reason preventing from using a single framework.} The same architecture is employed for both classifiers in Proposition~\ref{thm:w_training}. The network contains three hidden layers with 50 nodes per layer and employs the ReLU activation function.  Batch normalization is applied after each hidden layer, and a dropout layer with a rate of 0.1 is added after the second hidden layer. The output layer consists of a single node with a sigmoid activation function. Training uses the Adam optimizer with learning rate $\eta=0.001$ and the weighted binary cross-entropy loss. This classifier is trained for up to 1000 epochs with a batch size of 10{,}000, and early stopping with patience 10 is used. In addition, we train an ensemble of 10 networks with bootstrap resampling to reduce the variance of the estimated $\mathcal{W}$ function. The range of nuisance parameter $\theta$ used in training is set to be $[0.5,2.0]$.

\begin{figure}[htbp]
  \centering
  \includegraphics[width=0.55\linewidth]{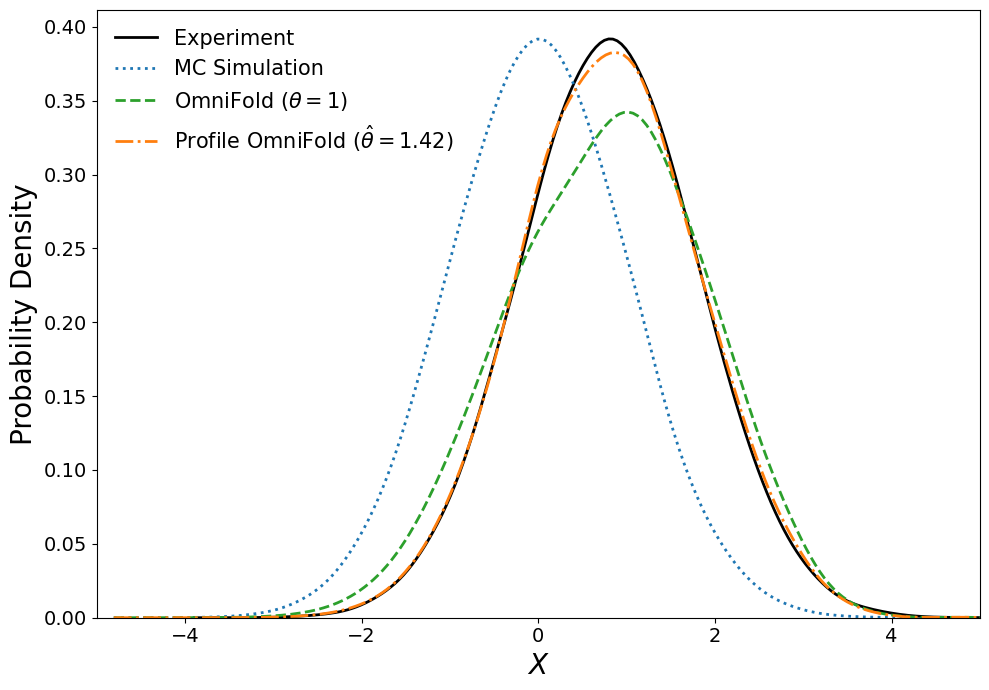}
  \includegraphics[width=0.4\linewidth]{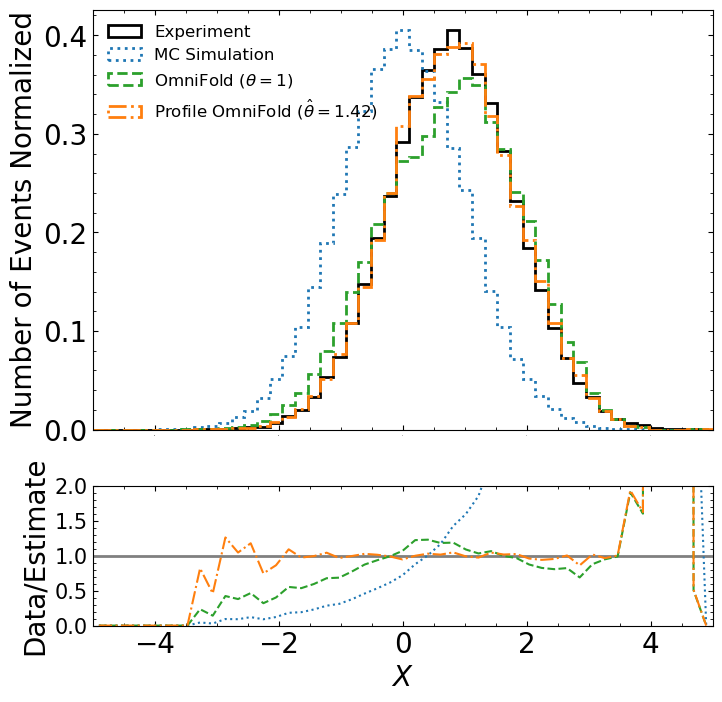}
  \caption{Results of unfolding the Gaussian example. Estimated $\mathcal{W}$ function is being used in the algorithm. \textbf{Left}: Particle-level kernel density estimates of the truth distribution (black), the MC distribution ({\color{blue}blue}), and the reweighted MC distributions obtained using the POF ({\color{orange}orange}) and OF ({\color{ForestGreen}green}) algorithms, each run for 10 iterations. \textbf{Top-right}: Histograms of the four corresponding spectra, aggregated into 50 bins. \textbf{Bottom-right}: The ratio of the truth spectrum to the unfolded spectra.} 
  \label{fig:Gaussian2DExample_X1_est_w}
\end{figure}

\begin{figure}[htbp]
    \centering
    \includegraphics[width=0.45\linewidth]{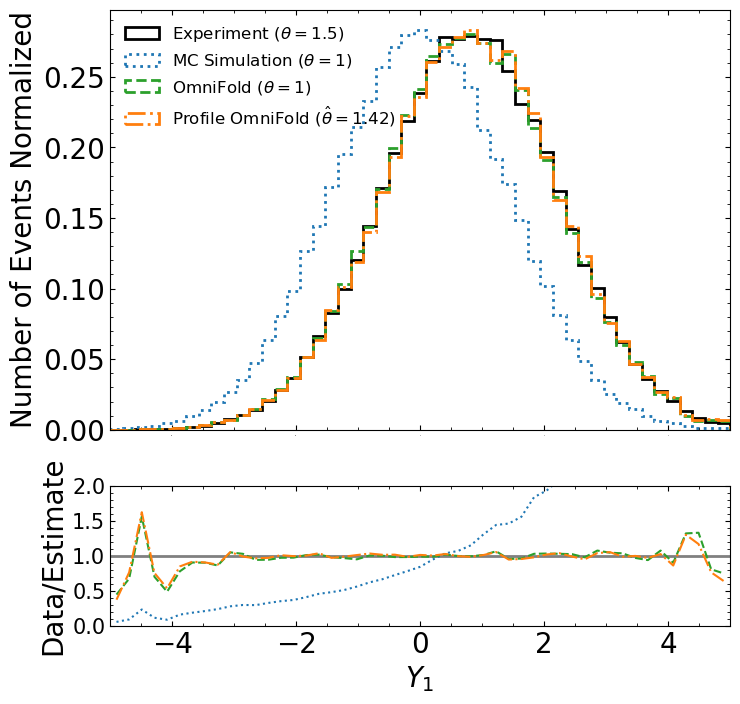}
    \includegraphics[width=0.45\linewidth]{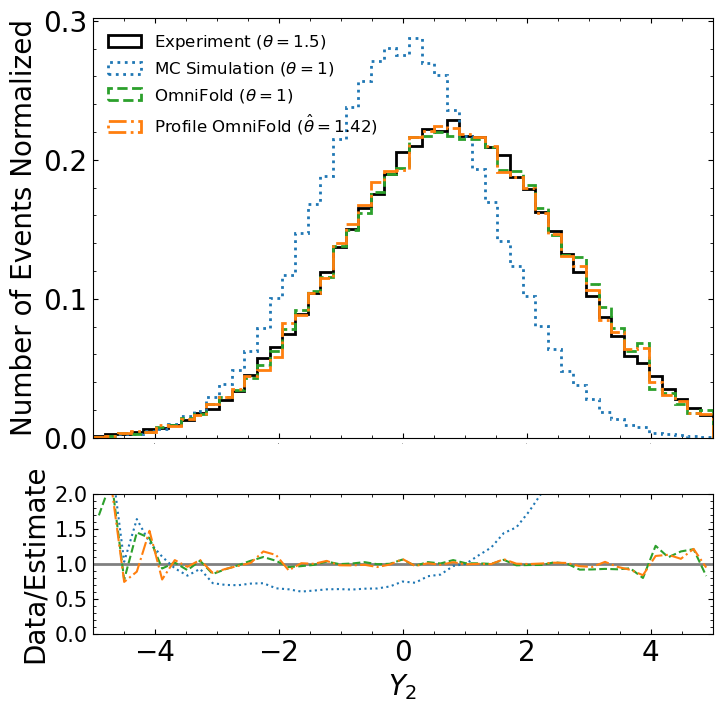}
    \caption{Results corresponding to Figure~\ref{fig:Gaussian2DExample_X1_est_w} in detector-level space. \textbf{Left}: Histograms of the corresponding spectra of $Y_1$. \textbf{Right}: Histograms of the corresponding spectra of $Y_2$.}
    \label{fig:Gaussian2DExample_Y1_Y2_est_w}
\end{figure}

\subsection{Results}

Figure \ref{fig:Gaussian2DExample_X1} illustrates the results of unfolding the Gaussian data using both the proposed POF algorithm and the original OF algorithm. In the unbinned solution, kernel density estimates are used to represent the simulation, data, and reweighted distributions, while the binned solution employs histograms with 50 bins. The blue curve is the Monte Carlo distribution for which the reweighting function $\nu(x)$ will be applied. The results show that the original OF solution (green) deviates significantly from the true distribution (black). This discrepancy arises because OF assumes $p(y|x)=q(y|x)$, which is invalid in the present setting. In addition to biasing the unfolded distribution, a misspecified response kernel would also invalidate uncertainty quantification without profiling over the nuisance parameters. In contrast, the POF algorithm simultaneously updates the nuisance parameter along with the reweighting function. The results show that the unfolded solution (orange) aligns closely with the truth and the estimated nuisance parameter is $\hat{\theta}=1.48$, which is close to the true value $\theta = 1.50$.

Figure~\ref{fig:Gaussian2DExample_Y1_Y2} shows the corresponding reweighted detector-level spectra of $Y_1$ and $Y_2$. The reweighted spectra are close to the experimental distribution (black) for both POF (orange) and OF (green) solutions. This is expected as both POF and OF are designed to reweight the detector-level MC distribution to match the experimental distribution as closely as possible. However, since the response kernel is not correctly specified in the original OF algorithm, the reweighted particle-level distribution does not match the true particle-level distribution, leading to a poor unfolding result.

\begin{figure}[htbp]
    \centering
    \includegraphics[width=\linewidth]{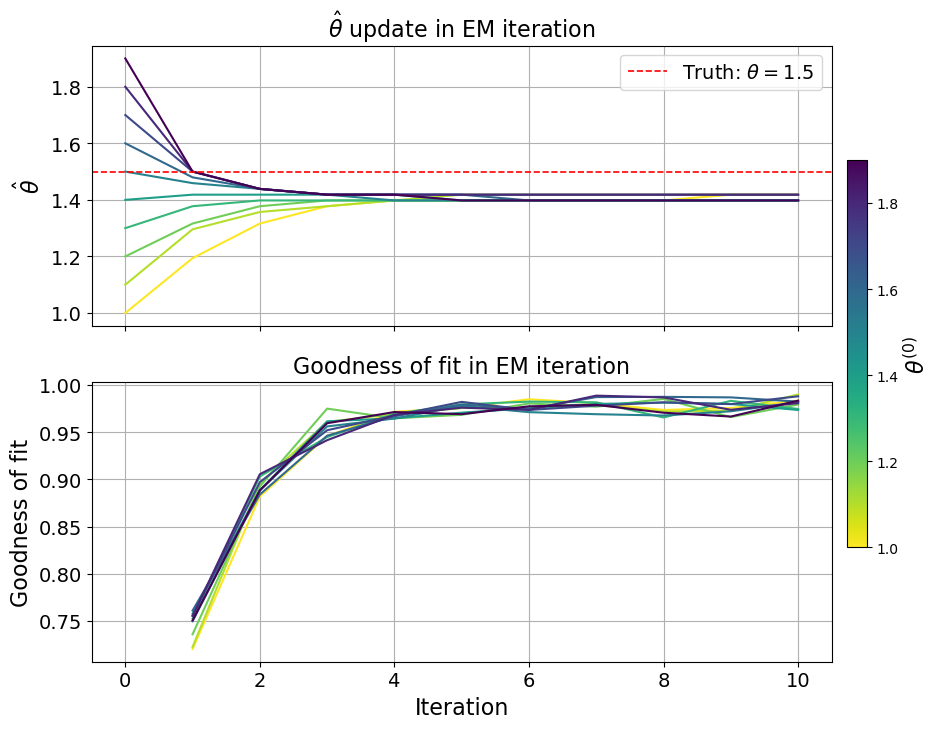}
    \caption{Evolution of the nuisance parameter and the step-1 classifier’s goodness-of-fit statistic for the POF algorithm under different initializations $\theta^{(0)}$. The results shown in Figures~\ref{fig:Gaussian2DExample_X1_est_w}--\ref{fig:Gaussian2DExample_Y1_Y2_est_w} correspond to the solution yielding the highest goodness-of-fit statistic. \textbf{Top}: Updated estimates $\hat{\theta}$ across iterations for different initializations $\theta^{(0)}$. \textbf{Bottom}: Goodness-of-fit statistic of the step-1 classifier at each iteration.}
    \label{fig:Gaussian2DExample_theta_acc_est_w}
\end{figure}

Moreover, Figure~\ref{fig:Gaussian2DExample_theta_acc} shows the evolution of the nuisance parameter $\theta$ and the goodness-of-fit statistic for the step-1 classifier, defined in Eq.~\eqref{eq:gof}. The top plot shows that the nuisance parameter converges to the true value within a few iterations, regardless of the initial value. The bottom plot shows the goodness-of-fit statistic converging to 1 for all initial values, indicating that the reweighted distribution $\tilde{q}(y)$ is close to the target distribution $p(y)$. As discussed in Section~\ref{sec:algorithm}, the results shown in Figures~\ref{fig:Gaussian2DExample_X1}--\ref{fig:Gaussian2DExample_Y1_Y2} correspond to the solution with the highest goodness-of-fit statistic, although in this case study all solutions give equally good fits after a few iterations.

Figure~\ref{fig:Gaussian2DExample_X1_est_w} illustrates the results obtained using the estimated $\mathcal{W}$ function instead of the analytic form. The results show that the POF solution (orange) still aligns closely with the truth (black). However, the estimated nuisance parameter is $\hat{\theta}=1.42$, which is slightly off from the true value $\theta=1.50$. Figure \ref{fig:Gaussian2DExample_theta_acc_est_w} shows that the estimated nuisance parameter converges to around 1.42, regardless of the initial value. The goodness-of-fit statistic converges to close to 1 (although slightly worse than the case with analytic $\mathcal{W}$) for all initial values, indicating that the reweighted distribution $\tilde{q}(y)$ is close to the target distribution $p(y)$. Figure~\ref{fig:Gaussian2DExample_Y1_Y2_est_w} shows that the reweighted detector-level spectra of $Y_1$ and $Y_2$ are also close to the experimental distribution (black) for both POF (orange) and OF (green) solutions.

In practice, we have observed that the estimated nuisance parameter is sensitive to the quality of the fitted $\mathcal{W}$ function. Even small changes in classifier training, such as a different range for simulating $\theta_i$, or simply training two neural networks with the same hyperparameters could lead to different estimates of the nuisance parameters. Such variability can be reduced by training an ensemble of neural networks, but it still remains a practical challenge when applying the algorithm. Nevertheless, the final unfolded density, which is the primary quantity of interest, appears to be relatively robust to variations in the fitted $\mathcal{W}$ function.

We have also experimented with different values of the true nuisance parameter $\theta$ in the data-generating process, and the results are qualitatively similar to those presented here. The details are included in the supplementary material.

\subsection{Extension to two dimensions at the particle level}
\label{sec:gaussian_2d}
{We now extend the Gaussian example to a two-dimensional particle-level quantity $(X_1, X_2)$ drawn from a bivariate Gaussian distribution with mean $\mu$ and covariance matrix $\Sigma$,
\begin{equation}
    (X_1, X_2) \sim \mathcal{N}(\mu, \Sigma).
\end{equation}
Each dimension is independently measured twice with different smearing resolutions, resulting in a four-dimensional detector-level representation
\begin{equation}
\begin{split}
    Y_{11} &= X_1 + Z_{11}, \qquad Y_{12} = X_1 + Z_{12}, \\
    Y_{21} &= X_2 + Z_{21}, \qquad Y_{22} = X_2 + Z_{22},
\end{split}
\end{equation}
where $Z_{11}\sim\mathcal{N}(0,1)$, $Z_{12}\sim\mathcal{N}(0,\sigma_1^2)$, $Z_{21}\sim\mathcal{N}(0,1)$, and $Z_{22}\sim\mathcal{N}(0,\sigma_2^2)$. Here $\sigma_1$ and $\sigma_2$ are the nuisance parameters governing the resolution of the second detector-level measurement in each dimension, so the full bivariate nuisance parameter is $\theta=(\sigma_1,\sigma_2)$. The POF algorithm is then applied to recover the underlying two-dimensional particle-level distribution and the nuisance parameter $\theta$.

\begin{figure}[htbp]
  \centering
  \includegraphics[width=0.96\linewidth]{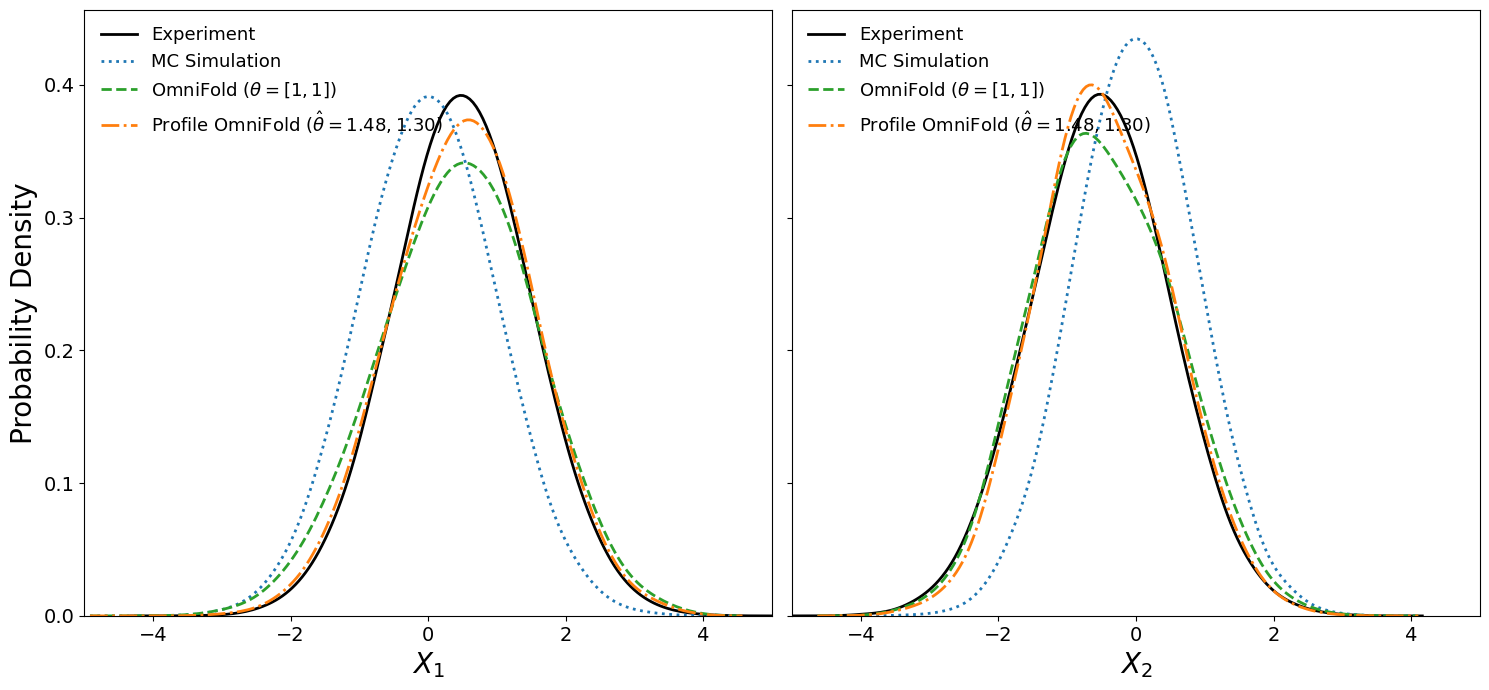}
  \includegraphics[width=0.48\linewidth]{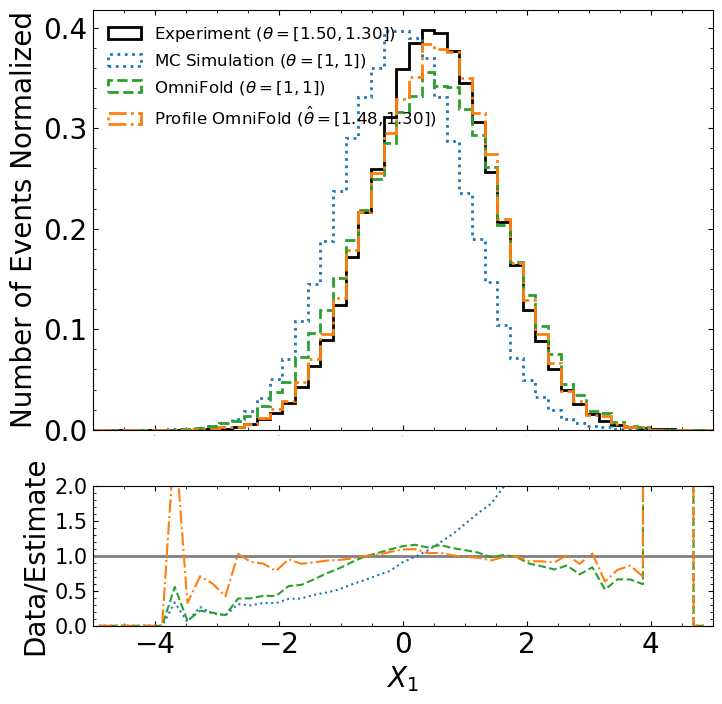}
  \includegraphics[width=0.48\linewidth]{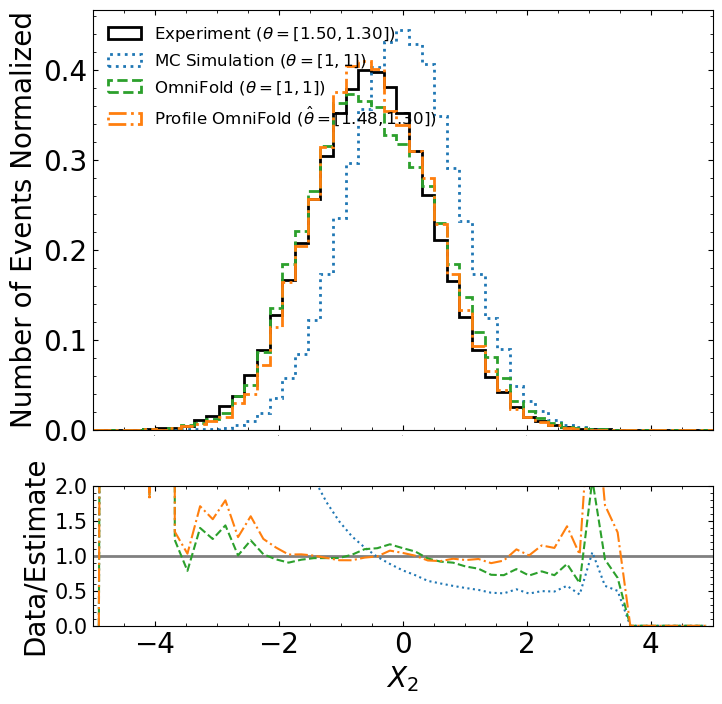}
  \caption{Particle-level unfolding results for the bivariate Gaussian example (analytic $\mathcal{W}$). \textbf{Top-left}: $X_1$ (unbinned). \textbf{Top-right}: $X_2$ (unbinned). \textbf{Bottom-left}: $X_1$ (binned). \textbf{Bottom-right}: $X_2$ (binned). Colors as in Figure~\ref{fig:Gaussian2DExample_X1}: truth (black), MC ({\color{blue}blue}), POF ({\color{orange}orange}), OF ({\color{ForestGreen}green}), each run for 10 iterations.}
  \label{fig:Gaussian4DExample_X1_X2}
\end{figure}

\begin{figure}[htbp]
    \centering
    \includegraphics[width=0.96\linewidth]{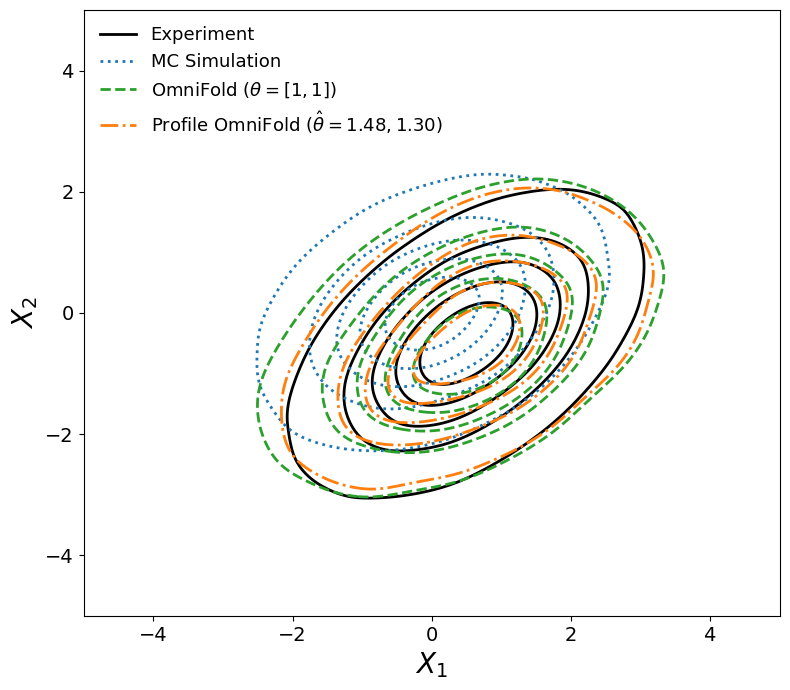}
    \caption{Contour plot of the particle-level distribution for the bivariate Gaussian example.}
    \label{fig:Gaussian4DExample_X_contour}
\end{figure}
The response kernel factorizes across dimensions as 
\begin{equation}
    p(y|x,\theta)=p(y_{11},y_{12}|x_1,\sigma_1)\cdot p(y_{21},y_{22}|x_2,\sigma_2),
\end{equation}
where each factor is a Gaussian density. Consequently, $\mathcal{W}(y,x,\theta)$ is available in closed form and is used directly in the algorithm.

Monte Carlo data are generated with $\mu=(0,0)$, $\Sigma=\begin{pmatrix}1.0&0.3\\0.3&0.8\end{pmatrix}$, and $\theta=(\sigma_1,\sigma_2)=(1,1)$. Experimental data are generated with $\mu=(0.5,-0.5)$, $\Sigma=\begin{pmatrix}1.0&0.5\\0.5&1.0\end{pmatrix}$, and $\theta=(1.5,1.3)$. We simulate $10^5$ events each for the MC data and the experimental data. The neural network classifier for estimating density ratios follows the same architecture and training procedure described in Section~\ref{sec:gaussian_neural_network}, with the input dimension increased to match the higher-dimensional data.

\begin{figure}[htbp]
    \centering
    \includegraphics[width=0.48\linewidth]{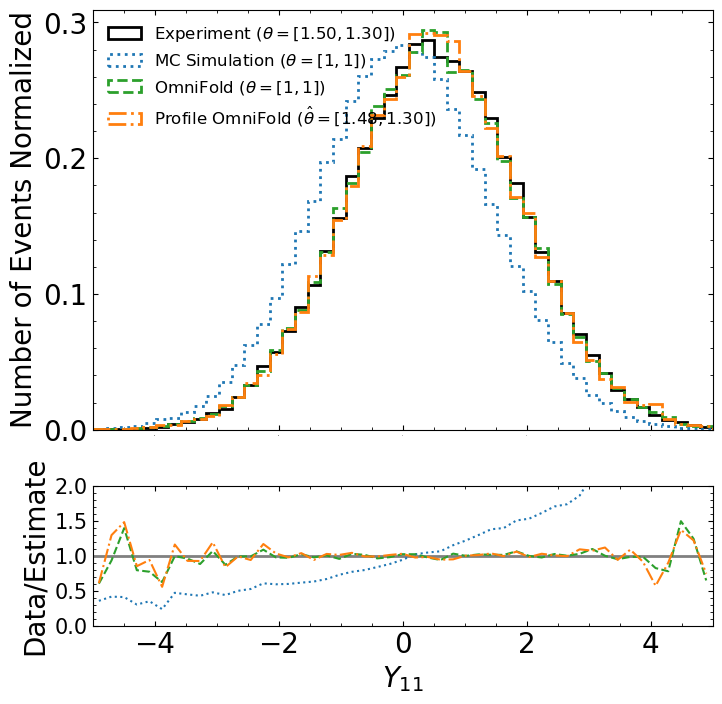}
    \includegraphics[width=0.48\linewidth]{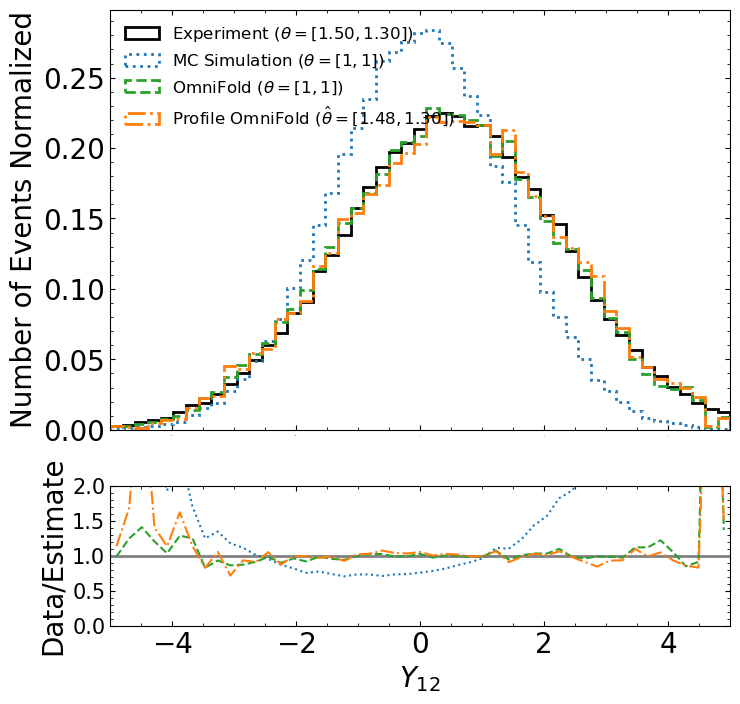} \\[4pt]
    \includegraphics[width=0.48\linewidth]{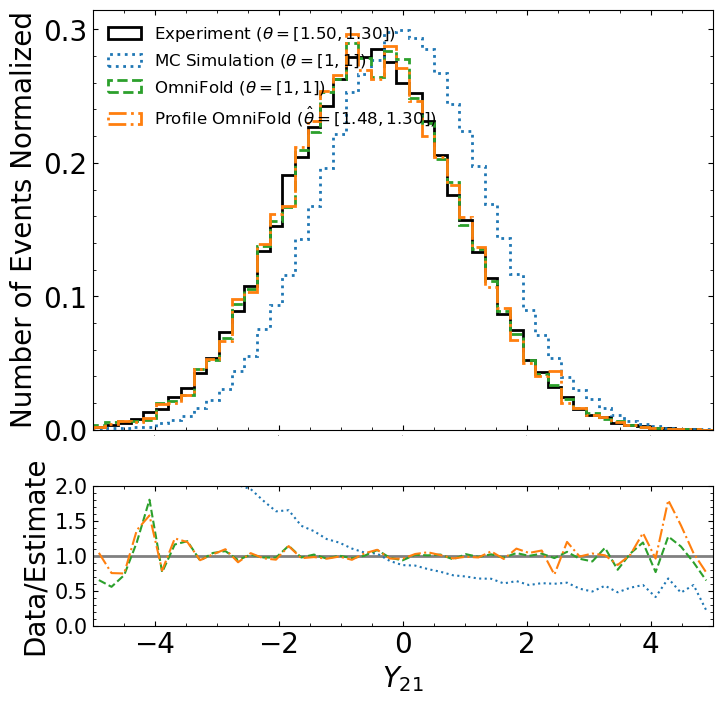}
    \includegraphics[width=0.48\linewidth]{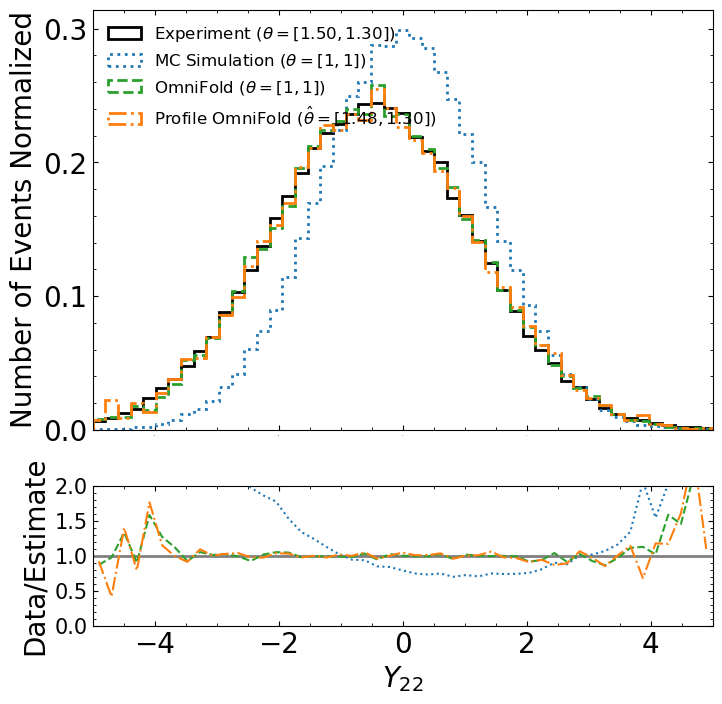}
    \caption{Detector-level spectra for the bivariate Gaussian example. \textbf{Top-left}: $Y_{11}$ (first measurement of $X_1$). \textbf{Top-right}: $Y_{12}$ (second measurement of $X_1$, nuisance-sensitive). \textbf{Bottom-left}: $Y_{21}$ (first measurement of $X_2$). \textbf{Bottom-right}: $Y_{22}$ (second measurement of $X_2$, nuisance-sensitive).}
    \label{fig:Gaussian4DExample_Y_grid}
\end{figure}

Figures~\ref{fig:Gaussian4DExample_X1_X2} and~\ref{fig:Gaussian4DExample_X_contour} show the unfolding results. The OF solution (green) deviates from the true particle-level distribution, both in the marginal distributions of $X_1$ and $X_2$ and in their joint distribution, due to the misspecified nuisance parameters. In contrast, the POF solution (orange) recovers the truth more closely. Notably, POF and OF provide comparably good fits to the observed smeared data, as shown in Figure~\ref{fig:Gaussian4DExample_Y_grid}. Therefore, the bias in OF would not be apparent from the smeared distributions alone. The estimated nuisance parameters are $\hat{\sigma}_1 = 1.48$ and $\hat{\sigma}_2 = 1.3$ (true values $\sigma_1=1.5$, $\sigma_2=1.3$).
Figure~\ref{fig:Gaussian4DExample_theta_acc} shows the evolution of $\theta=(\sigma_1,\sigma_2)$ and the goodness-of-fit statistic across iterations. Both components of $\theta$ converge to values close to the truth, and the goodness-of-fit statistic converges to nearly 1, indicating that the reweighted detector-level distribution matches the experimental distribution well. These results confirm that POF extends naturally to multidimensional settings, and it can successfully recover the underlying particle-level distribution while simultaneously estimating multiple nuisance parameters.

\begin{figure}[htbp]
    \centering
    \includegraphics[width=0.96\linewidth]{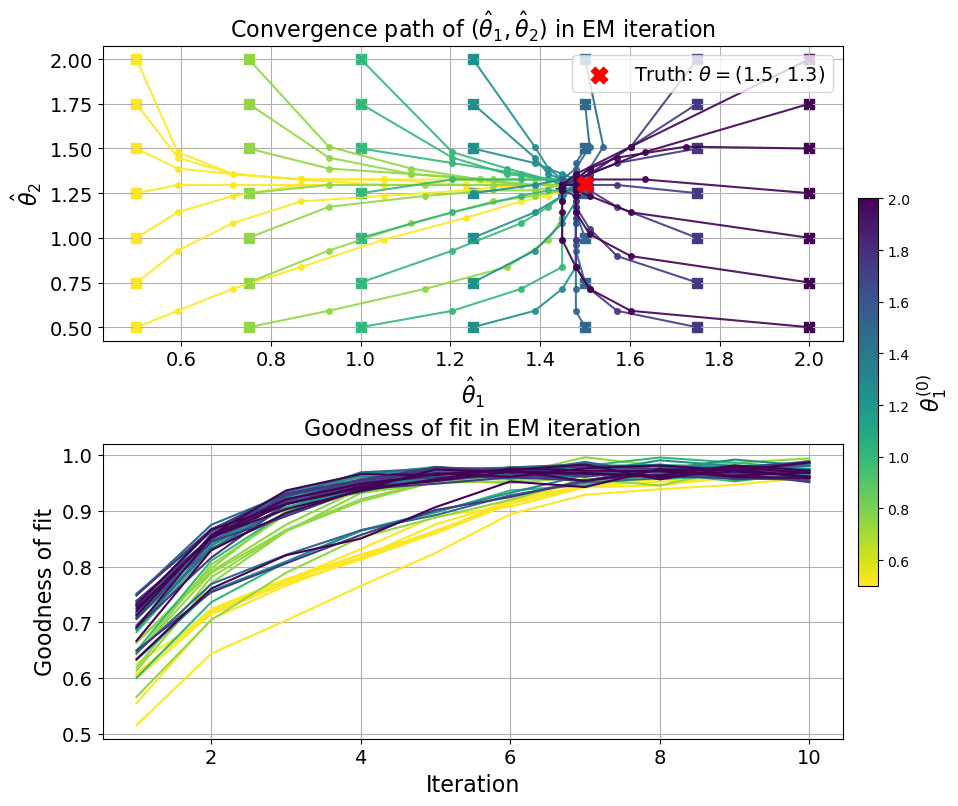}
    \caption{Evolution of the nuisance parameters and goodness-of-fit statistic for the bivariate Gaussian example. \textbf{Top}: Updated estimates $(\hat{\theta}_1, \hat{\theta}_2)$ across iterations for different initializations. \textbf{Bottom}: Goodness-of-fit statistic of the step-1 classifier at each iteration.}
    \label{fig:Gaussian4DExample_theta_acc}
\end{figure}

\section{Simulated Public Collider Data by the CMS Experiment}
\label{sec:open_data}
We now study the important process of generic quark and gluon scattering at high-energy particle colliders. The outgoing particles radiate and produce collimated streams of particles called jets.  The inclusive jet energy spectrum is useful for studies of the strong force at small distance scales, for searches for new, fundamental interactions, and for developing physics models that enable other analyses with jets as a background process. The most likely type of event consists of two, high-energy jets that have nearly the same momentum transverse to the collision axis due to conservation of momentum. The detector acts locally in space, so the measured momenta of the two jets are independently smeared. The amount of smearing is approximately known from simulations. These simulations and their data-based calibrations contain a number of nuisance parameters.  In our example, we are only sensitive to the effective jet energy resolution, which governs the overall amount of momentum smearing.  We introduce a nuisance parameter that is a multiplicative factor determining how much the jet energy resolution in data differs from simulation.  Due to the symmetries of the problem, the difference in the jet momenta is particularly sensitive to this nuisance parameter without being sensitive to the underlying momentum spectrum.  Thus, we measure the joint distribution of the sum and difference in jet momenta per event while simultaneously constraining the nuisance parameter.

\subsection{Dataset}

To demonstrate this setup in practice, we use high-fidelity simulations from the CMS Experiment at the Large Hadron Collider. In particular, the data were generated with \textsc{Pythia}~6.426~\citep{Sjostrand:2006za} using the Z2 tune~\citep{Chatrchyan:2011id} and interfaced with a \textsc{Geant4}-based~\citep{Agostinelli:2002hh,1610988,Allison:2016lfl} detailed detector simulation of the CMS experiment~\citep{Chatrchyan:2008aa}.  This dataset comes from the CMS Open Data Portal~\citep{CMS:QCDsim1000-1400,CMS:QCDsim1400-1800,CMS:QCDsim1800} and is processed into an MIT Open Data format~\citep{komiske_patrick_2019_3341502,komiske_patrick_2019_3341770,komiske_patrick_2019_3341772,Komiske:2019jim}.  We use the events as both ``simulation'' and ``data'' in order to have a known target for testing. 

The datasets from this collection are sorted by the parton-level hard-scattering scale $\hat{p}_T$ from \textsc{Pythia}, which is in general different from the jet-level transverse momentum $p_T$ we are interested in studying.  For simplicity, we consider one slice of the collection with 600 GeV $<\hat{p}_T<$ 800 GeV.  This slice corresponds to sufficiently high momentum jets that effects from the triggering system are not relevant. Particles (at truth level) or particle flow candidates (at reconstructed/detector level) are used as inputs to jet clustering, implemented using \textsc{FastJet}~3.2.1~\citep{Cacciari:2011ma,Cacciari:2005hq} and the anti-$k_t$ algorithm~\citep{Cacciari:2008gp} with radius parameter $R=0.5$. The total number of events is 43,892. We consider the two highest $p_T$ jets at both particle (truth) and detector (reconstructed) levels, denoted as $p_{T,1}^{\text{truth}},p_{T,2}^{\text{truth}}, p_{T,1}^{\text{reco}}, p_{T,2}^{\text{reco}}$. Then our observables are
\begin{equation}
\begin{split}
    Y_{1} &= p_{T,1}^{\text{truth}} + \theta(p_{T,1}^{\text{reco}}-p_{T,1}^{\text{truth}}) + p_{T,2}^{\text{truth}} + \theta(p_{T,2}^{\text{reco}}-p_{T,2}^{\text{truth}}), \\
    Y_{2} &= p_{T,1}^{\text{truth}} + \theta(p_{T,1}^{\text{reco}}-p_{T,1}^{\text{truth}}) - p_{T,2}^{\text{truth}} - \theta(p_{T,2}^{\text{reco}}-p_{T,2}^{\text{truth}}),
\end{split}
\end{equation}
and the target quantity is
\begin{equation}
    X = p_{T,1}^{\text{truth}} + p_{T,2}^{\text{truth}}.
\end{equation}
Here $\theta$ adjusts the jet energy resolution in data relative to simulation. The true value of $\theta$ in data is 1.7 while the nominal value in simulation is 1.0, indicating that the jet energy resolution in data is 70\% larger than that in simulation. To induce a mismatch in the marginal distribution between the simulation and the data, we construct the MC sample by applying weighted sampling to the events.

\subsection{Neural network architecture and training}
The neural network architecture and training procedure used in the POF algorithm follow the same setup as in the Gaussian example. For training the $\mathcal{W}$ function, the same architecture is employed for both classifiers as in Section~\ref{sec:gaussian_neural_network}. The only difference is that the $f_1$ classifier is trained with an early-stopping patience of 30 epochs, as we found that it requires more iterations to converge in practice. The range of nuisance parameter $\theta$ used in training is set to be $[0.5,2.0]$.

\subsection{Results}
Figure~\ref{fig:OpenDataExample_X1} presents the unfolded CMS Open Data results obtained using both the proposed POF algorithm and the original OF algorithm. In the unbinned solution, kernel density estimates are used to represent the simulation, data, and reweighted distributions, while the binned solution employs histograms with 50 bins. Consistent with the Gaussian example, the original OF solution (green) deviates substantially from the truth (black) because it assumes the nuisance parameter is fixed and correctly specified at $\theta = 1.0$. In contrast, the POF solution (orange) closely matches the truth, with the estimated nuisance parameter of $\hat{\theta} = 1.62$ (true value $\theta = 1.7$).

\begin{figure}[htbp]
  \centering
  \includegraphics[width=0.55\linewidth]{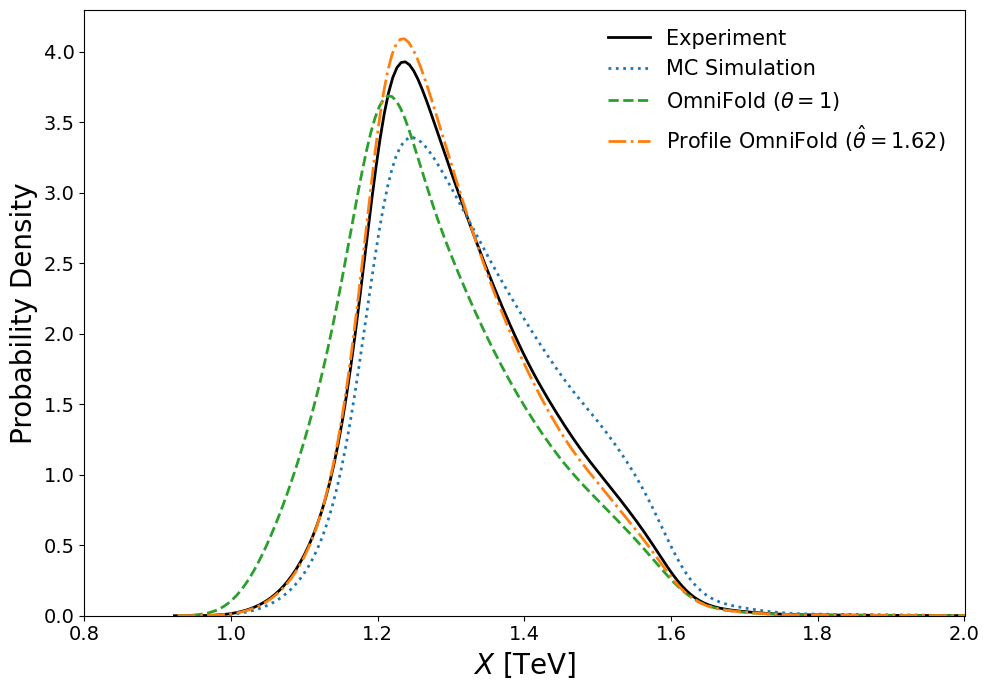}
  \includegraphics[width=0.4\linewidth]{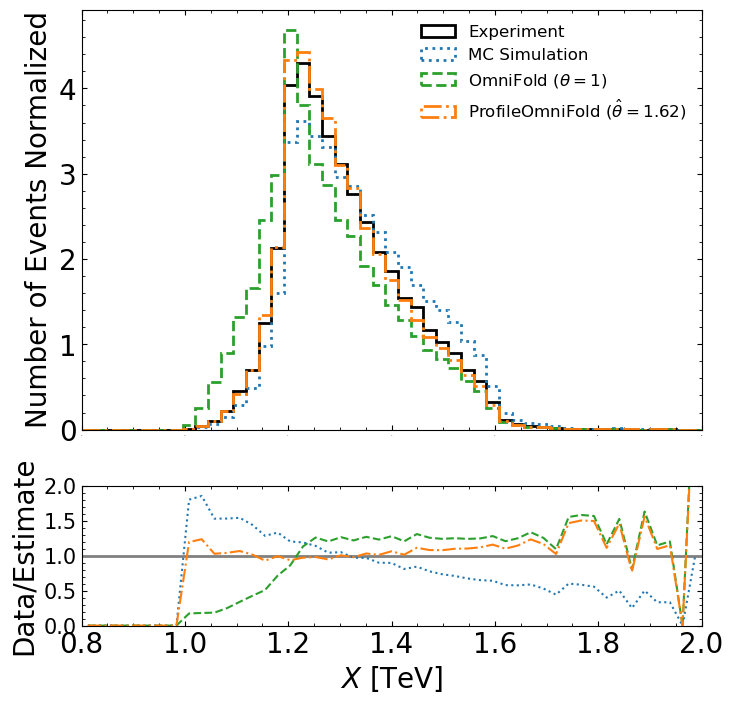}
  \caption{Results of unfolding the CMS open data. \textbf{Left}: Particle-level kernel density estimates of the truth distribution (black), the MC distribution ({\color{blue}blue}), and the reweighted MC distributions obtained using the POF ({\color{orange}orange}) and OF ({\color{ForestGreen}green}) algorithms, each run for 10 iterations. \textbf{Top-right}: Histograms of the four corresponding spectra, aggregated into 50 bins. \textbf{Bottom-right}: The ratio of the truth spectrum to the unfolded spectra.} 
  \label{fig:OpenDataExample_X1}
\end{figure}

\begin{figure}[htbp]
    \centering
    \includegraphics[width=0.45\linewidth]{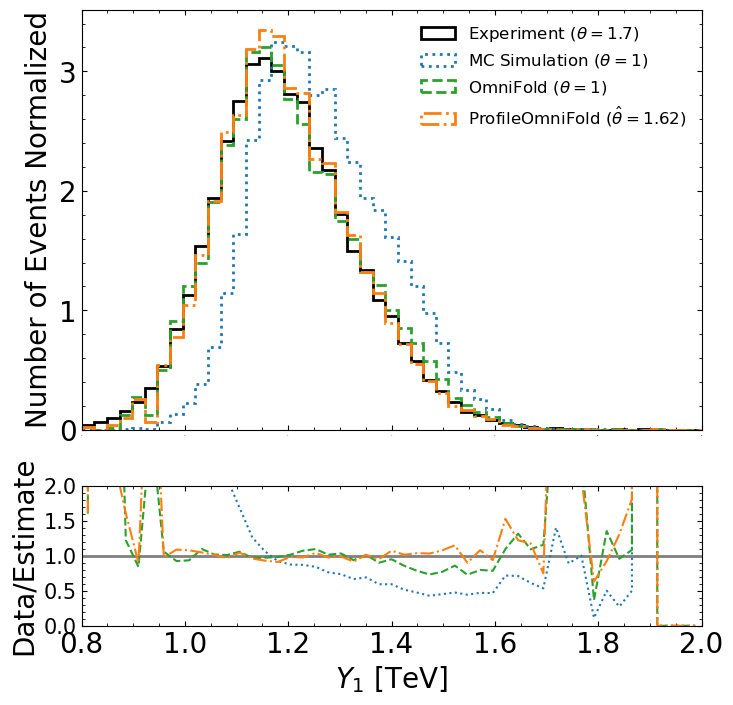}
    \includegraphics[width=0.45\linewidth]{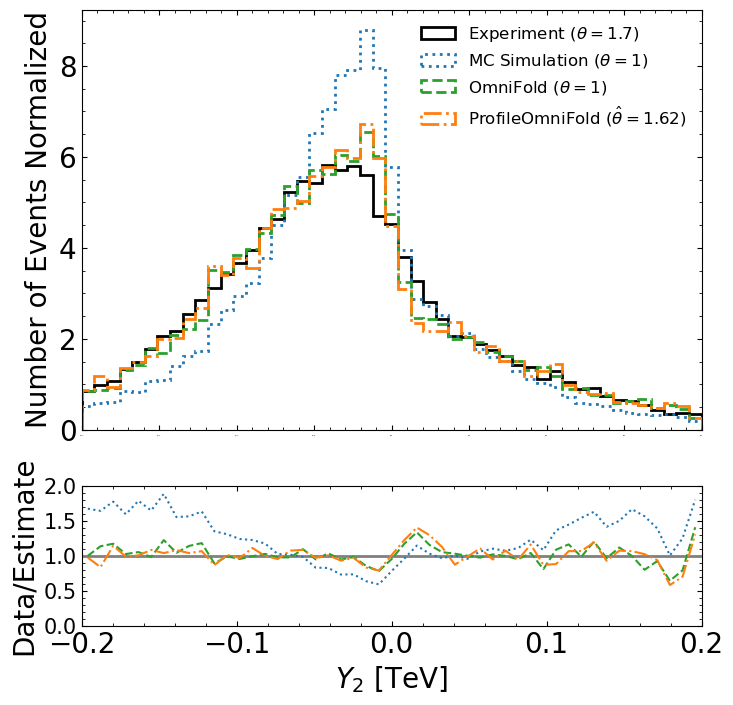}
    \caption{Results corresponding to Figure~\ref{fig:OpenDataExample_X1} in detector-level space. \textbf{Left}: Histograms of the corresponding spectra of $Y_1$. \textbf{Right}: Histograms of the corresponding spectra of $Y_2$.}
    \label{fig:OpenDataExample_Y1_Y2}
\end{figure}
 
Figure \ref{fig:OpenDataExample_Y1_Y2} shows the corresponding reweighted detector-level spectra for $Y_1$ and $Y_2$. Although the reweighted distributions for both POF (orange) and OF (green) generally follow the experimental data (black), some noticeable discrepancies appear—for example, around the peak of the $Y_1$ distribution and near zero in the $Y_2$ distribution. These deviations are plausibly attributed to mismatches in support between the experimental and MC detector-level distributions, which can induce large or unstable weights. Notably, this behavior is not specific to the POF approach and is also observed in the original OF algorithm. Nevertheless, the unfolded distribution obtained via POF remains in close agreement with the truth, indicating that the impact of these weight instabilities on the final unfolded solution is limited.

Moreover, Figure~\ref{fig:OpenDataExample_theta_evolution} shows that the nuisance parameter does not necessarily converge to a good estimate. In particular, when initialized at $\theta^{(0)} = 1.0$ or $1.1$, the estimated value stabilizes around $\hat{\theta} \approx 1.35$. In these cases, the corresponding goodness-of-fit statistic also fails to approach 1, indicating that the reweighted distribution $\tilde{q}(y)$ does not adequately match the observed distribution $p(y)$. This behavior suggests that the algorithm may converge to a local, rather than global, maximum of the likelihood if the starting value $\theta^{(0)}$ is far from the optimal value. The results reported in Figures~\ref{fig:OpenDataExample_X1}--\ref{fig:OpenDataExample_Y1_Y2} are obtained by selecting, among all runs, the solution achieving the highest goodness-of-fit statistic, thereby mitigating this issue. Additional experiments with different jet energy resolution are provided in the supplementary material. These experiments yield consistent results: POF reliably recovers the true particle-level distribution, whereas OF fails in the presence of a misspecified nuisance parameter. However, the choice of the initial nuisance parameter remains important for POF, as starting too far from the true value can lead to suboptimal estimates. For both OF and POF, the reweighted detector-level distributions may also fail to perfectly match the experimental data, particularly when the discrepancy between the true $\theta$ and the nominal $\theta$ used in simulation is large. These findings highlight the importance of employing multiple initializations and selecting the best solution based on the goodness-of-fit statistic.

\begin{figure}[htbp]
    \centering
    \includegraphics[width=\linewidth]{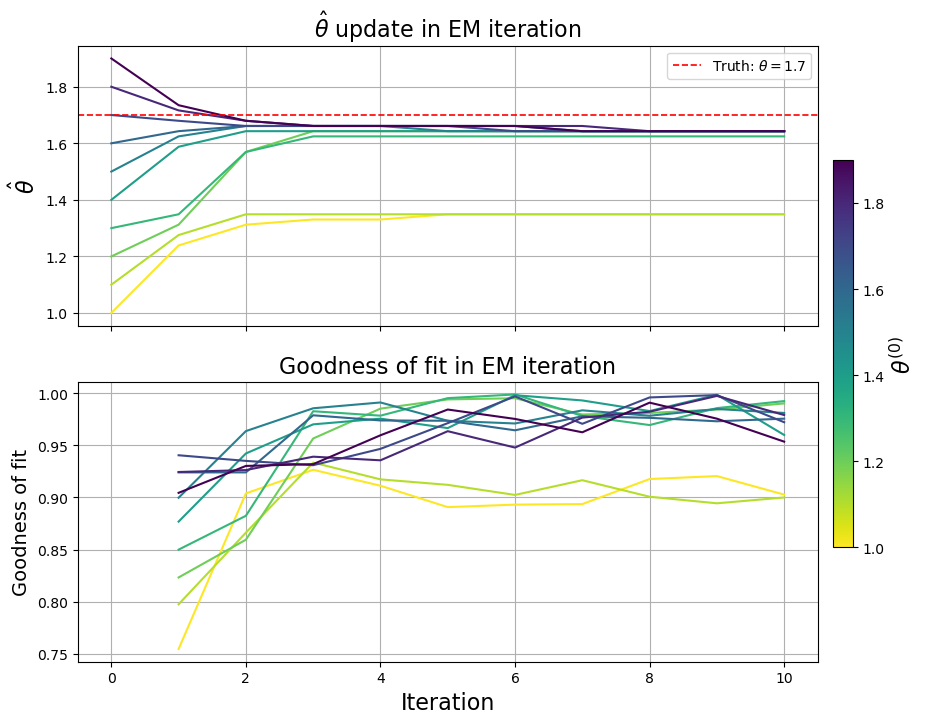}
    \caption{Evolution of the nuisance parameter and the step-1 classifier’s goodness-of-fit statistic for the POF algorithm under different initializations $\theta^{(0)}$. The results shown in Figures~\ref{fig:OpenDataExample_X1}--\ref{fig:OpenDataExample_Y1_Y2} correspond to the solution yielding the highest goodness-of-fit statistic.  \textbf{Top}: Updated estimates $\hat{\theta}$ across iterations for different initializations $\theta^{(0)}$. \textbf{Bottom}: Goodness-of-fit statistic of the step-1 classifier at each iteration.}
    \label{fig:OpenDataExample_theta_evolution}
\end{figure}

\section{Discussion}
\label{sec:discussion}
In this paper, we have proposed a new simulation-based unfolding algorithm, Profile \textsc{OmniFold}, which extends the original \textsc{OmniFold} algorithm to the case where the forward model is not completely specified. POF iteratively updates the reweighting function and nuisance parameters to maximize the population log-likelihood. The algorithm is an EM algorithm that shares similar steps as in \textsc{OmniFold}, which allows for easy implementation while preserving many of its benefits, such as being able to use machine-learning classifiers to estimate density ratios during the iteration. 

The results from the simple Gaussian example and an open dataset from the CMS Experiment demonstrate the algorithm's promising performance. In the case of an incorrectly specified forward model, the POF algorithm is able to accurately estimate the true particle-level distribution, whereas the original \textsc{OmniFold} algorithm fails.

One limitation for the current method is the requirement of training the $\mathcal{W}$ function, which is the conditional density ratio of the smearing kernel parametrized by the nuisance parameter and the Monte Carlo simulation, i.e., $\mathcal{W}(y,x,\theta)=p(y|x,\theta)/q(y|x)$. We found empirically that the estimate nuisance parameters are rather sensitive to the estimated $\mathcal{W}$ function. Different training configurations, such as the number of epochs, early stopping, and range of the nuisance parameter for the training data, can affect the convergence of the nuisance parameter and the final unfolding results. One direction for future work is to explore potential ways to improve the robustness of the $\mathcal{W}$ function estimation, or even to avoid the need of estimating the $\mathcal{W}$ function altogether.

Another important direction for future work is uncertainty quantification of the unfolding results. The current POF algorithm does not provide uncertainty estimates for either the unfolded distribution or the nuisance parameters. Even in the original \textsc{OmniFold} algorithm, it is unclear how to propagate the uncertainty in the classifier-based density ratio estimates to the final unfolded result. One possible (computationally expensive) solution is to use bootstrap resampling \citep{Canelli2026UnbinnedUnfolding} to estimate the uncertainty of the unfolded distribution and nuisance parameters. However, rigorous analysis of uncertainty quantification still remains an open problem in this setting.

\section*{Significance Statement}
Modern particle and nuclear physics experiments rely on unfolding methods to infer the underlying particle-level spectrum of physical quantities from detector-level observations that have been distorted by detector effects. This problem is central for comparing measurements across experiments and for enabling detector-independent scientific interpretation, but it is challenging because detector responses are complex, high-dimensional, and typically known only through simulations. Existing machine learning-based unfolding methods, including \textsc{OmniFold}, assume that the simulated detector response is correctly specified; when this assumption fails, the unfolded results can be biased. This paper addresses this limitation by introducing Profile \textsc{OmniFold}, an unbinned, classifier-based method that jointly estimates the particle-level spectrum and the nuisance parameters governing the detector response. By accounting for the forward-model uncertainty during the unfolding procedure, the proposed method improves over existing approaches that ignore such uncertainty. Through both a controlled Gaussian example and an application motivated by CMS Open Data, we show that accounting for nuisance parameters can substantially reduce bias in the unfolded solution and recover more reliable particle-level results when the nominal simulation is misspecified. These findings highlight the importance of incorporating forward-model uncertainty into simulation-based inverse problems and provide a practical path for applying machine learning-based unfolding methods more robustly in high-energy physics and related scientific fields.

%%%%%%%%%%%%%%%%%%%%%%%%%%%%%%%%%%%%%%%%%%%%%%
%% Appendix---Please move all appendices to %%
%% a Supplementary file.                    %%
%%%%%%%%%%%%%%%%%%%%%%%%%%%%%%%%%%%%%%%%%%%%%%
%% Support information, if any,             %%
%% should be provided in the                %%
%% Acknowledgements section.                %%
%%%%%%%%%%%%%%%%%%%%%%%%%%%%%%%%%%%%%%%%%%%%%%
\begin{acks}[Acknowledgments]
 We thank Wahid Bhimji for co-hosting HZ during his internship at LBNL.
 We acknowledge the computing resources provided by National Energy Research Scientific Computing Center (NERSC) for performing the experiments in this manuscript. 
 We thank members of the Statistical Methods for the Physical Sciences (STAMPS) Research Center at CMU for insightful discussions and feedback on this work.
\end{acks}
%%%%%%%%%%%%%%%%%%%%%%%%%%%%%%%%%%%%%%%%%%%%%%
%% Funding information, if any,             %%
%% should be provided in the                %%
%% funding section.                         %%
%%%%%%%%%%%%%%%%%%%%%%%%%%%%%%%%%%%%%%%%%%%%%%
\begin{funding}
BN, KD, and HZ were supported by the U.S. Department of Energy (DOE), Office of Science under contracts DE-AC02-05CH11231 and DE-AC02-76SF00515. VM was supported by JST EXPERT-J, Japan Grant Number JPMJEX2509. MK, LW and HZ were supported by NSF grants DMS-2053804 and DMS-2310632.
\end{funding}

%%  imsart-nameyear.bst  will be used to                   %%
%%  MR numbers will be added by VTeX.                      %%
%%                                                         %%
%%  Use \cite{...} to cite references in text.             %%
%%                                                         %%
%%%%%%%%%%%%%%%%%%%%%%%%%%%%%%%%%%%%%%%%%%%%%%%%%%%%%%%%%%%%%
\newpage
%% if your bibliography is in bibtex format, uncomment commands:
\bibliographystyle{imsart-nameyear} % Style BST file
\bibliography{myrefs}       % Bibliography file (usually '*.bib')

%% or include bibliography directly:
% \begin{thebibliography}{}
% \bibitem[\protect\citeauthoryear{???}{???}]{b1}
% \end{thebibliography}

%%%%%%%%%%%%%%%%%%%%%%%%%%%%%%%%%%%%%%%%%%%%%%
%% Supplementary Material, including data   %%
%% sets and code, should be provided in     %%
%% {supplement} environment with title      %%
%% and short description. It cannot be      %%
%% available exclusively as external link.  %%
%% All Supplementary Material must be       %%
%% available to the reader on Project       %%
%% Euclid with the published article.       %%
%%%%%%%%%%%%%%%%%%%%%%%%%%%%%%%%%%%%%%%%%%%%%%
\newpage
\renewcommand{\appendixname}{Supplementary Material}
\appendix
\input{supplement_revised}

\end{document}

%% file: supplement_revised.tex
\section{Additional Experimental Results}
\subsection{Gaussian Data}
In this section, we include additional results for Gaussian data with different nuisance parameters. The nuisance parameter $\theta$ takes values in $\{0.6,0.8,1.2,1.4\}$.

\begin{enumerate}
    \item $\theta=0.6$
    \begin{figure}[H]
      \centering
      \includegraphics[width=0.55\linewidth]{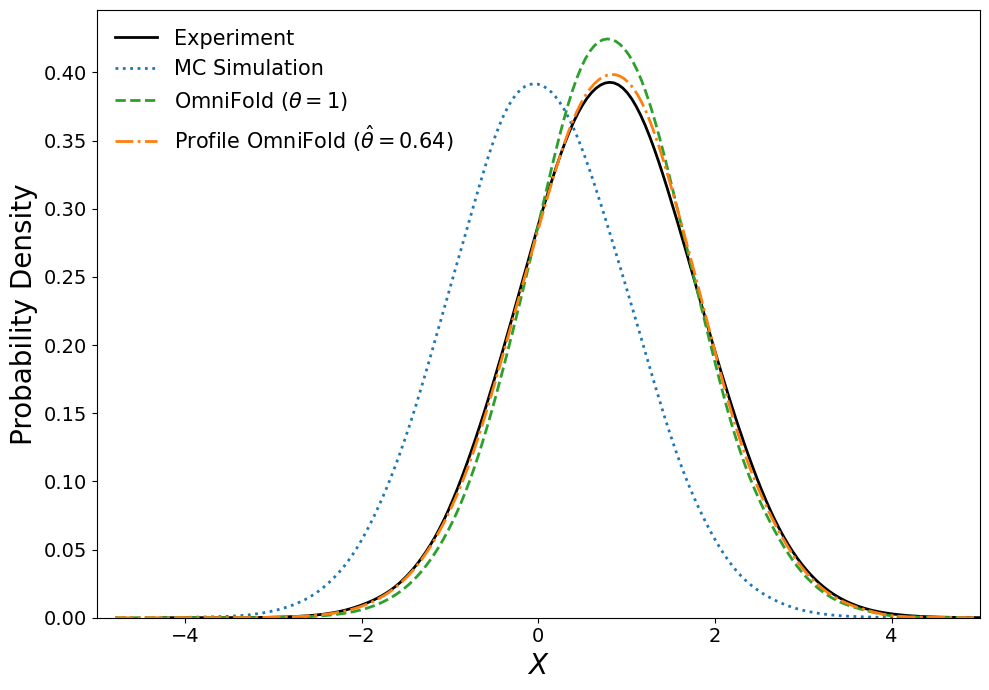}
      \includegraphics[width=0.4\linewidth]{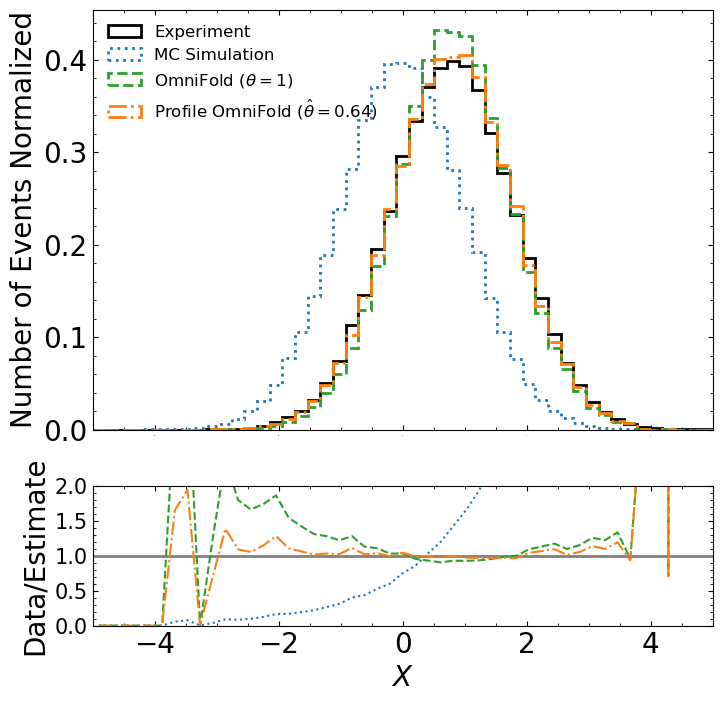}
      \caption{Results of unfolding the Gaussian example. Analytic $\mathcal{W}$ function is being used in the algorithm. \textbf{Left}: Particle-level kernel density estimates of the truth distribution (black), the MC distribution ({\color{blue}blue}), and the reweighted MC distributions obtained using the POF ({\color{orange}orange}) and OF ({\color{ForestGreen}green}) algorithms, each run for 10 iterations. \textbf{Top-right}: Histograms of the four corresponding spectra, aggregated into 50 bins. \textbf{Bottom-right}: The ratio of the truth spectrum to the unfolded spectra.} 
      \label{fig:gaussian_data_x_theta=0.6}
    \end{figure}
    
    \begin{figure}[H]
        \centering
        \includegraphics[width=0.45\linewidth]{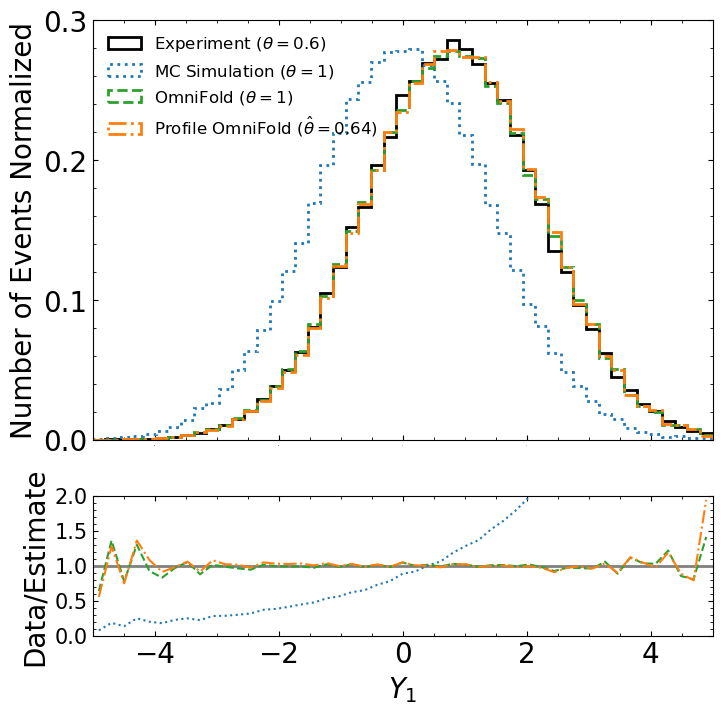}
        \includegraphics[width=0.45\linewidth]{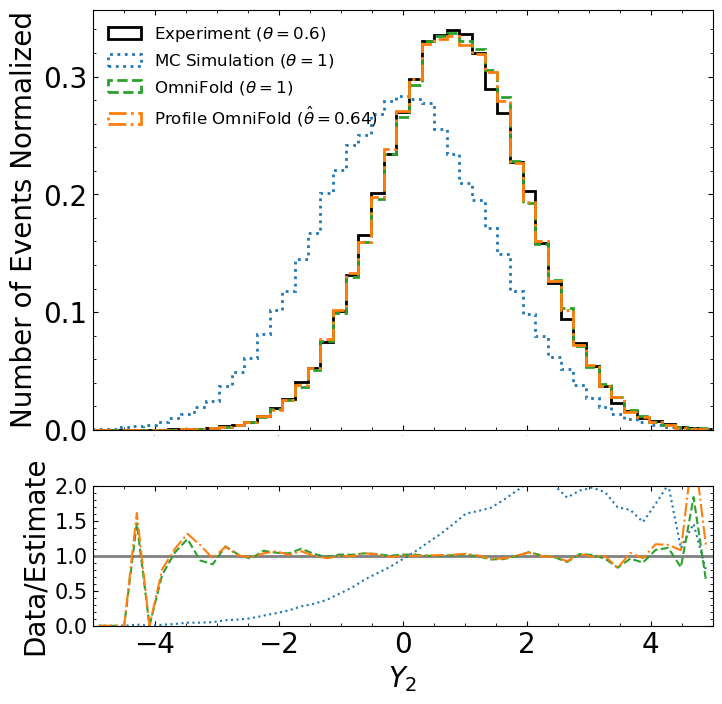}
        \caption{Results corresponding to Figure~\ref{fig:gaussian_data_x_theta=0.6} in detector-level space. \textbf{Left}: Histograms of the corresponding spectra of $Y_1$. \textbf{Right}: Histograms of the corresponding spectra of $Y_2$.}
        \label{fig:gaussian_data_y_theta=0.6}
    \end{figure}
    
    \begin{figure}[H]
        \centering
        \includegraphics[width=\linewidth]{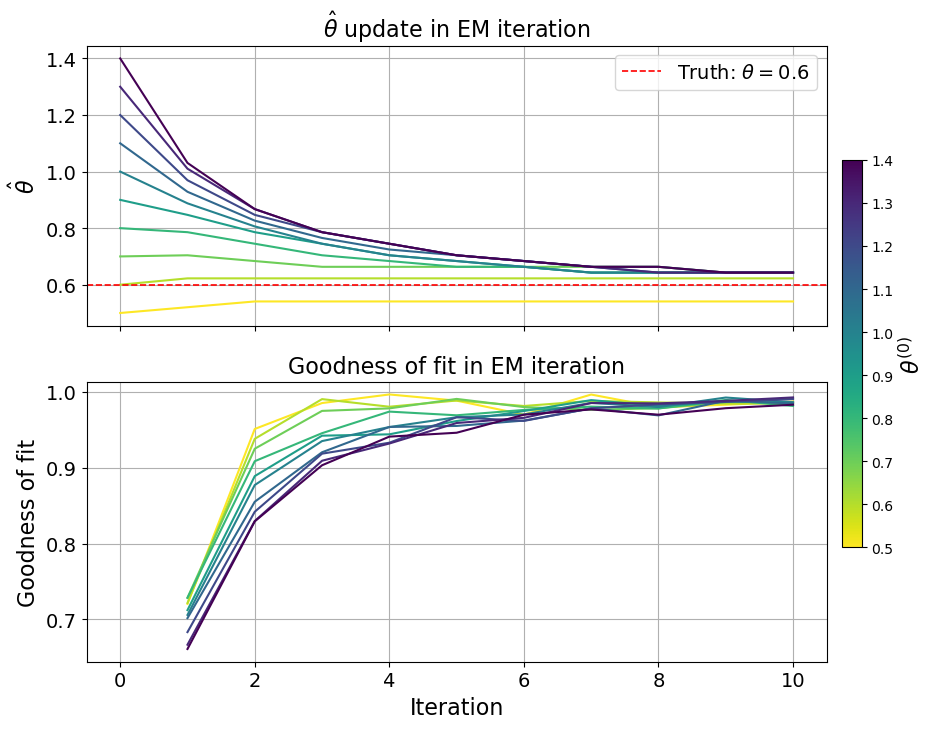}
        \caption{Evolution of the nuisance parameter and the step-1 classifier’s goodness-of-fit statistic for the POF algorithm under different initializations $\theta^{(0)}$. The results shown in Figures~\ref{fig:gaussian_data_x_theta=0.6}--\ref{fig:gaussian_data_y_theta=0.6} correspond to the solution yielding the highest goodness-of-fit statistic.  \textbf{Top}: Updated estimates $\hat{\theta}$ across iterations for different initializations $\theta^{(0)}$. \textbf{Bottom}: Goodness-of-fit statistic of the step-1 classifier at each iteration.}
        \label{fig:gaussian_data_theta_acc_evolution_theta=0.6}
    \end{figure}

    \newpage
    \item $\theta=0.8$
    
    \begin{figure}[H]
      \centering
      \includegraphics[width=0.55\linewidth]{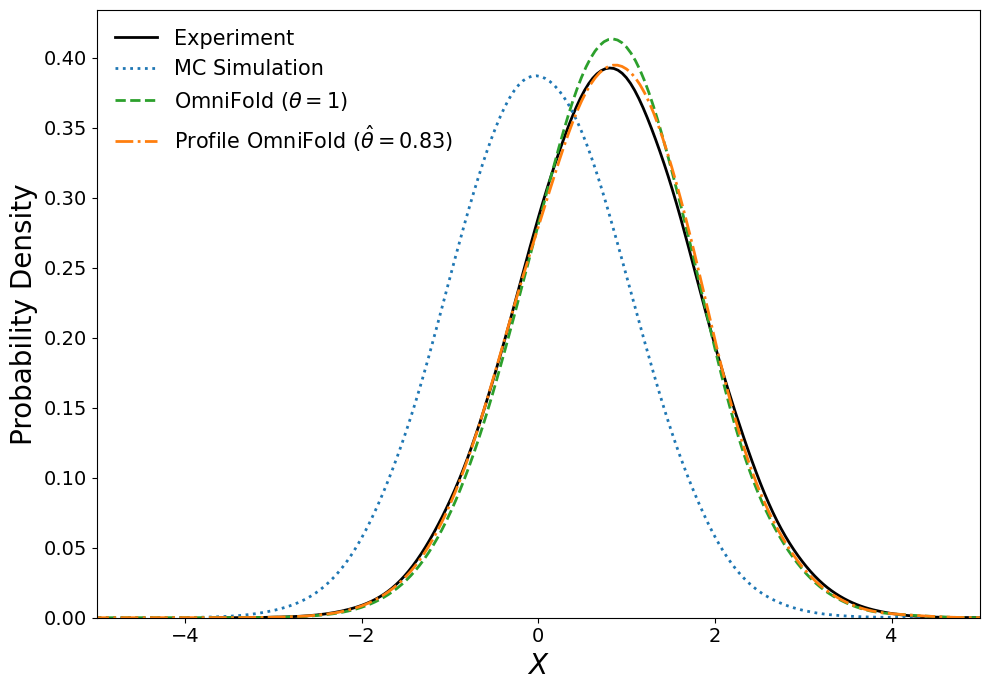}
      \includegraphics[width=0.4\linewidth]{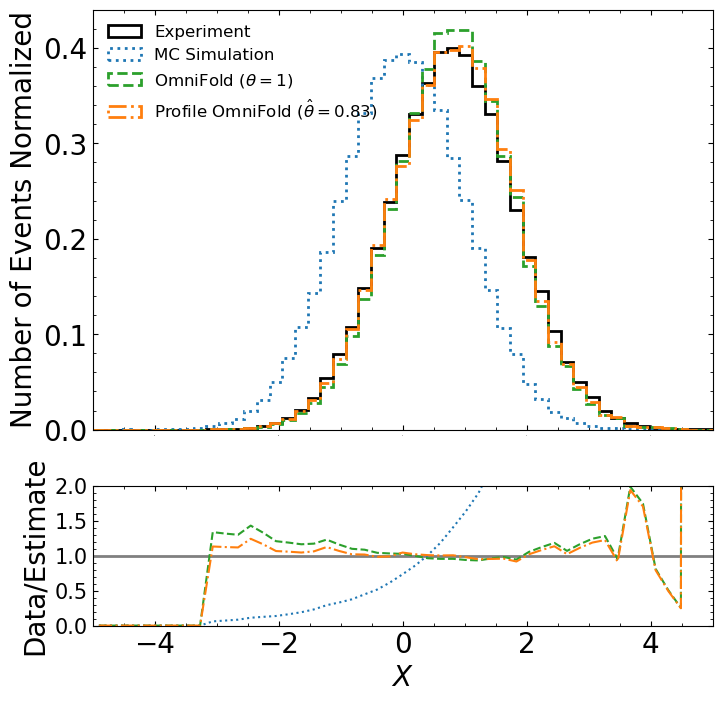}
      \caption{Results of unfolding the Gaussian example. Analytic $\mathcal{W}$ function is being used in the algorithm. \textbf{Left}: Particle-level kernel density estimates of the truth distribution (black), the MC distribution ({\color{blue}blue}), and the reweighted MC distributions obtained using the POF ({\color{orange}orange}) and OF ({\color{ForestGreen}green}) algorithms, each run for 10 iterations. \textbf{Top-right}: Histograms of the four corresponding spectra, aggregated into 50 bins. \textbf{Bottom-right}: The ratio of the truth spectrum to the unfolded spectra.} 
      \label{fig:gaussian_data_x_theta=0.8}
    \end{figure}

    \begin{figure}[H]
        \centering
        \includegraphics[width=0.45\linewidth]{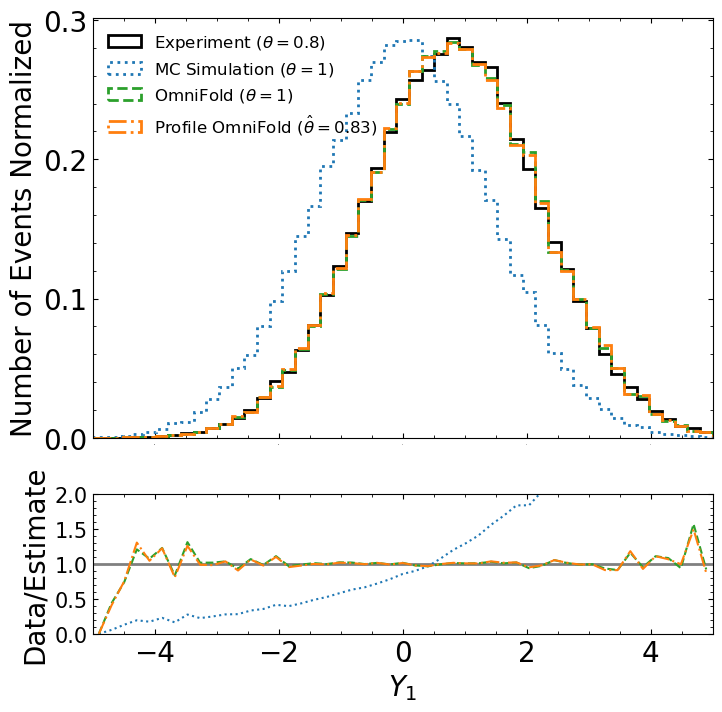}
        \includegraphics[width=0.45\linewidth]{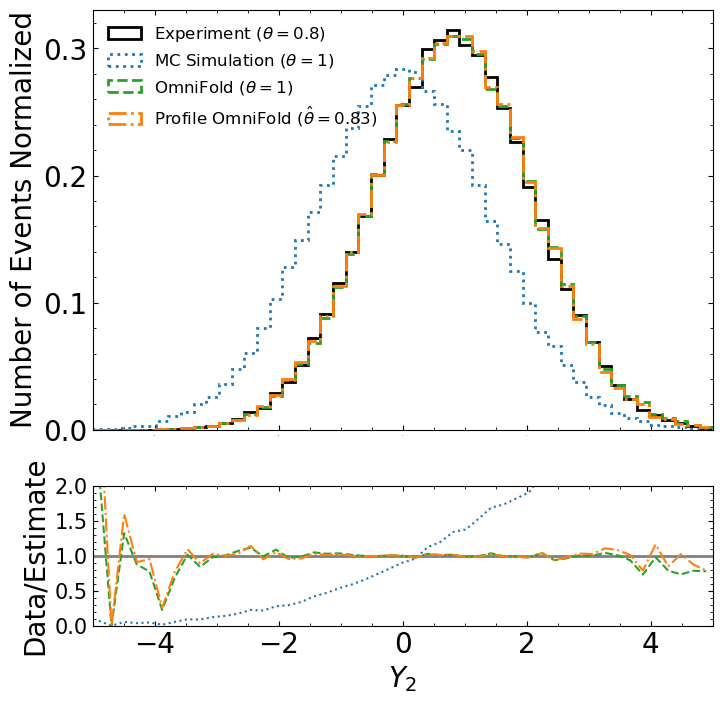}
        \caption{Results corresponding to Figure~\ref{fig:gaussian_data_x_theta=0.8} in detector-level space. \textbf{Left}: Histograms of the corresponding spectra of $Y_1$. \textbf{Right}: Histograms of the corresponding spectra of $Y_2$.}
        \label{fig:gaussian_data_y_theta=0.8}
    \end{figure}

    \begin{figure}[H]
        \centering
        \includegraphics[width=\linewidth]{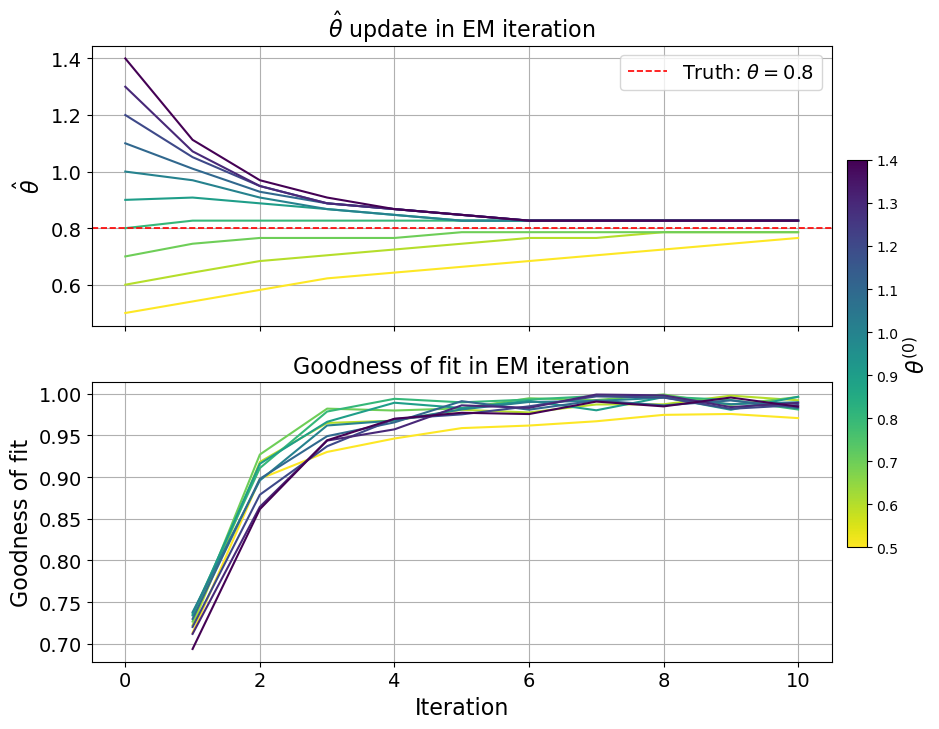}
        \caption{Evolution of the nuisance parameter and the step-1 classifier’s goodness-of-fit statistic for the POF algorithm under different initializations $\theta^{(0)}$. The results shown in Figures~\ref{fig:gaussian_data_x_theta=0.8}--\ref{fig:gaussian_data_y_theta=0.8} correspond to the solution yielding the highest goodness-of-fit statistic. \textbf{Top}: Updated estimates $\hat{\theta}$ across iterations for different initializations $\theta^{(0)}$. \textbf{Bottom}: Goodness-of-fit statistic of the step-1 classifier at each iteration.}
        \label{fig:gaussian_data_theta_acc_evolution_theta=0.8}
    \end{figure}
\newpage
    \item $\theta=1.2$
    \begin{figure}[H]
      \centering
      \includegraphics[width=0.55\linewidth]{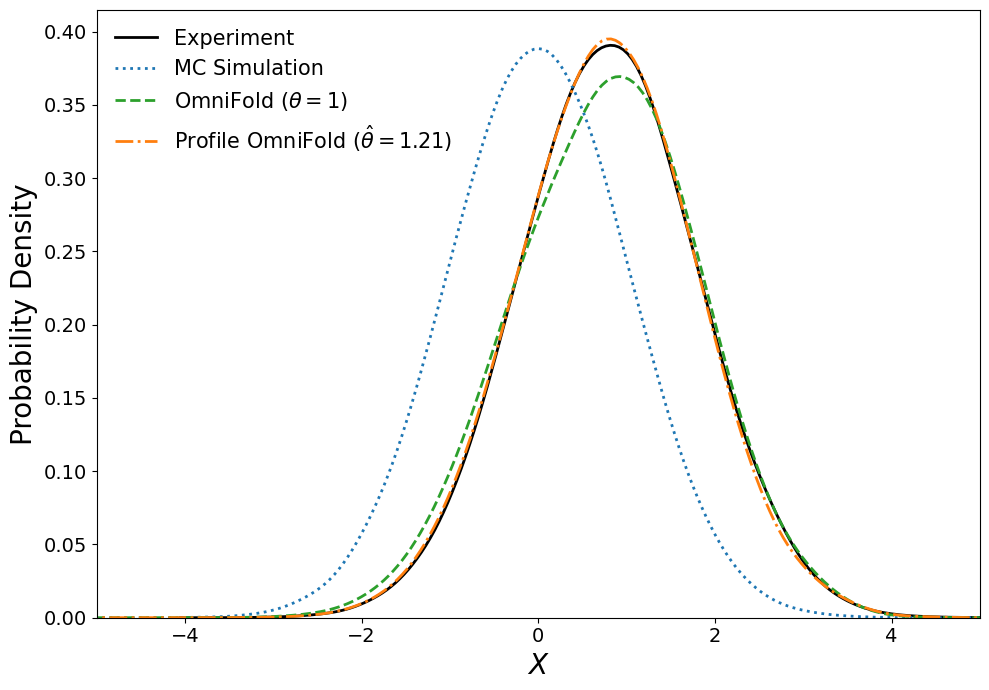}
      \includegraphics[width=0.4\linewidth]{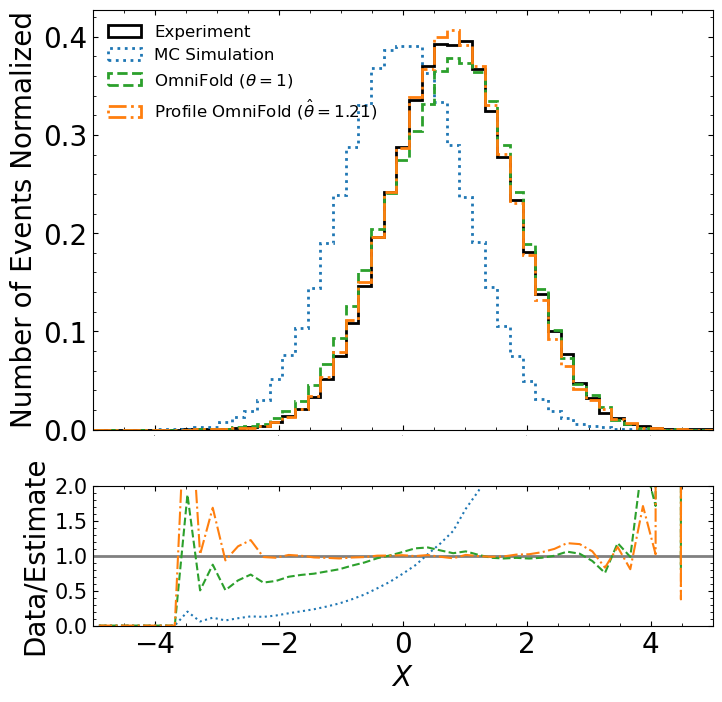}
      \caption{Results of unfolding the Gaussian example. Analytic $\mathcal{W}$ function is being used in the algorithm. \textbf{Left}: Particle-level kernel density estimates of the truth distribution (black), the MC distribution ({\color{blue}blue}), and the reweighted MC distributions obtained using the POF ({\color{orange}orange}) and OF ({\color{ForestGreen}green}) algorithms, each run for 10 iterations. \textbf{Top-right}: Histograms of the four corresponding spectra, aggregated into 50 bins. \textbf{Bottom-right}: The ratio of the truth spectrum to the unfolded spectra.} 
      \label{fig:gaussian_data_x_theta=1.2}
    \end{figure}

    \begin{figure}[H]
        \centering
        \includegraphics[width=0.45\linewidth]{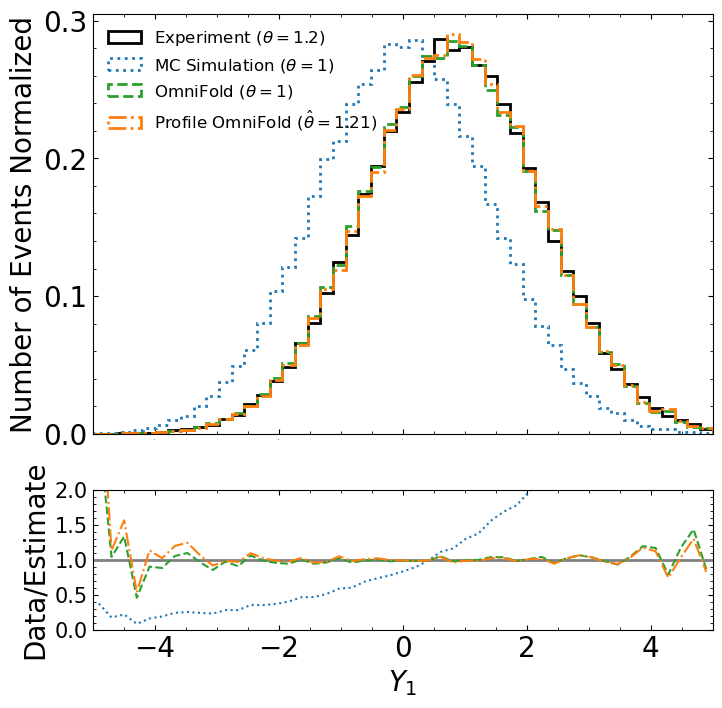}
        \includegraphics[width=0.45\linewidth]{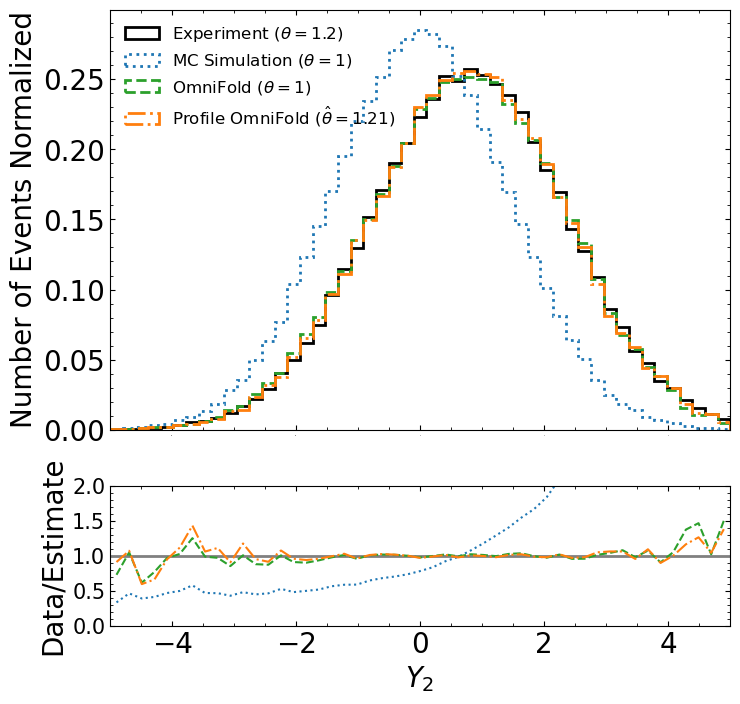}
        \caption{Results corresponding to Figure~\ref{fig:gaussian_data_x_theta=1.2} in detector-level space. \textbf{Left}: Histograms of the corresponding spectra of $Y_1$. \textbf{Right}: Histograms of the corresponding spectra of $Y_2$.}
        \label{fig:gaussian_data_y_theta=1.2}
    \end{figure}

    \begin{figure}[H]
        \centering
        \includegraphics[width=\linewidth]{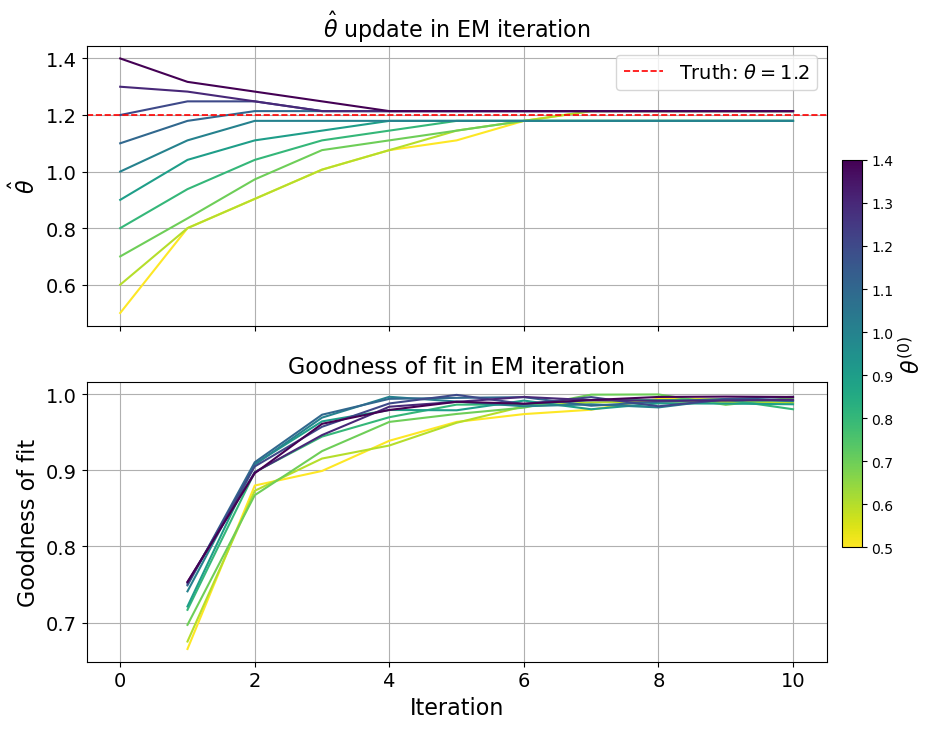}
        \caption{Evolution of the nuisance parameter and the step-1 classifier’s goodness-of-fit statistic for the POF algorithm under different initializations $\theta^{(0)}$. The results shown in Figures~\ref{fig:gaussian_data_x_theta=1.2}--\ref{fig:gaussian_data_y_theta=1.2} correspond to the solution yielding the highest goodness-of-fit statistic. \textbf{Top}: Updated estimates $\hat{\theta}$ across iterations for different initializations $\theta^{(0)}$. \textbf{Bottom}: Goodness-of-fit statistic of the step-1 classifier at each iteration.}
        \label{fig:gaussian_data_theta_acc_evolution_theta=1.2}
    \end{figure}

    \newpage
    \item $\theta=1.4$
    
    \begin{figure}[H]
      \centering
      \includegraphics[width=0.55\linewidth]{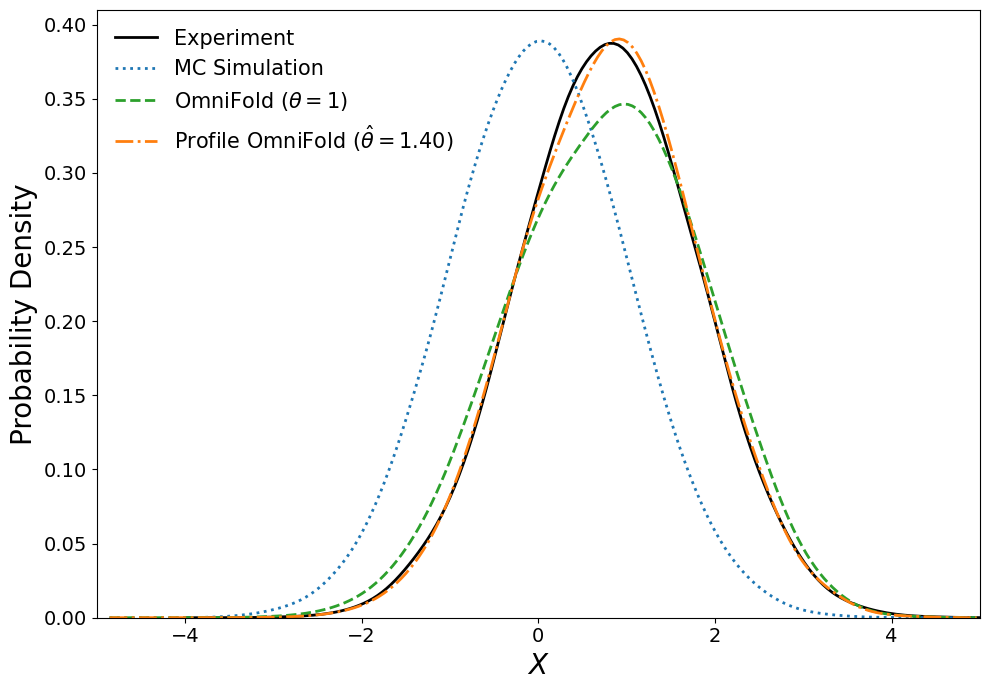}
      \includegraphics[width=0.4\linewidth]{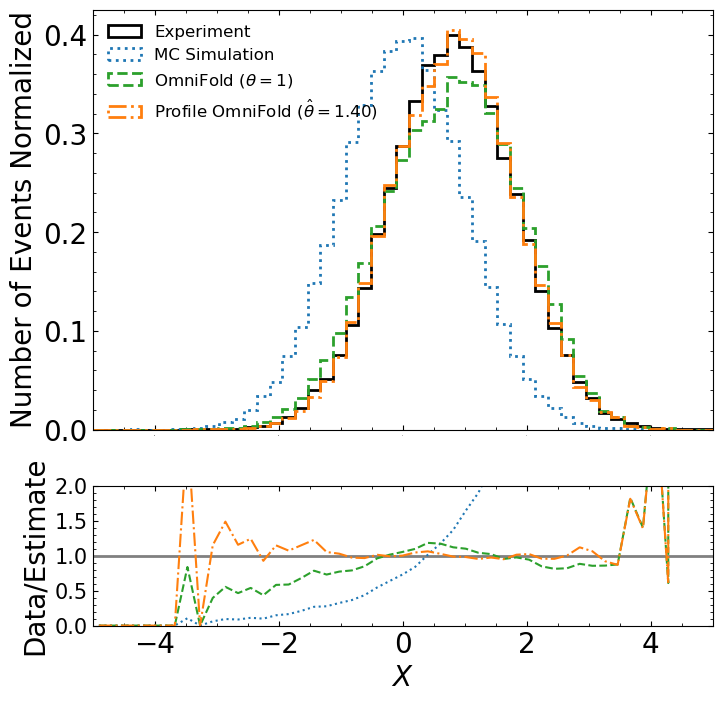}
      \caption{Results of unfolding the Gaussian example. Analytic $\mathcal{W}$ function is being used in the algorithm. \textbf{Left}: Particle-level kernel density estimates of the truth distribution (black), the MC distribution ({\color{blue}blue}), and the reweighted MC distributions obtained using the POF ({\color{orange}orange}) and OF ({\color{ForestGreen}green}) algorithms, each run for 10 iterations. \textbf{Top-right}: Histograms of the four corresponding spectra, aggregated into 50 bins. \textbf{Bottom-right}: The ratio of the truth spectrum to the unfolded spectra.} 
      \label{fig:gaussian_data_x_theta=1.4}
    \end{figure}

    \begin{figure}[H]
        \centering
        \includegraphics[width=0.45\linewidth]{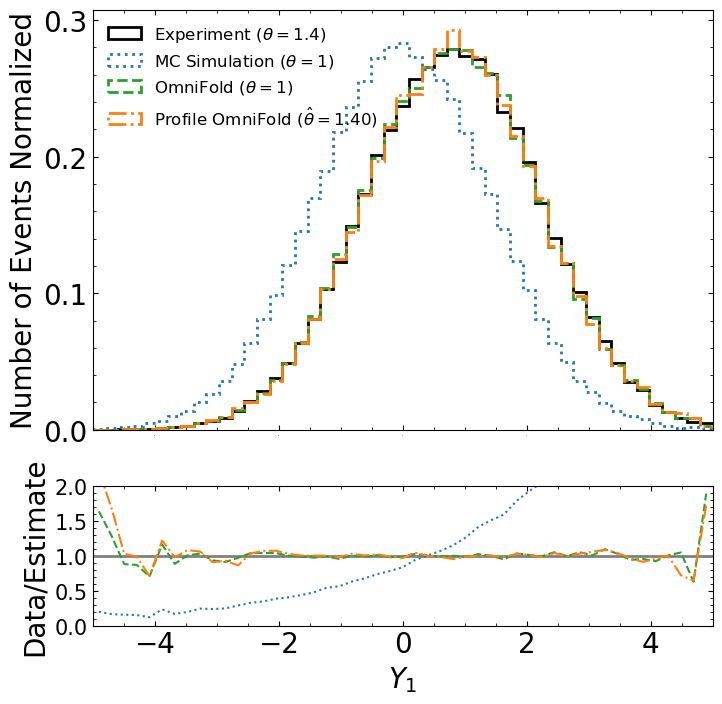}
        \includegraphics[width=0.45\linewidth]{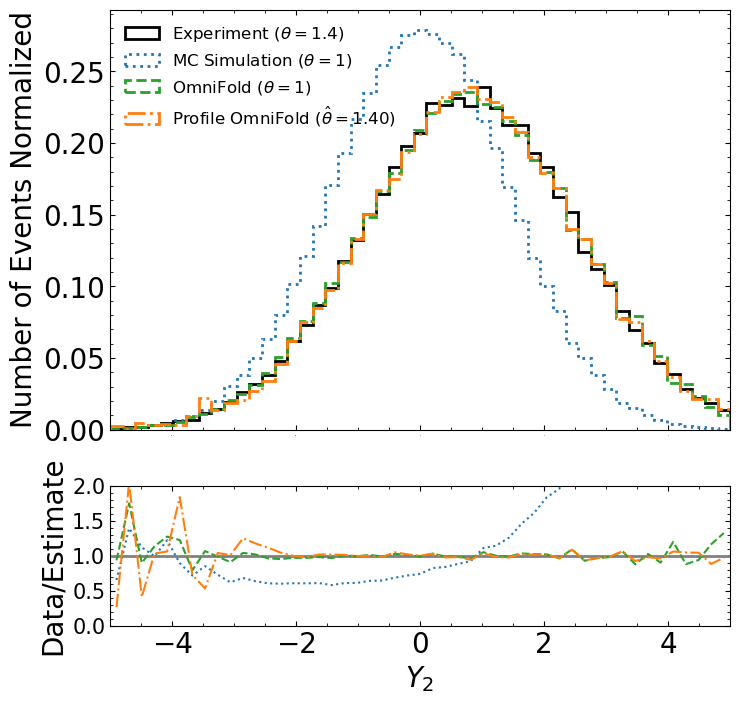}
        \caption{Results corresponding to Figure~\ref{fig:gaussian_data_x_theta=1.4} in detector-level space. \textbf{Left}: Histograms of the corresponding spectra of $Y_1$. \textbf{Right}: Histograms of the corresponding spectra of $Y_2$.}
        \label{fig:gaussian_data_y_theta=1.4}
    \end{figure}

    \begin{figure}[H]
        \centering
        \includegraphics[width=\linewidth]{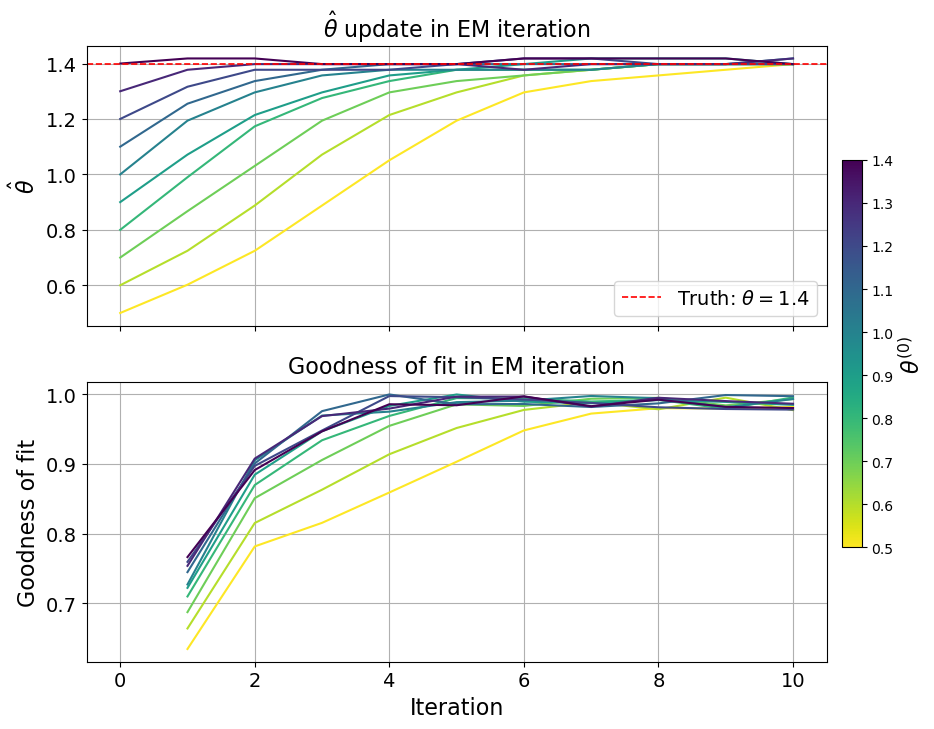}
        \caption{Evolution of the nuisance parameter and the step-1 classifier’s goodness-of-fit statistic for the POF algorithm under different initializations $\theta^{(0)}$. The results shown in Figures~\ref{fig:gaussian_data_x_theta=1.4}--\ref{fig:gaussian_data_y_theta=1.4} correspond to the solution yielding the highest goodness-of-fit statistic. \textbf{Top}: Updated estimates $\hat{\theta}$ across iterations for different initializations $\theta^{(0)}$. \textbf{Bottom}: Goodness-of-fit statistic of the step-1 classifier at each iteration.}
        \label{fig:gaussian_data_theta_acc_evolution_theta=1.4}
    \end{figure}

\end{enumerate}

\newpage

\subsection{CMS Open Data}
In this section, we include extended empirical results for the CMS open data. The nuisance parameter $\theta$ takes values in $\{0.6,0.8,1.2,1.4\}$. The range of nuisance parameter $\theta$ used in training $\mathcal{W}$ function is set to be $[0.5,1.5]$.

\begin{enumerate}
    \item $\theta=0.6$
    \begin{figure}[H]
      \centering
      \includegraphics[width=0.55\linewidth]{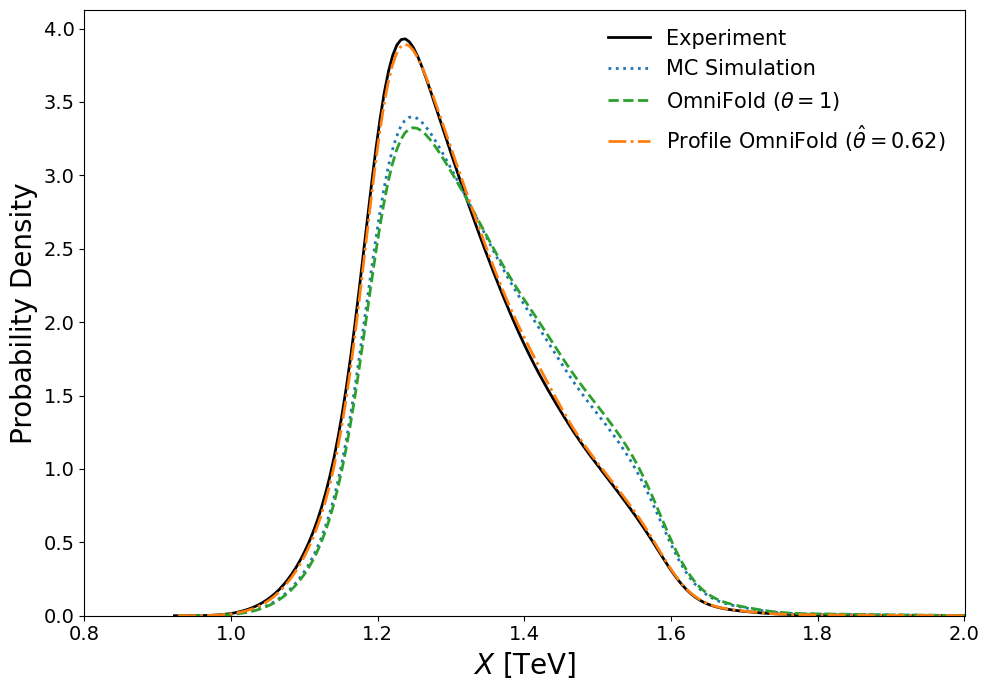}
      \includegraphics[width=0.4\linewidth]{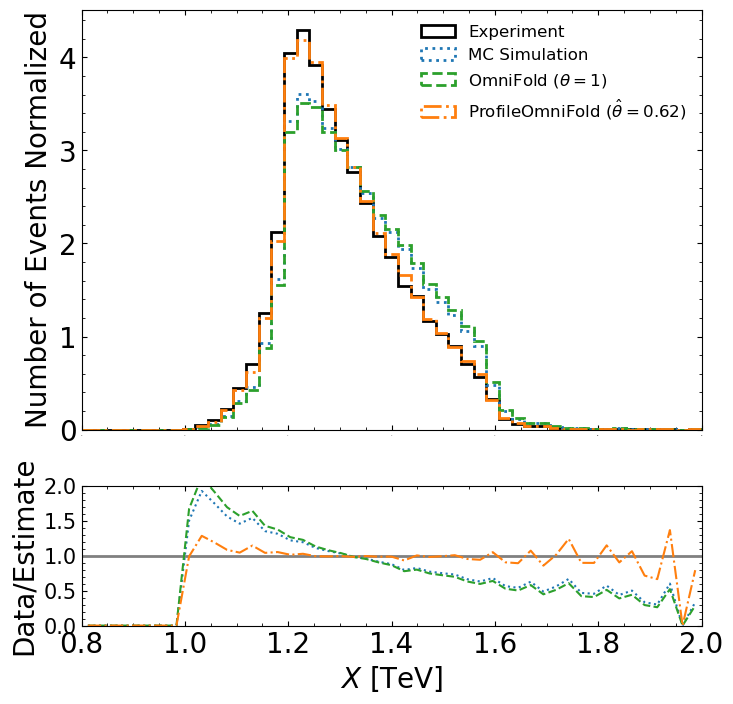}
      \caption{Results of unfolding the CMS open data. \textbf{Left}: Particle-level kernel density estimates of the truth distribution (black), the MC distribution ({\color{blue}blue}), and the reweighted MC distributions obtained using the POF ({\color{orange}orange}) and OF ({\color{ForestGreen}green}) algorithms, each run for 10 iterations. \textbf{Top-right}: Histograms of the four corresponding spectra, aggregated into 50 bins. \textbf{Bottom-right}: The ratio of the truth spectrum to the unfolded spectra.} 
      \label{fig:open_data_x_theta=0.6}
    \end{figure}
    
    \begin{figure}[H]
        \centering
        \includegraphics[width=0.45\linewidth]{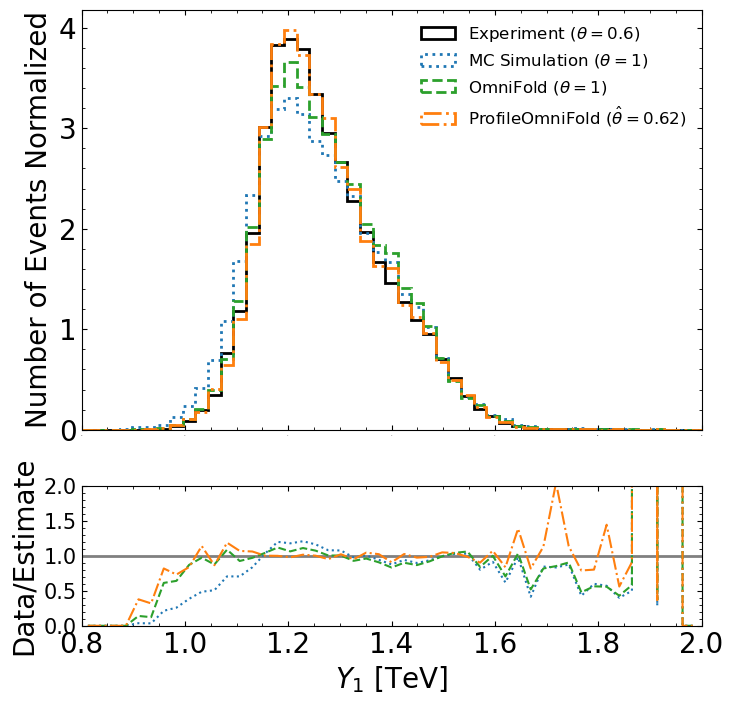}
        \includegraphics[width=0.45\linewidth]{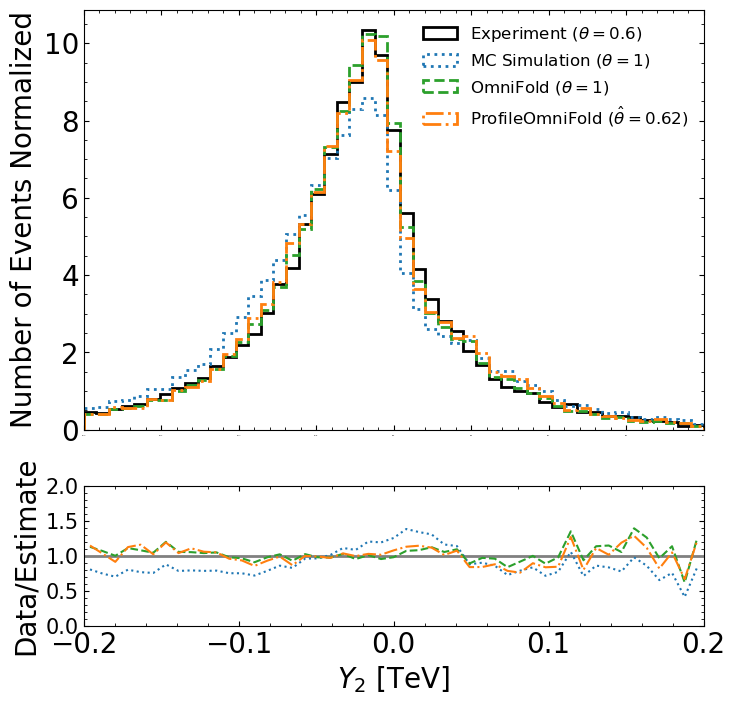}
        \caption{Results corresponding to Figure~\ref{fig:open_data_x_theta=0.6} in detector-level space. \textbf{Left}: Histograms of the corresponding spectra of $Y_1$. \textbf{Right}: Histograms of the corresponding spectra of $Y_2$.}
        \label{fig:open_data_y_theta=0.6}
    \end{figure}
    
    \begin{figure}[H]
        \centering
        \includegraphics[width=\linewidth]{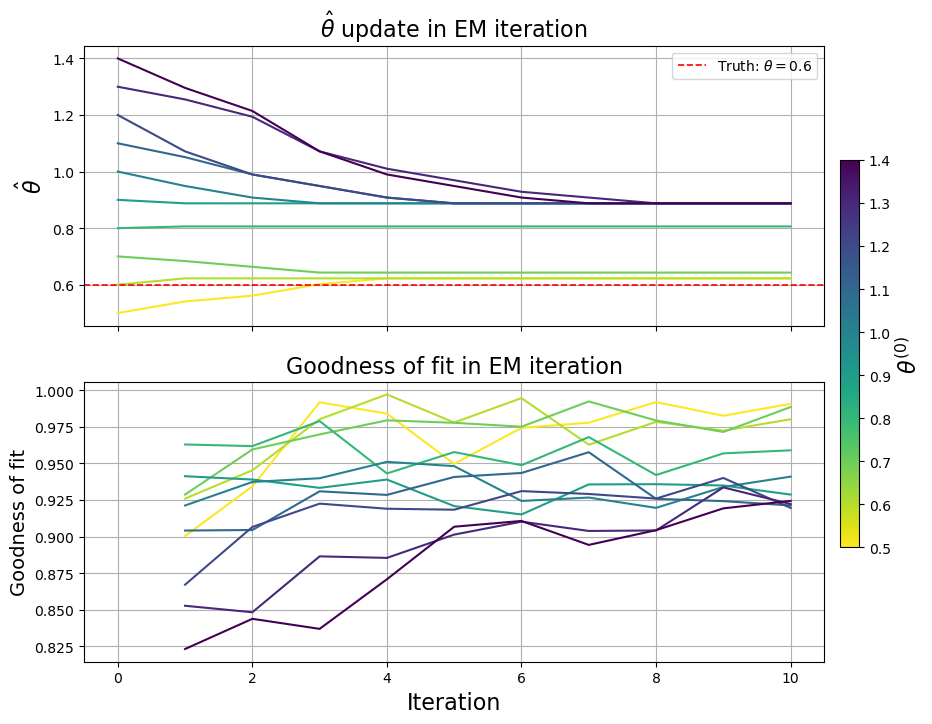}
        \caption{Evolution of the nuisance parameter and the step-1 classifier’s goodness-of-fit statistic for the POF algorithm under different initializations $\theta^{(0)}$. The results shown in Figures~\ref{fig:open_data_x_theta=0.6}--\ref{fig:open_data_y_theta=0.6} correspond to the solution yielding the highest goodness-of-fit statistic. \textbf{Top}: Updated estimates $\hat{\theta}$ across iterations for different initializations $\theta^{(0)}$. \textbf{Bottom}: Goodness-of-fit statistic of the step-1 classifier at each iteration.}
        \label{fig:open_data_theta_acc_evolution_theta=0.6}
    \end{figure}
\newpage
    \item $\theta=0.8$
    \begin{figure}[H]
      \centering
      \includegraphics[width=0.55\linewidth]{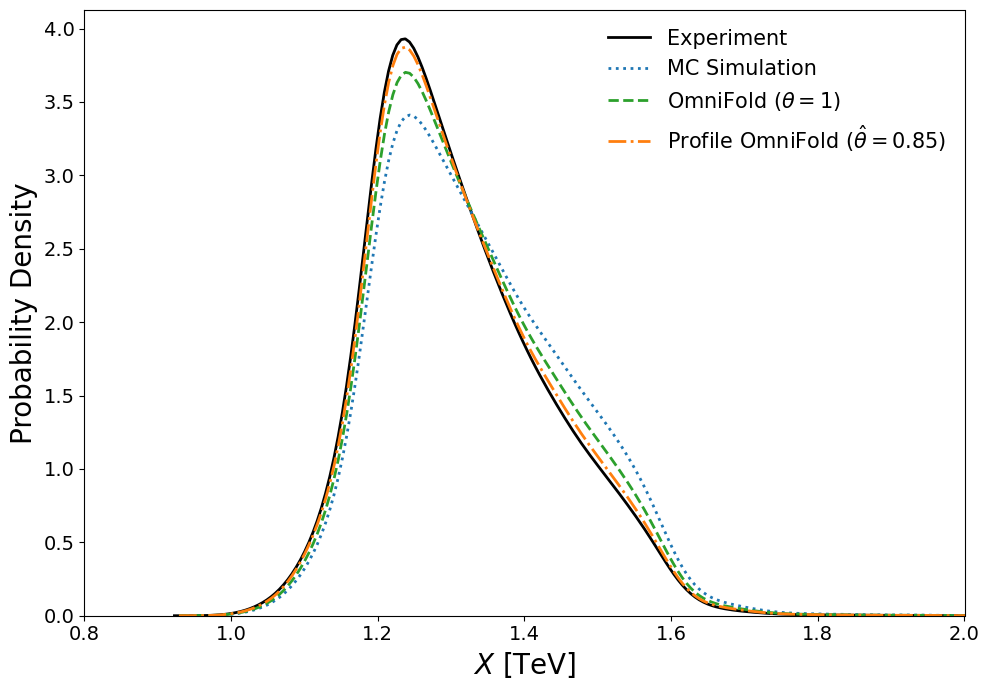}
      \includegraphics[width=0.4\linewidth]{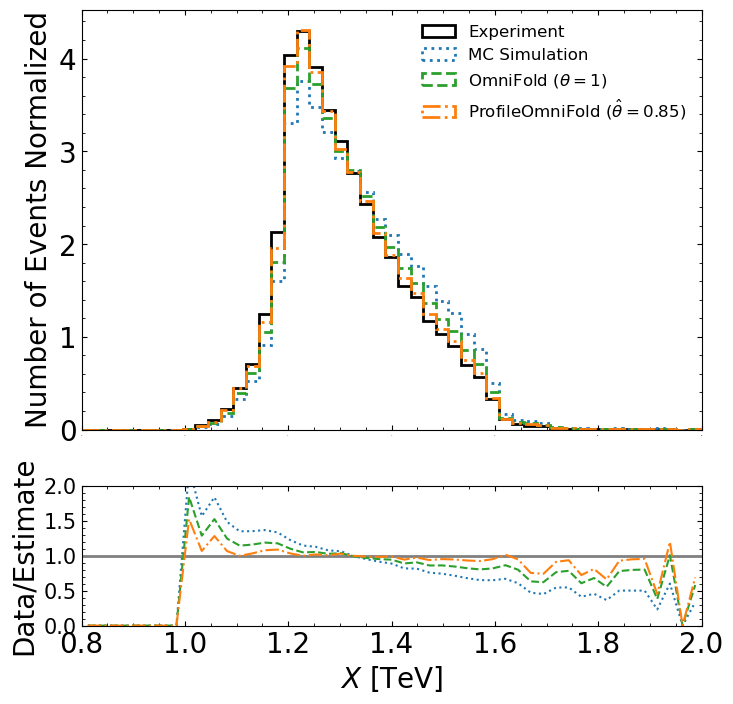}
      \caption{Results of unfolding the CMS open data. \textbf{Left}: Particle-level kernel density estimates of the truth distribution (black), the MC distribution ({\color{blue}blue}), and the reweighted MC distributions obtained using the POF ({\color{orange}orange}) and OF ({\color{ForestGreen}green}) algorithms, each run for 10 iterations. \textbf{Top-right}: Histograms of the four corresponding spectra, aggregated into 50 bins. \textbf{Bottom-right}: The ratio of the truth spectrum to the unfolded spectra.} 
      \label{fig:open_data_x_theta=0.8}
    \end{figure}
    
    \begin{figure}[H]
        \centering
        \includegraphics[width=0.45\linewidth]{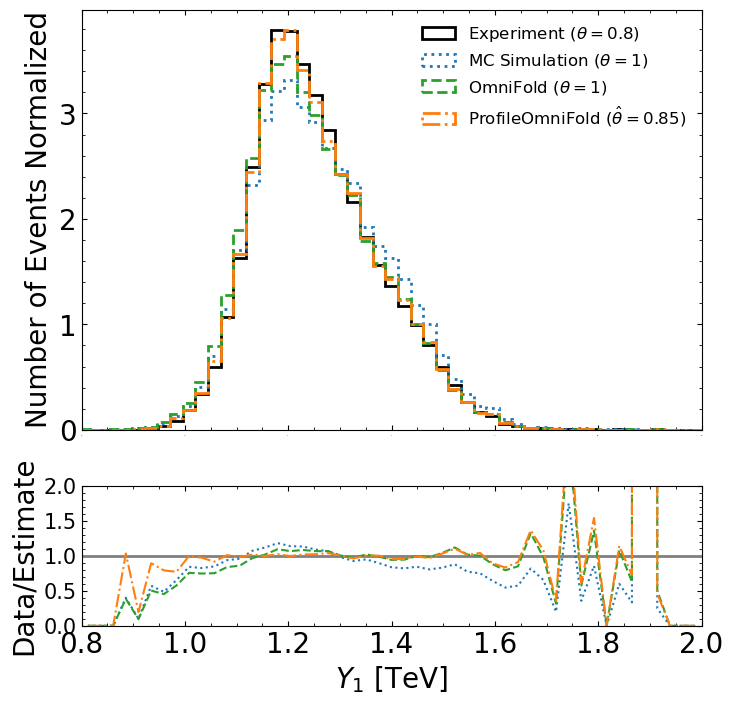}
        \includegraphics[width=0.45\linewidth]{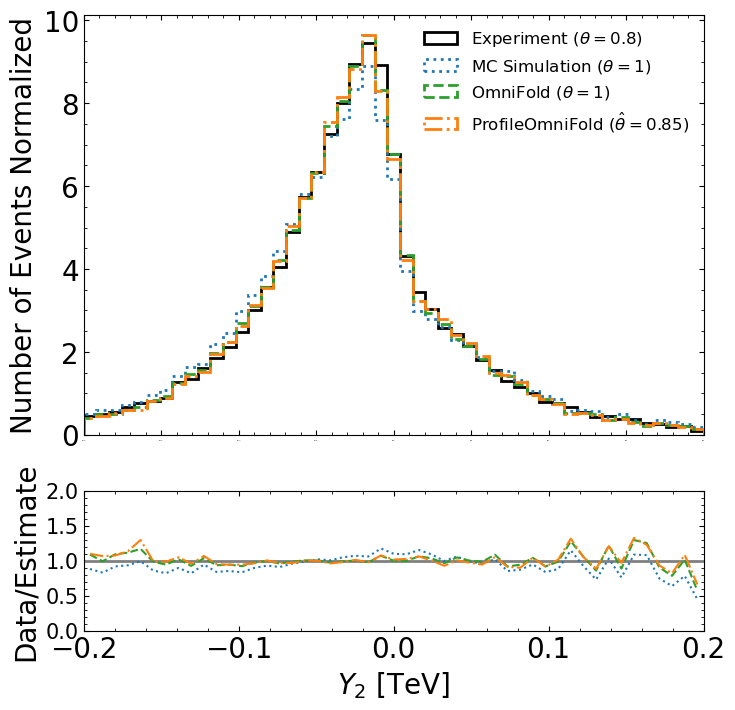}
        \caption{Results corresponding to Figure~\ref{fig:open_data_x_theta=0.8} in detector-level space. \textbf{Left}: Histograms of the corresponding spectra of $Y_1$. \textbf{Right}: Histograms of the corresponding spectra of $Y_2$.}
        \label{fig:open_data_y_theta=0.8}
    \end{figure}
    
    \begin{figure}[H]
        \centering
        \includegraphics[width=\linewidth]{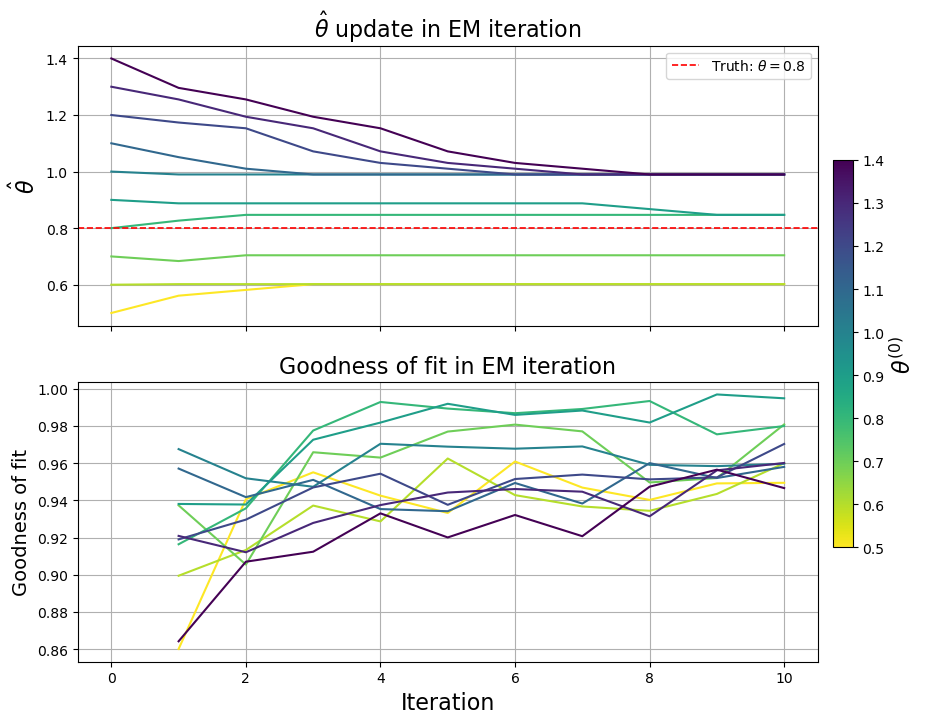}
        \caption{Evolution of the nuisance parameter and the step-1 classifier’s goodness-of-fit statistic for the POF algorithm under different initializations $\theta^{(0)}$. The results shown in Figures~\ref{fig:open_data_x_theta=0.8}--\ref{fig:open_data_y_theta=0.8} correspond to the solution yielding the highest goodness-of-fit statistic. \textbf{Top}: Updated estimates $\hat{\theta}$ across iterations for different initializations $\theta^{(0)}$. \textbf{Bottom}: Goodness-of-fit statistic of the step-1 classifier at each iteration.}
        \label{fig:open_data_theta_acc_evolution_theta=0.8}
    \end{figure}

    \newpage
    \item $\theta=1.2$
    \begin{figure}[H]
      \centering
      \includegraphics[width=0.55\linewidth]{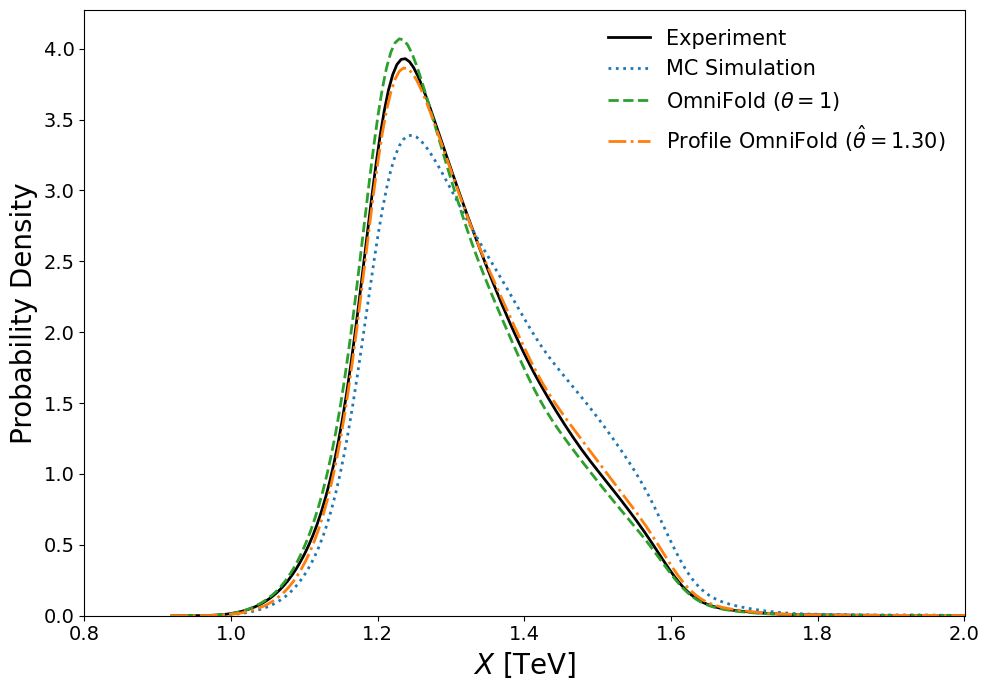}
      \includegraphics[width=0.4\linewidth]{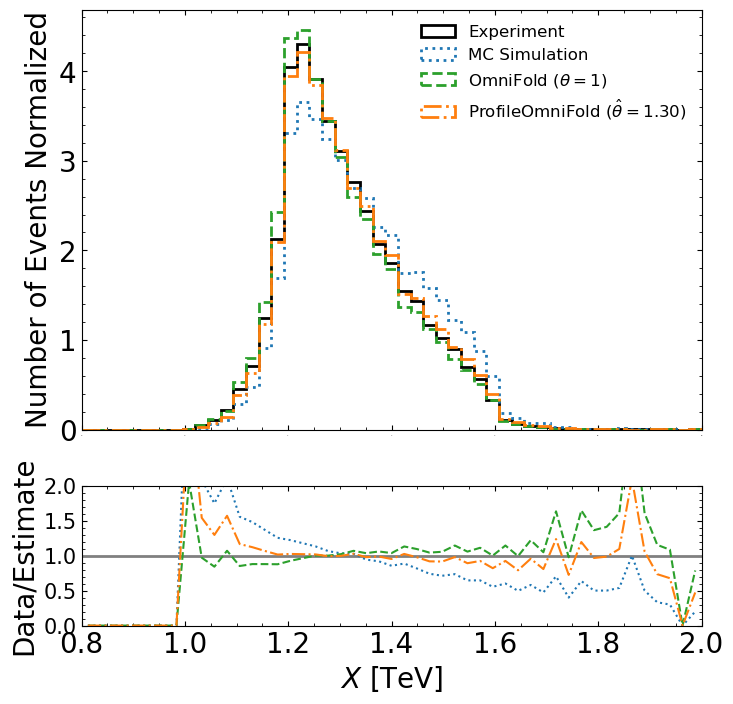}
      \caption{Results of unfolding the CMS open data. \textbf{Left}: Particle-level kernel density estimates of the truth distribution (black), the MC distribution ({\color{blue}blue}), and the reweighted MC distributions obtained using the POF ({\color{orange}orange}) and OF ({\color{ForestGreen}green}) algorithms, each run for 10 iterations. \textbf{Top-right}: Histograms of the four corresponding spectra, aggregated into 50 bins. \textbf{Bottom-right}: The ratio of the truth spectrum to the unfolded spectra.} 
      \label{fig:open_data_x_theta=1.2}
    \end{figure}
    
    \begin{figure}[H]
        \centering
        \includegraphics[width=0.45\linewidth]{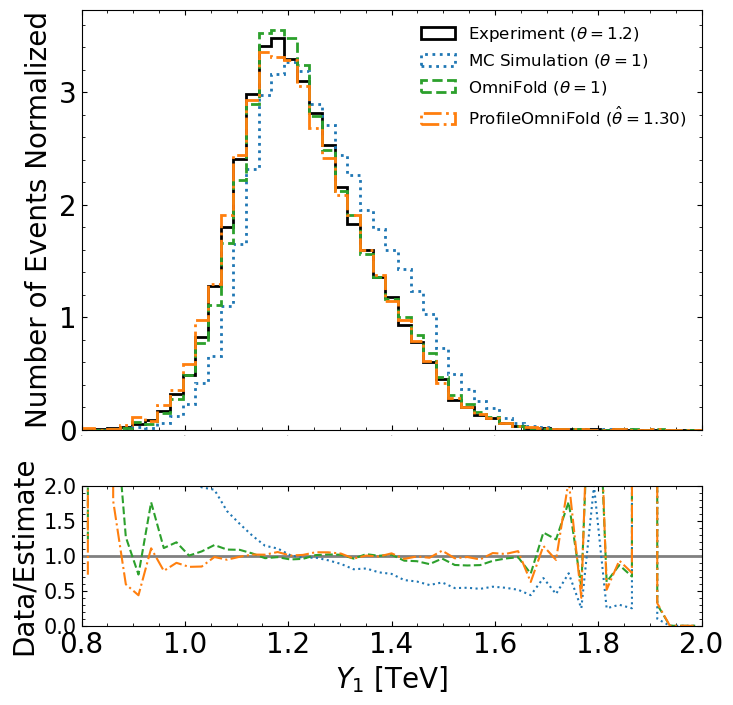}
        \includegraphics[width=0.45\linewidth]{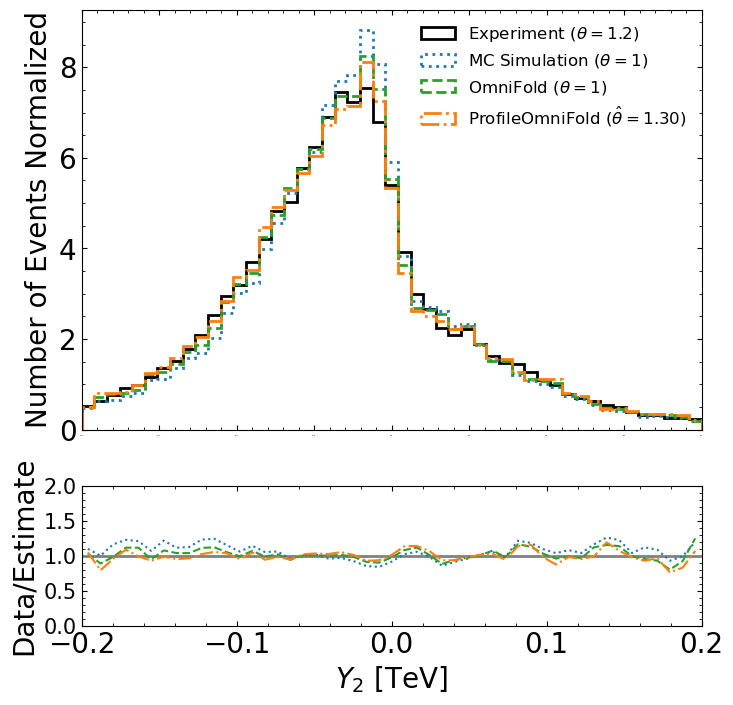}
        \caption{Results corresponding to Figure~\ref{fig:open_data_x_theta=1.2} in detector-level space. \textbf{Left}: Histograms of the corresponding spectra of $Y_1$. \textbf{Right}: Histograms of the corresponding spectra of $Y_2$.}
        \label{fig:open_data_y_theta=1.2}
    \end{figure}

    \begin{figure}[H]
        \centering
        \includegraphics[width=\linewidth]{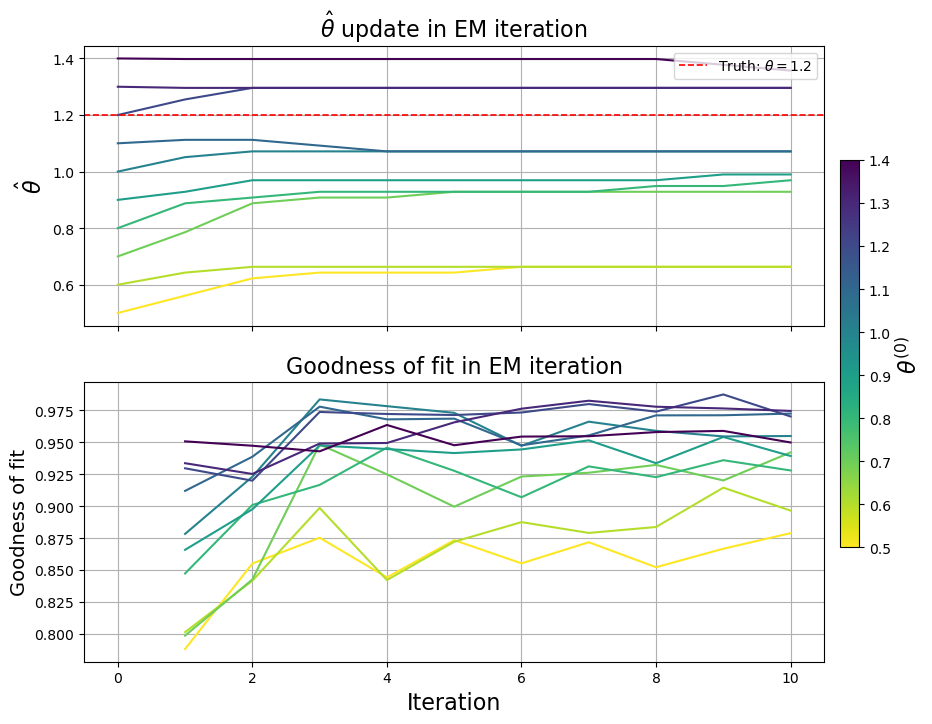}
        \caption{Evolution of the nuisance parameter and the step-1 classifier’s goodness-of-fit statistic for the POF algorithm under different initializations $\theta^{(0)}$. The results shown in Figures~\ref{fig:open_data_x_theta=1.2}--\ref{fig:open_data_y_theta=1.2} correspond to the solution yielding the highest goodness-of-fit statistic. \textbf{Top}: Updated estimates $\hat{\theta}$ across iterations for different initializations $\theta^{(0)}$. \textbf{Bottom}: Goodness-of-fit statistic of the step-1 classifier at each iteration.}
        \label{fig:open_data_theta_acc_evolution_theta=1.2}
    \end{figure}
\newpage
    \item $\theta=1.4$
    \begin{figure}[H]
      \centering
      \includegraphics[width=0.55\linewidth]{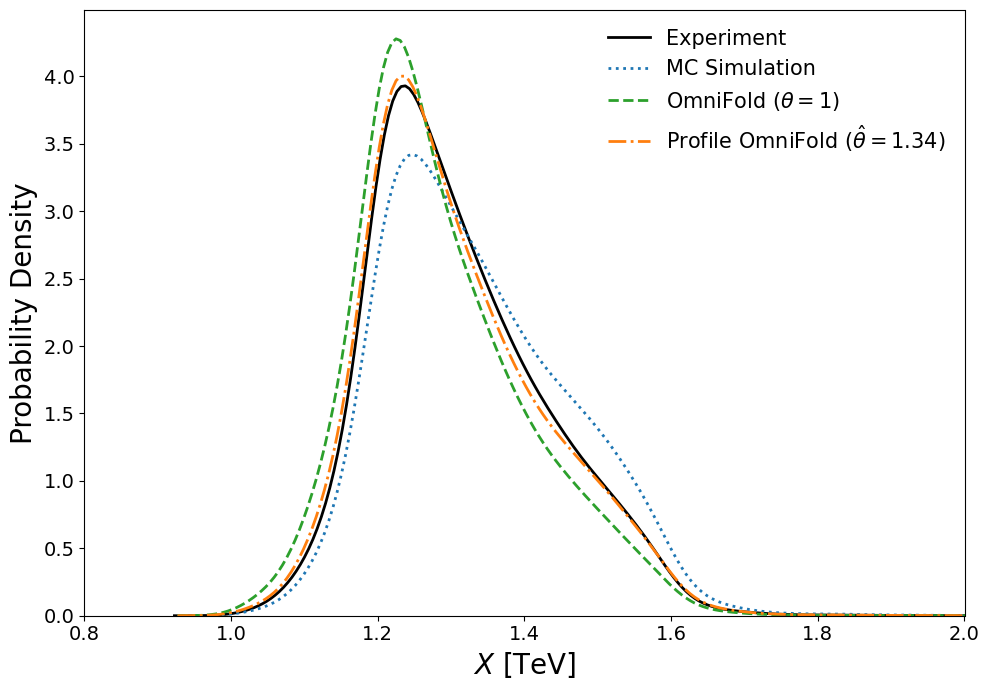}
      \includegraphics[width=0.4\linewidth]{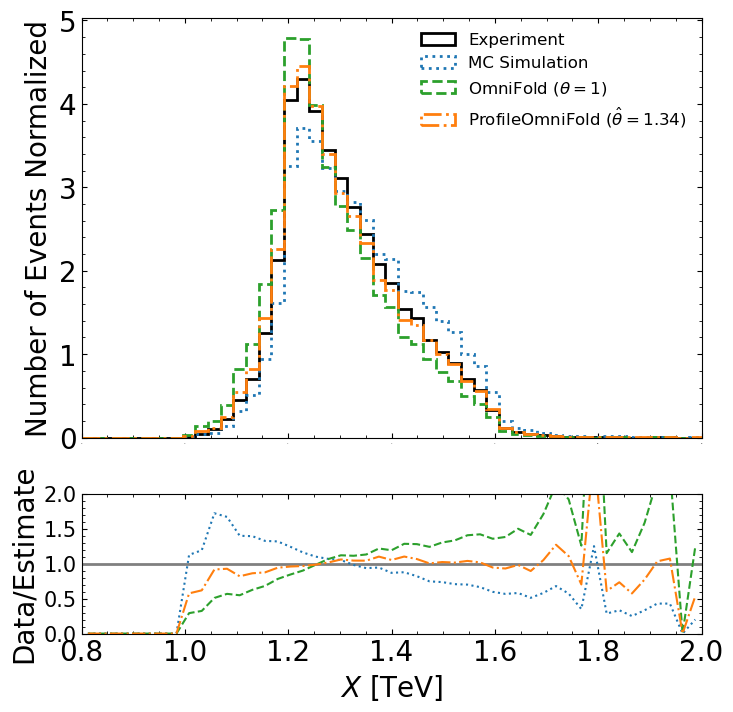}
      \caption{Results of unfolding the CMS open data. \textbf{Left}: Particle-level kernel density estimates of the truth distribution (black), the MC distribution ({\color{blue}blue}), and the reweighted MC distributions obtained using the POF ({\color{orange}orange}) and OF ({\color{ForestGreen}green}) algorithms, each run for 10 iterations. \textbf{Top-right}: Histograms of the four corresponding spectra, aggregated into 50 bins. \textbf{Bottom-right}: The ratio of the truth spectrum to the unfolded spectra.} 
      \label{fig:open_data_x_theta=1.4}
    \end{figure}
    
    \begin{figure}[H]
        \centering
        \includegraphics[width=0.45\linewidth]{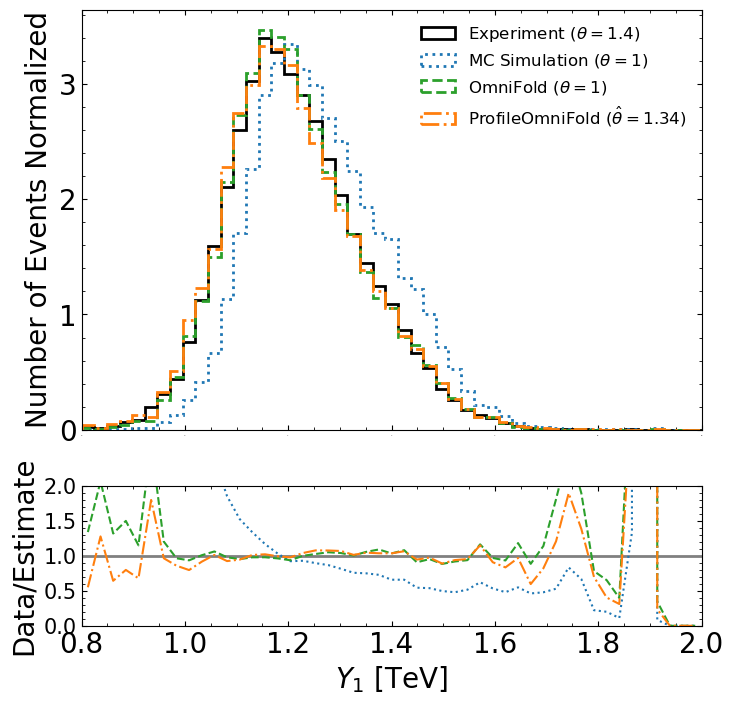}
        \includegraphics[width=0.45\linewidth]{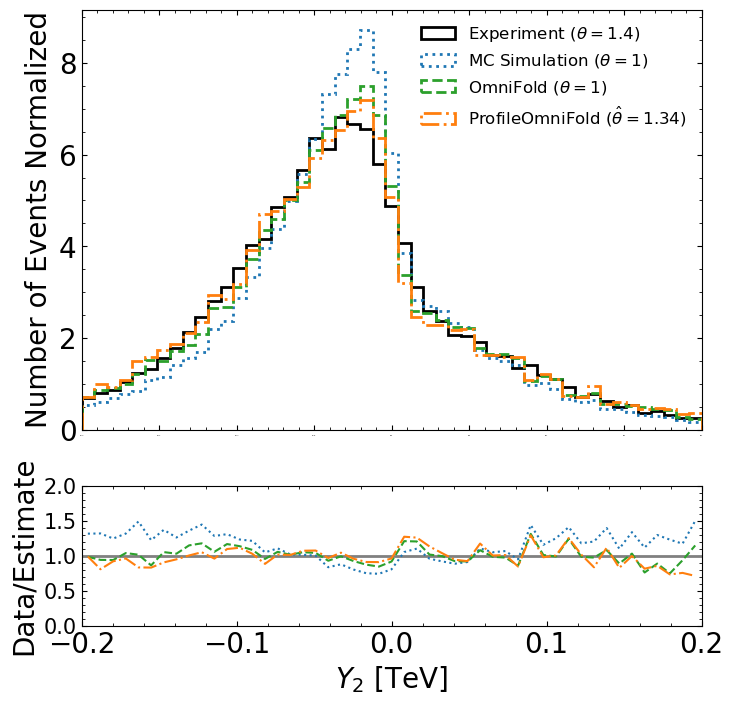}
        \caption{Results corresponding to Figure~\ref{fig:open_data_x_theta=1.4} in detector-level space. \textbf{Left}: Histograms of the corresponding spectra of $Y_1$. \textbf{Right}: Histograms of the corresponding spectra of $Y_2$.}
        \label{fig:open_data_y_theta=1.4}
    \end{figure}

    \begin{figure}[H]
        \centering
        \includegraphics[width=\linewidth]{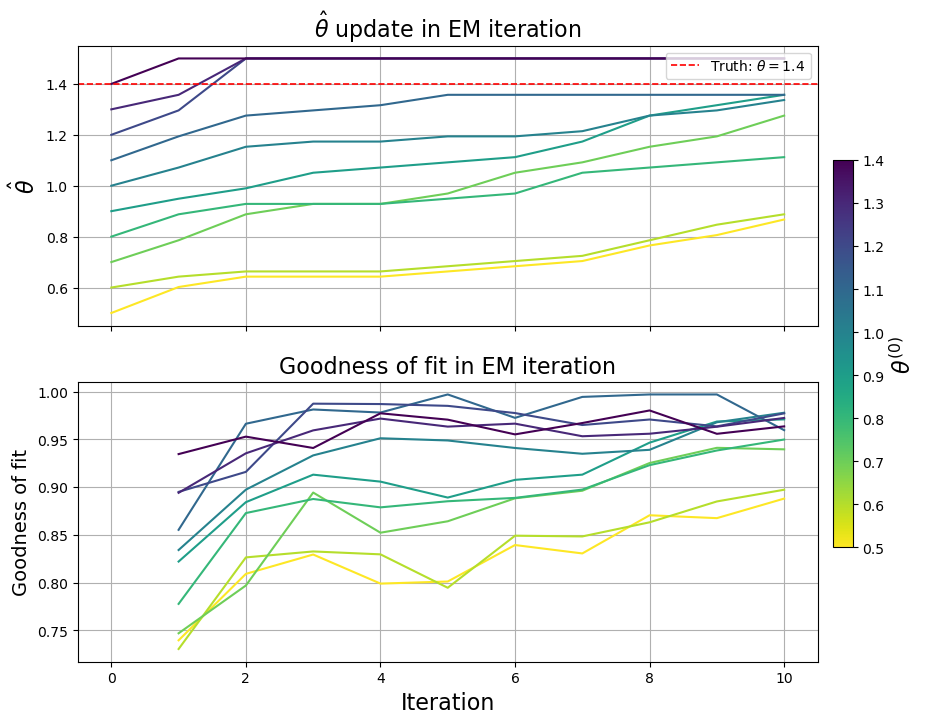}
        \caption{Evolution of the nuisance parameter and the step-1 classifier’s goodness-of-fit statistic for the POF algorithm under different initializations $\theta^{(0)}$. The results shown in Figures~\ref{fig:open_data_x_theta=1.4}--\ref{fig:open_data_y_theta=1.4} correspond to the solution yielding the highest goodness-of-fit statistic. \textbf{Top}: Updated estimates $\hat{\theta}$ across iterations for different initializations $\theta^{(0)}$. \textbf{Bottom}: Goodness-of-fit statistic of the step-1 classifier at each iteration.}
        \label{fig:open_data_theta_acc_evolution_theta=1.4}
    \end{figure}

\end{enumerate}

\newpage

\section{Technical Derivations}

\subsection{EM Derivation}
In this section, we provide a detailed derivation of the EM algorithm presented in Section~\ref{sec:EM_unfolding}.
\begin{proof}[Proof of Proposition 1]
\label{proof:proposition1}
    First, recall the population-level log-likelihood of the density function is 
\begin{align*}
    l(f) &= \int p(y)\log \left(\int p(y|x)f(x)dx\right)dy, \\ &\text{subject to } \int f(x)dx=1.
\end{align*}
The corresponding Q-function is
\begin{align*}
    Q(f,f^{(k)}) &= \int p(y)\int p(x|y,f^{(k)})\log p(x,y|f)dxdy, \\
    &\text{subject to } \int f(x)dx=1.
\end{align*}
To solve for $f^{(k+1)}=\arg\max_{f}Q(f,f^{(k)})$, we solve the problem in the Lagrangian form
\begin{align*}
    \Tilde{Q}(f,f^{(k)}) &= \int p(y)\int p(x|y,f^{(k)})\log p(x,y|f)dxdy - \lambda\left(\int f(x)dx-1\right).
\end{align*}
The Gâteaux derivative $\frac{\delta\tilde{Q}}{\delta f}$ satisfies
\begin{align*}
    \int\frac{\delta\tilde{Q}}{\delta f(x)}\phi(x)dx &= \left[\frac{d}{d\epsilon}\tilde{Q}(f+\epsilon\phi)\right]_{\epsilon=0} \\
    &=\left[\frac{d}{d\epsilon}\int p(y)\int p(x|y,f^{(k)})\log [p(y|x)(f(x)+\epsilon\phi(x))]dxdy - \lambda\int(f(x)+\epsilon\phi(x))dx-\lambda\right]_{\epsilon=0} \\
    &=\left[\int p(y)\int p(x|y,f^{(k)})\frac{p(y|x)\phi(x)}{p(y|x)(f(x)+\epsilon\phi(x))}dxdy - \lambda\int\phi(x)dx\right]_{\epsilon=0} \\
    &=\int p(y)\int p(x|y,f^{(k)})\frac{p(y|x)\phi(x)}{p(y|x)f(x)}dxdy - \lambda\int\phi(x)dx \\
    &=\int \phi(x) \left[\int p(y) p(x|y,f^{(k)})\frac{1}{f(x)}dy - \lambda\right] dx.
\end{align*}
Therefore, this shows that
\begin{align*}
    \frac{\delta\tilde{Q}}{\delta f(x)} &= \int p(y) p(x|y,f^{(k)})\frac{1}{f(x)}dy - \lambda.
\end{align*}
Setting the derivative to be 0,
\begin{align*}
    \frac{\delta\tilde{Q}}{\delta f(x)} &= \int \frac{p(x|y,f^{(k)})}{f(x)}p(y)dy - \lambda = 0 \\
    \lambda &= \int\frac{1}{f(x)}p(x|y,f^{(k)})p(y)dy \\
    &=\int\frac{1}{f(x)}\frac{p(y|x)f^{(k)}(x)}{\int p(y|x')f^{(k)}(x')dx'}p(y)dy.
\end{align*}
Integrating both sides by $\int f(x)dx$ yields that $\lambda=1$. Therefore, the stationary point satisfies
\begin{align*}
    f(x) = \int\frac{p(y|x)f^{(k)}(x)}{\int p(y|x')f^{(k)}(x')dx'}p(y)dy.
\end{align*}
Moreover, the second order derivative $\frac{\delta\tilde{Q}}{\delta f(x)\delta f(x')}$ satisfies
\begin{align*}
    \int\int\frac{\delta\tilde{Q}}{\delta f(x)\delta f(x')}\phi(x)\phi(x')dxdx' &= \left[\frac{d^2}{d\epsilon^2}\tilde{Q}(f+\epsilon\phi)\right]_{\epsilon=0} \\
    &=\left[\frac{d}{d\epsilon}\int p(y)\int p(x|y,f^{(k)})\frac{\phi(x)}{(f(x)+\epsilon\phi(x))}dxdy - \lambda\int\phi(x)dx\right]_{\epsilon=0} \\
    &=\left[-\int p(y)\int p(x|y,f^{(k)})\frac{\phi^2(x)}{\left[(f(x)+\epsilon\phi(x))\right]^2}dxdy \right]_{\epsilon=0} \\
    &=\int\frac{\phi^2(x)}{f^2(x)}\left(-\int p(y)p(x|y,f^{(k)})dy\right)dx.
\end{align*}
Therefore, this implies that
\begin{align*}
    \frac{\delta\tilde{Q}}{\delta f(x)\delta f(x')} = -\frac{\delta(x-x')}{f^2(x)}\int p(y)p(x|y,f^{(k)})dy \leq 0
\end{align*}
which shows that $\tilde{Q}$ is concave in $f$.

\end{proof}

The proof works similarly for \textsc{OmniFold} update by reparameterizing $f(x)=\nu(x)q(x)$. See details in \cite{Andreassen2020}.

\subsection{Profile \textsc{OmniFold} Derivation}
In this section, we provide a detailed derivation of the Profile \textsc{OmniFold} (POF) algorithm presented in Section~\ref{sec:pof}.
\begin{proof}[Proof of Proposition 2]
\label{proof:proposition2}
The Q-function in the POF algorithm is
\begin{align*}
    Q(\nu,\theta|\nu^{(k)},\theta^{(k)}) &= \int p(y)\int p(x|y,\nu^{(k)},\theta^{(k)})\log p(x,y|\nu,\theta)dxdy + \log p_0(\theta) \\
    &=\int p(y)\int p(x|y,\nu^{(k)},\theta^{(k)})\log[\mathcal{W}(y,x,\theta)q(y|x)\nu(x)q(x)]dxdy + \log p_0(\theta) \\
    &\text{subject to }\int \nu(x)q(x)dx=1.
\end{align*}
The Q-function can be decomposed into two parts that depend on $\nu$ and $\theta$ separately, i.e.
\begin{align*}
    Q(\nu,\theta|\nu^{(k)},\theta^{(k)}) &= \int p(y)\int p(x|y,\nu^{(k)},\theta^{(k)})\log[\nu(x)q(x)q(y|x)]dxdy \\
    &+ \int p(y)\int p(x|y,\nu^{(k)},\theta^{(k)})\log[\mathcal{W}(y,x,\theta)]dxdy + \log p_0(\theta) \\
    &= Q_1(\nu|\nu^{(k)},\theta^{(k)}) + Q_2(\theta|\nu^{(k)},\theta^{(k)}).
\end{align*}
Therefore, we can maximize $Q$ by maximizing $Q_1$ and $Q_2$ separately.
Write $Q_1$ in its Lagrangian form
\begin{align*}
    \tilde{Q}_1(\nu,|\nu^{(k)},\theta^{(k)}) &= Q_1(\nu|\nu^{(k)},\theta^{(k)}) - \lambda\left(\int \nu(x)q(x)dx-1\right). 
\end{align*}
Take derivative of $\tilde{Q}_1$ with respect to $\nu(x)$ and set it to be 0,
\begin{align*}
    \frac{\delta}{\delta\nu(x)}\tilde{Q}_1(\nu,|\nu^{(k)},\theta^{(k)})=\frac{\int p(y)p(x|y,\nu^{(k)},\theta^{(k)})dy}{\nu(x)} - \lambda q(x) = 0.
\end{align*}
Integrating both sides over $\int \nu(x)dx$ yields that $\lambda=1$.
Therefore, the stationary condition for $\nu(x)$ satisfies
\begin{equation*}
\label{eq:nu_stationary}
\begin{split}
    \nu(x) &= \frac{\int p(y)p(x|y,\nu^{(k)},\theta^{(k)})dy}{q(x)} \\
    &= \int \frac{p(y)\mathcal{W}(y,x,\theta^{(k)})q(y|x)\nu^{(k)}(x)dy}{\int \mathcal{W}(y,x',\theta^{(k)})q(y|x')\nu^{(k)}(x')q(x')dx'} \\
    &= \nu^{(k)}(x)\int\frac{p(y)}{\tilde{q}^{(k)}(y)} \mathcal{W}(y,x,\theta^{(k)})q(y|x)dy,
\end{split}
\end{equation*}
where $\tilde{q}^{(k)}(y)=\int \mathcal{W}(y,x',\theta^{(k)})q(y|x')\nu^{(k)}(x')q(x')dx'$. Moreover, since 
\begin{align*}
    \frac{\delta}{\nu(x)\nu(x')}\tilde{Q}(\nu,\theta^{(k)}|\nu^{(k)},\theta^{(k)})=-\frac{\delta(x-x')}{\nu^2(x)}\int p(y)p(x|y,\nu^{(k)},\theta^{(k)})dy \leq 0,
\end{align*}
the stationary point is a global maximum.
Now for $Q_2$, we have
\begin{align*}
    Q_2(\theta|\nu^{(k)},\theta^{(k)}) &= \int p(y)\int p(x|y,\nu^{(k)},\theta^{(k)})\log[\mathcal{W}(y,x,\theta)]dxdy + \log p_0(\theta).
\end{align*}
Hence,
\begin{align*}
    \text{argmax}_\theta Q_2(\theta|\nu^{(k)},\theta^{(k)}) &= \text{argmax}_\theta \int p(y)\int p(x|y,\nu^{(k)},\theta^{(k)})\log[\mathcal{W}(y,x,\theta)]dxdy + \log p_0(\theta) \\
    &= \text{argmax}_\theta \int \int p(y)\frac{\mathcal{W}(y,x,\theta^{(k)})q(y|x)\nu^{(k)}(x)q(x)}{\tilde{q}^{(k)}(y)}\log[\mathcal{W}(y,x,\theta)]dxdy + \log p_0(\theta) \\
    &= \text{argmax}_\theta \int \int q(x,y)\nu^{(k)}\mathcal{W}(y,x,\theta^{(k)})\frac{p(y)}{\tilde{q}^{(k)}(y)}\log[\mathcal{W}(y,x,\theta)]dxdy + \log p_0(\theta).
\end{align*}
\end{proof}

\subsection{$\mathcal{W}$ Function Training Through Classifiers}
In this section, we present a detailed proof of the classifier-based estimation of the $\mathcal{W}$ function introduced in Section~\ref{sec:pof} for the POF algorithm.
\begin{proof}[Proof of Proposition 3]
    First, rewrite
\begin{align*}
    \mathcal{W}(y,x,\theta)&=\frac{p(y|x,\theta)}{q(y|x)}\\
    &= \frac{p(x,y|\theta)}{p(x|\theta)}\cdot\frac{q(x,y)}{q(x)}\\
    &=\frac{p(x,y|\theta)}{q(x,y)}\cdot\frac{q(x)}{p(x|\theta)}.
\end{align*}
For the first ratio, let $f_1:\mathcal{X}\times\mathcal{Y}\times\Theta\rightarrow[0,1]$ be the Bayes optimal classifier to distinguish dataset $\mathcal{D}_1=\{X_i,Y_i,\theta_i\}$ from $\mathcal{D}_2=\{X'_i,Y'_i,\theta_i'\}$. Then the learned ratio from the classifier is
\begin{align*}
    \frac{f_1(x,y,\theta)}{1-f_1(x,y,\theta)}=\frac{p(x,y,\theta)}{q(x,y,\theta)} &= \frac{p(x,y|\theta)p(\theta)}{q(x,y|\theta)q(\theta)} \\
    &= \frac{p(x,y|\theta)p(\theta)}{q(x,y)q(\theta)}
\end{align*}
where the last line follows since $q(x,y|\theta)=q(x,y)$, i.e. the joint distribution of $X_i',Y_i'$ does not depend on $\theta_i'$. Similarly, for the second ratio, let $f_2:\mathcal{X}\times\Theta\rightarrow[0,1]$ be the Bayes optimal classifier to distinguish dataset $\tilde{\mathcal{D}}_2=\{X'_i,\theta'_i\}$ from $\tilde{\mathcal{D}}_1=\{X_i,\theta_i\}$. Then the learned ratio from the classifier is
\begin{align*}
    \frac{f_2(x,\theta)}{1-f_2(x,\theta)}=\frac{q(x,\theta)}{p(x,\theta)} &= \frac{q(x|\theta)q(\theta)}{p(x|\theta)p(\theta)} \\
    &= \frac{q(x)q(\theta)}{p(x|\theta)p(\theta)}
\end{align*}
where again the last line follows since the distribution of $X_i'$ does not depend on $\theta_i'$. Therefore, combining these, we have
\begin{align*}
    \frac{f_1(x,y,\theta)f_2(x,y,\theta)}{(1-f_1(x,y,\theta))(1-f_2(x,y,\theta))} &= \frac{p(x,y|\theta)p(\theta)}{q(x,y)q(\theta)} \cdot \frac{q(x)q(\theta)}{p(x|\theta)p(\theta)} \\
    &= \frac{p(x,y|\theta)}{q(x,y)}\cdot\frac{q(x)}{p(x|\theta)}.
\end{align*}
\end{proof}

\section{Data and Codes Access}
The data and code used to produce the results are available in the following GitHub repository: \url{https://github.com/richardzhs/ProfileOmnifold}